# Lo Gnomone Clementino

# Astronomia Meridiana in Basilica

Costantino Sigismondi
(ICRANet e Observatório Nacional, Rio de Janeiro)

## Prefazione

Costruito per chiara volontà del papa 70 anni dopo il caso Galileo, lo Gnomone Clementino è un grande telescopio solare che non fa uso di lenti a 92 anni dall'invenzione del cannocchiale. Queste due caratteristiche basterebbero da sole a giustificare l'interesse verso questo strumento.
L'astronomia meridiana è alla base dell'astrometria e dell'astrofisica moderna. Lo Gnomone Clementino sta oggi all'astronomia, come il veliero "Amerigo Vespucci" sta alla Marina Italiana. E' possibile svolgere ogni genere di osservazione e studio su questo strumento, e dal 2002 vi tengo lezioni teorico-pratiche del corso di Storia dell'Astronomia e La Terra nel Sistema Solare della Sapienza, Università di Roma, Facoltà di Lettere e Filosofia. Questo testo aggiunge alcuni tasselli alla ricerca storica sulla meridiana, appoggiandosi, com'è naturale, sulle spalle di giganti che mi hanno preceduto in questi studi.

In particolare la misura dell'azimut della meridiana, ed il suo inquadramento tra gli strumenti simili ed alcuni studi di astrometria sui dati del 1701-1703 di Bianchini, che sono apparsi fin'ora soltanto su riviste specializzate ed in Inglese vengono qui proposti in Italiano e semplificati, per valorizzare sempre più questa straordinaria opera d'arte e di scienza.

# Introduzione

In occasione dell'anno internazionale dell'astronomia, il 2009, in cui ricorre il quarto centenario delle prime osservazioni di Galileo Galilei con il cannocchiale, è opportuno tornare ad esaminare uno strumento di eccezionale versatilità, come lo Gnomone Clementino, inaugurato nel 1702.

Lo strumento fu fortemente voluto e poi finanziato da papa Clemente XI, tra i suoi primi atti da pontefice. Egli era stato ordinato sacerdote solo il 27 settembre 1700; il 6 ottobre 1700 aveva celebrato la sua prima messa solenne nella Basilica di Santa Maria degli Angeli, allora sede della Certosa, nel giorno di S. Bruno, fondatore dei Certosini. Eletto papa il 23 novembre chiese subito a Francesco Bianchini, astronomo e archeologo veronese, allievo di Montanari, di realizzare lo gnomone, e Bianchini cominciò i lavori il primo gennaio 1701, primo giorno utile per osservare entrambi i transiti meridiani della Polare durante la lunga notte invernale.

La storia che ne precede la realizzazione è densa di sorprese, ed in questo volumetto vogliamo avvicinare i lettori ai vari aspetti sia tecnici che culturali, rimasti sino ad ora appannaggio solo di pochi addetti ai lavori e cattedratici.

Lo Gnomone Clementino, noto anche come Meridiana di Santa Maria degli Angeli, è uno strumento che consente di raggiungere precisioni di un secondo d'arco nella posizione del Sole, pur non essendo provvisto di alcuna ottica.

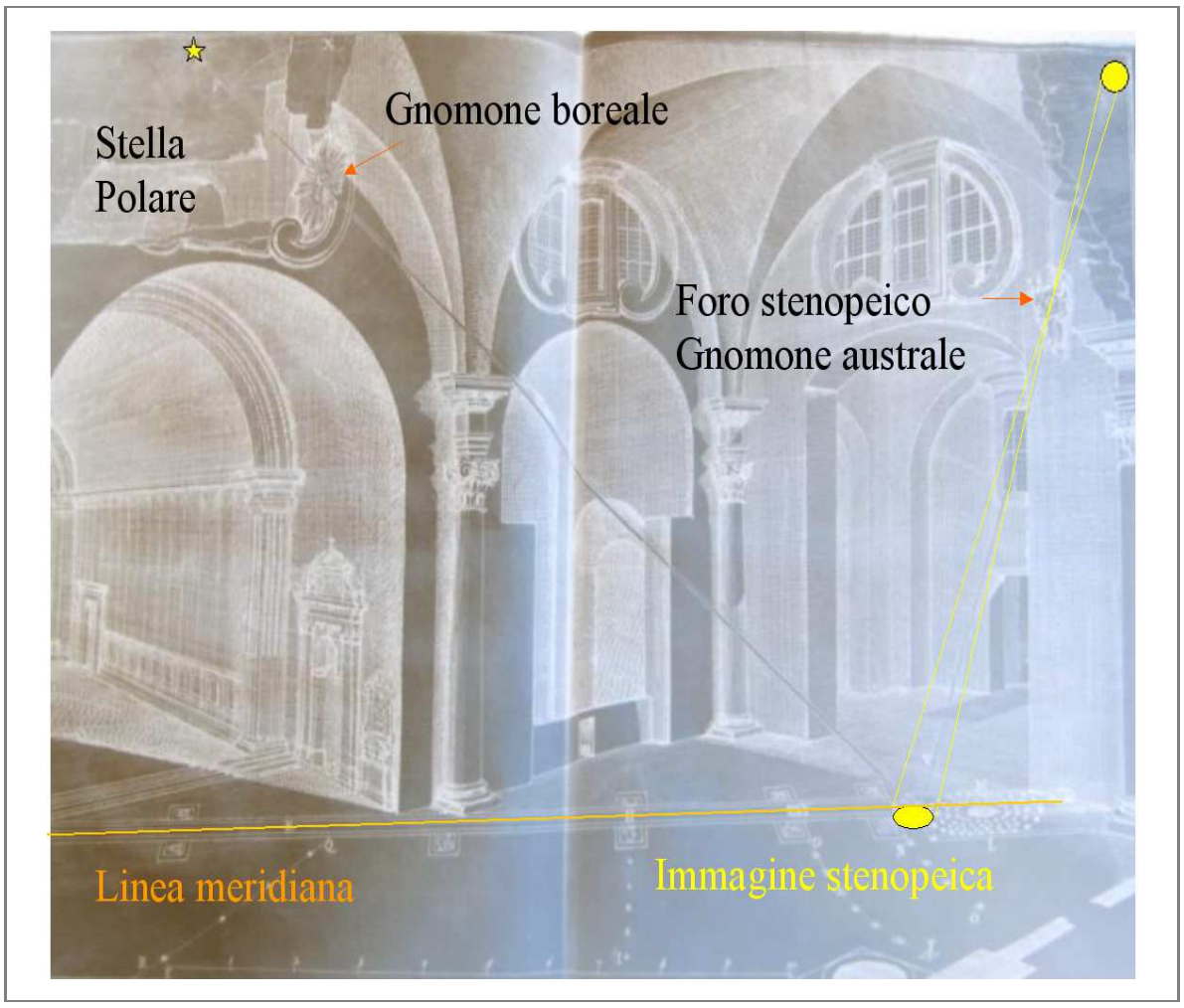

Potrebbe sembrare strano che proprio nella città dove lavoravano gli ottici più all'avanguardia del XVII secolo: Eustachio Divini e Giuseppe Campani, che aveva il suo atélier a Piazza Navona, si decida di costruire un grande telescopio solare senza lenti. Eppure ci sono delle ragioni pratiche molto forti, che andiamo ad esaminare.

D'altro canto nei telescopi solari moderni sono state adottate soluzioni tecnologiche spesso molto diverse dai telescopi dedicati alle osservazioni notturne, prima fra tutte la costruzione di Torri solari, dove l'obiettivo fosse il più lontano possibile dal suolo che si riscalda durante il giorno e genera turbolenza atmosferica, possibilmente circondate da uno specchio d'acqua.

Le premesse teorico-pratiche per lo sviluppo di uno strumento del genere risalgono ad Aristotele e alla sua cerchia quanto al fenomeno dell'immagine del Sole prodotta, a sufficiente distanza, da un foro di qualunque forma (Pseudo-Aristotele, Problemata III, 15). Ulugh Begh a Costantinopoli costruì nella Basilica di Santa Sofia il primo gnomone a foro stenopeico nel 1437 per uso astronomico. Non si può escludere che l'orologio solare di Gerberto (poi divenuto papa Silvestro II 999-1003) realizzato a Magdeburgo potesse essere dello stesso genere. Paolo del Pozzo Toscanelli a Firenze realizzò nel 1475 lo gnomone nella lanterna del Duomo a 90 metri d'altezza, che resta ancora oggi il più alto del Mondo. Un secolo dopo Egnazio Danti, domenicano, poi vescovo di Alatri, realizzò strumenti simili a Santa Maria Novella a Firenze e a S. Petronio a Bologna, dove nel 1655 Cassini tracciò la sua celebre meridiana con cui "domò i cavalli del Sole" e si conquistò la fiducia del Re Sole che lo chiamò a dirigere l'Osservatorio di Parigi.

La fortuna degli strumenti a foro stenopeico, in principio realizzabili in qualunque epoca storica, ebbe un sussulto subito dopo l'invenzione del telescopio galileiano proprio per l'osservazione delle macchie solari. Keplero stesso osservò le macchie solari senza l'uso di lenti, ma solo con un foro stenopeico che proiettava l'immagine in una camera oscura.

Per questo l'etimologia della parola telescopio, strumento con cui si può "guardare lontano" consente di applicare questo termine sia a strumenti dotati di parti ottiche rifrangenti o riflettenti, sia al nostro gnomone che non ne ha.

Mentre il termine gnomone ha la stessa radice di gnosi, "conoscenza", cioè è uno strumento mediante il quale si può conoscere il moto del Sole.

Infine possiamo citare anche il termine astrolabio, il cui significato è strumento che "prende" gli astri, misurandone quindi altezza e azimut, e calcolandone orari di levata e tramonto.

Il foro stenopeico, etimologicamente "apertura stretta", è il cuore dello strumento, che raccoglie i raggi del Sole e ne proietta l'immagine fino al lontano tropico, "punto di svolta" del Capricorno, che nella Basilica di Santa Maria degli Angeli sta a quasi 50 metri di distanza.

La linea meridiana materializza sul pavimento la "Via Solis" che il Sole percorre per sei mesi in una direzione e per gli altri sei nella direzione opposta. Il successo della teoria Copernicana o Kepleriana nei confronti del sistema di Tolomeo poteva essere provato solo da strumenti come questo, capaci di misurare la durata dell'anno tropico, con la precisione di un secondo e di vagliare, quindi, la bontà della riforma Gregoriana del calendario.

Con questi obbiettivi in mente, e la misura della variazione dell'obliquità dell'orbita terrestre, ovvero l'inclinazione dell'asse terrestre sul piano dell'orbita, Francesco Bianchini si mise a realizzare uno strumento, che tra tutte le meridiane a foro stenopeico, resta un *unicum* sia come soluzioni tecniche che per realizzazione artistica. Per questo da tutto il Mondo arrivano in continuazione turisti e studiosi per ammirarlo.

# Gnomone Clementino

Questo è il nome che Francesco Bianchini diede allo strumento. Il ruolo chiave svolto dal papa Clemente XI, Gianfrancesco Albani, non era stato ancora messo abbastanza in evidenza nei testi pubblicati fino ad ora.

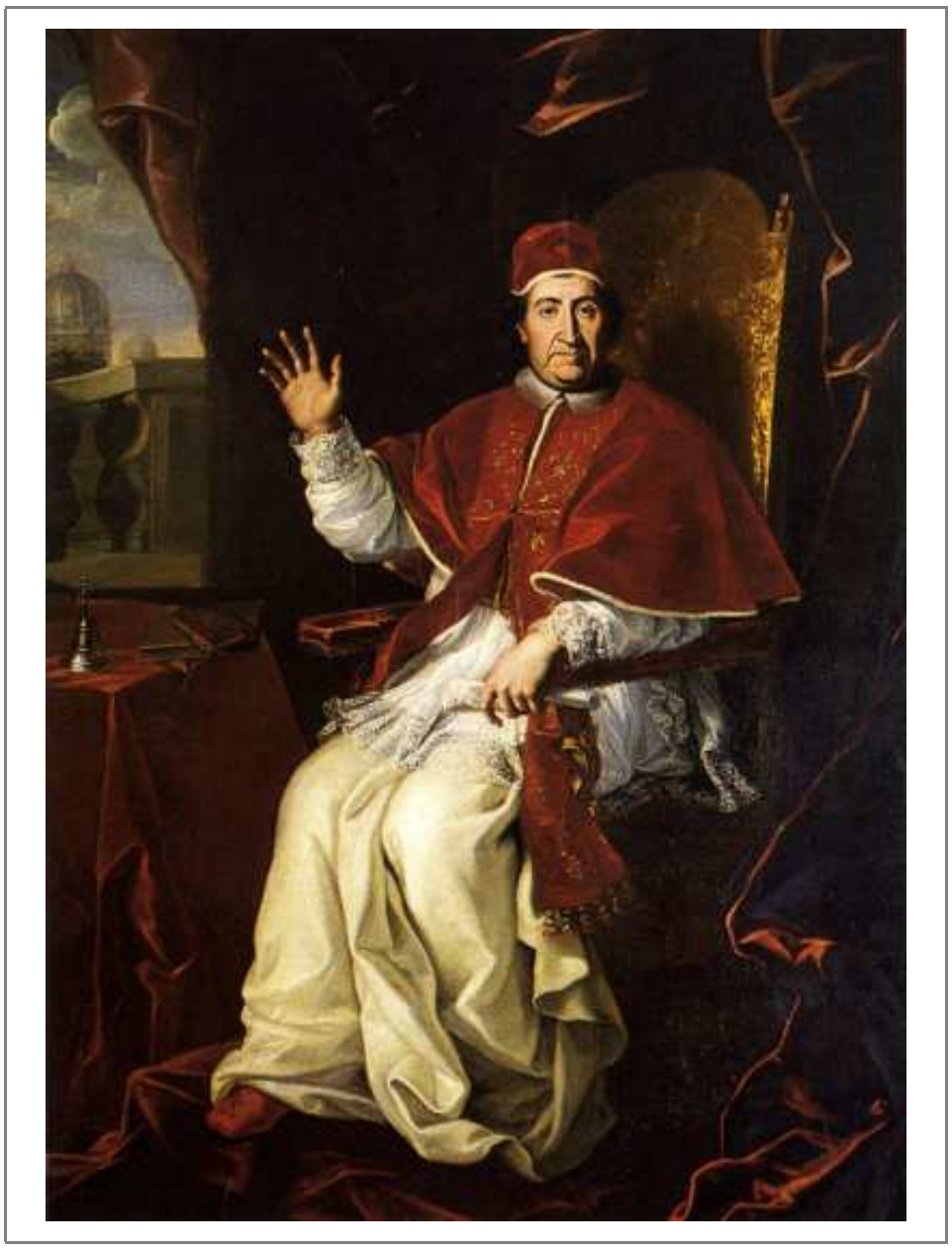

Il legame tra l'Albani e Santa Maria degli Angeli è ancora più forte di quello che oggi Benedetto XVI mostra di avere con la Basilica di S. Paolo dove sono i monaci Benedettini. Il 6 ottobre è la festa di S. Bruno, e proprio in quella data il papa venne ad inaugurare la meridiana nel 1702, secondo anniversario della sua prima messa solenne ivi celebrata. Si tratta di qualcosa di più che puro mecenatismo illuminato. Queste notizie sono state desunte dai carteggi del Bianchini conservati alla biblioteca Vallicelliana in Roma.

# Francesco Bianchini

Veronese (1662-1729), studiò a Padova, svolse il ruolo di soprintendente dei beni archeologici di Roma. Conobbe Newton, ed oltre alle misure con lo gnomone clementino, pubblicò studi su Venere e contribuì alla realizzazione della prima carta geografica dello stato pontificio.

Insieme alle osservazioni fatte allo gnomone bolognese poté correggere le longitudini relative di Bologna e Roma, "raddrizzando" l'Italia tolemaica che come tutta l'Ecumène si estendeva molto di più del vero in longitudine: 14° contro 10°.

Infatti in tutte le rappresentazioni antiche dell'Italia, l'asse dello stivale è molto inclinato nella direzione Est-Ovest.

Bianchini fu nominato canonico di Santa Maria Maggiore, dove è sepolto, prese solo gli ordini minori di suddiacono.

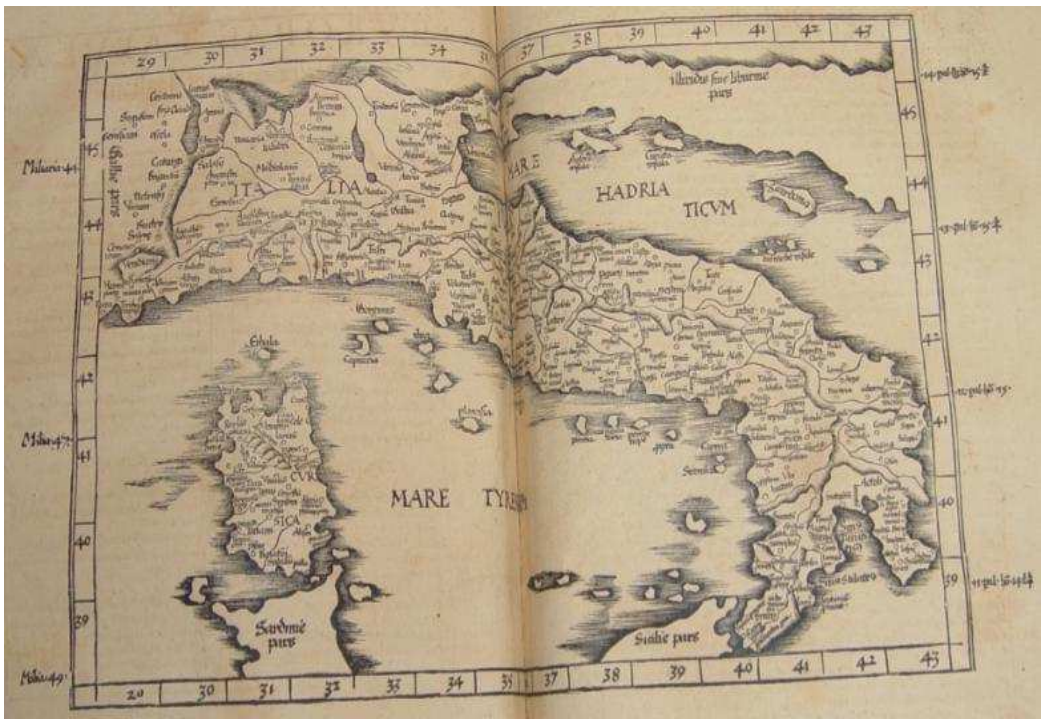

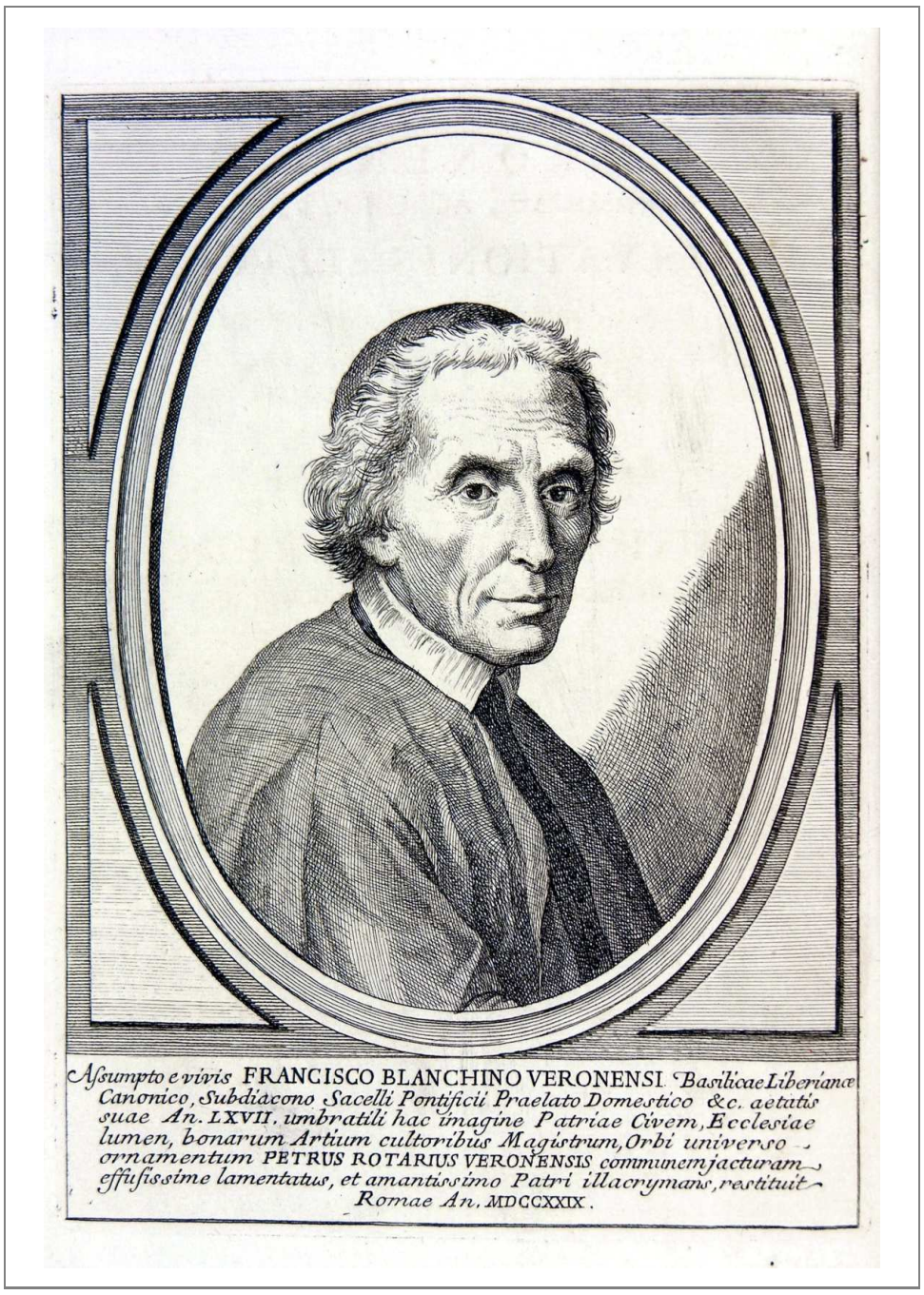

# Principi di funzionamento

Il cuore dello Gnomone è il foro stenopeico. Lì passano i raggi del Sole che producono l'immagine che si muove sul pavimento e che viene studiata dagli astronomi. A Roma già Augusto aveva provveduto nel 10 a. C. a realizzare un grande gnomone nel Campo Marzio, l'Horologium Augusti, dove era l'ombra prodotta dalla punta dell'Obeliscus Solis a fornire le indicazioni astronomiche.

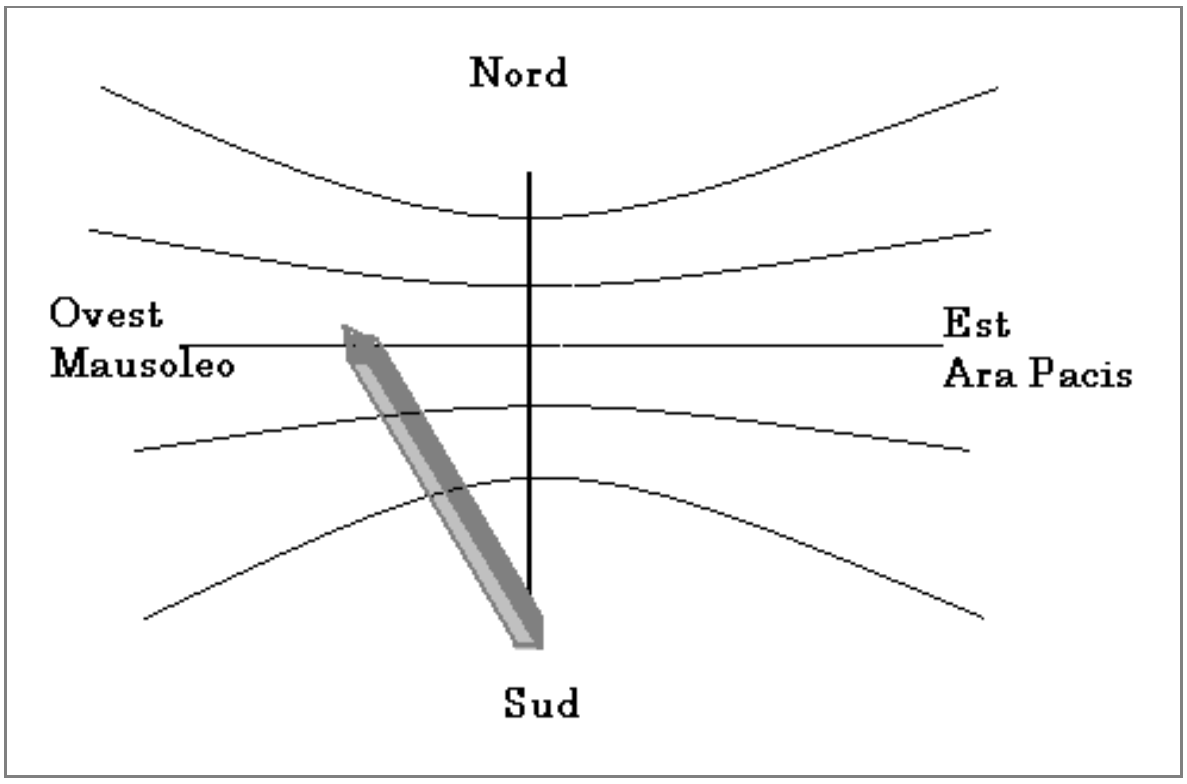

Questo venerando strumento, menzionato da Plinio nella Naturalis Historia, andò progressivamente in rovina, fino a cadere, forse in occasione del terremoto del 1048. Fu rinvenuto sotto papa Benedetto XIV Lambertini nel 1748 e ricollocato davanti alla Curia Innocenziana (palazzo di Montecitorio) da papa Pio VI Braschi nel 1792, con tanto di linea meridiana.

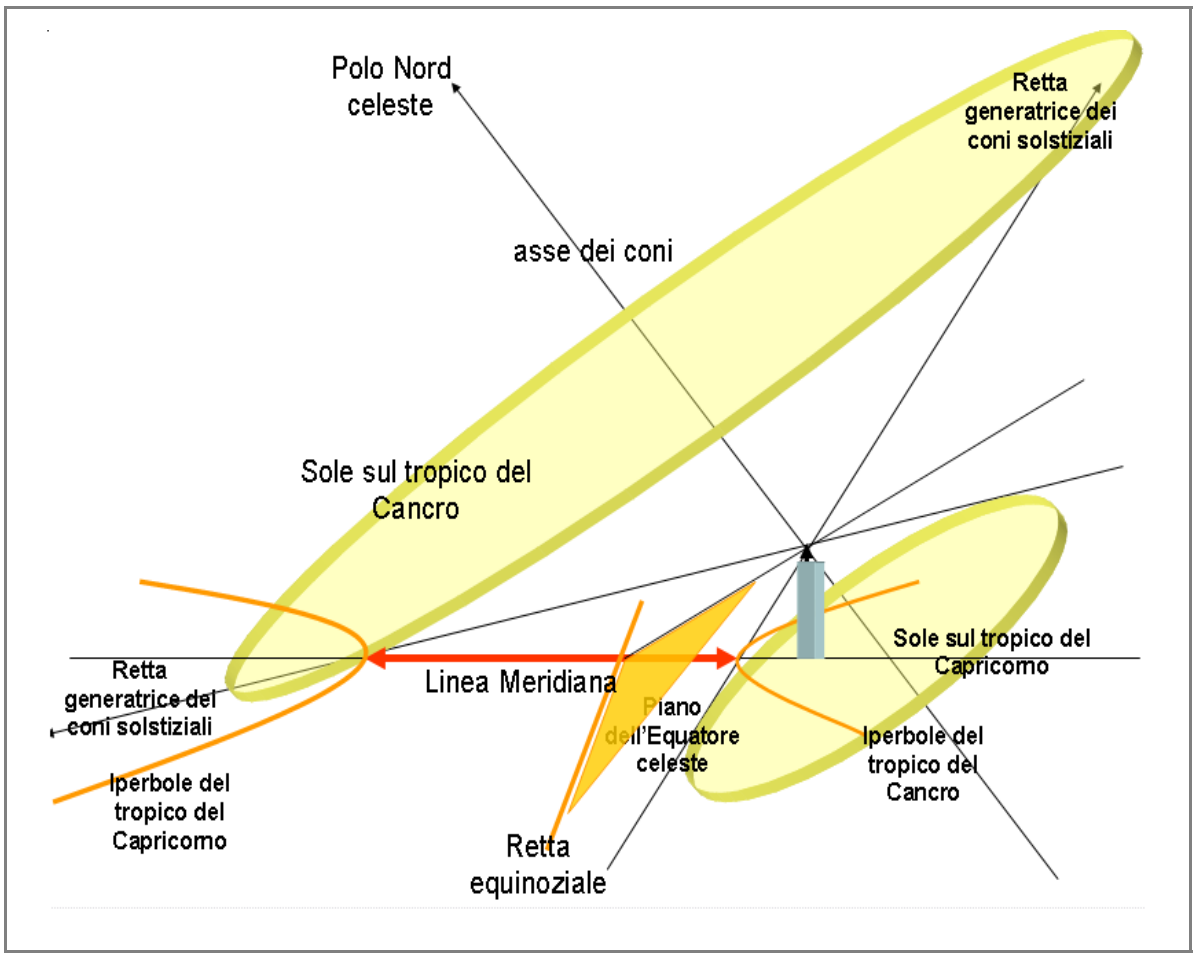

Questo schema mostra come attraverso lo Gnomone, che è sia la punta dell'obelisco "ad deprehendedas umbras" o il foro stenopeico si proiettino a terra i circoli della sfera celeste: il meridiano locale che diventa la linea meridiana, i tropici che definiscono i solstizi e l'equatore celeste che diventa una linea perpendicolare alla meridiana.

I percorsi quotidiani dell'immagine del Sole nelle chiese, o dell'ombra della punta dell'obelisco, sono rami di iperbole con concavità opposte sopra e sotto l'equatore celeste.

Sul pavimento della basilica sono stati riportati con sentieri di stelle le iperboli corrispondenti a Sirio, Arturo e al Sole il 20 agosto 1702 quando Clemente XI venne a visitare lo Gnomone.

# Come è stata tracciata la meridiana

La scelta della Basilica di Santa Maria degli Angeli fu dettata anche da motivi di ordine pratico. La Basilica era abbastanza grande da poter ospitare una linea lunga 45 metri e le mura, che appartenevano alle Terme di Diocleziano, dopo 14 secoli non avevano più movimenti di assestamento.

Giandomenico Cassini a Bologna aveva dovuto ricollimare dopo quaranta anni il suo famoso gnomone poiché la posizione del foro stenopeico, posto nella volta di una campata laterale di San Petronio, era leggermente mutata.

Leonardo Ximenes, gesuita, che ripristinò nel 1761 con gli auspici del granduca di Toscana lo gnomone del Toscanelli, verificò gli stessi problemi per la cupola di Santa Maria del Fiore: questa si muoveva leggermente nel corso delle stagioni, quanto bastava per influire sulle misure fatte con lo gnomone.

Per tracciare la meridiana, una volta collocato il foro stenopeico in un vano apposito a 20.344 m di quota sul pavimento della chiesa, Bianchini doveva conoscere la verticale del foro, la direzione Nord – Sud e la latitudine del luogo.

Per la verticale usò il filo a piombo, smorzandone nell'acqua le piccole oscillazioni.

# Azimut

Per determinare la direzione Nord – Sud, Bianchini si valse di una meridiana ausiliaria che era a Palazzo Venezia, controllata dall'astronomo francese Giacomo Filippo Maraldi e quando su questa avveniva il transito del Sole egli fissava sul pavimento della Basilica la posizione del Sole. In questo modo, se la meridiana ausiliaria aveva un errore di orientamento questo venne trasferito anche a quella di Santa Maria degli Angeli.

Il metodo usato da Cassini a Bologna, quello di adoperare il transito del Sole su cerchi di uguale altezza, non era praticabile a Santa Maria degli Angeli poiché sia la conformazione del vano dove è alloggiato il foro stenopeico, sia la posizione della linea meridiana rispetto alla crociera e alle colonne, non consentono di vedere più l'immagine del Sole passati alcuni minuti dopo il mezzogiorno locale.

Il metodo dei cerchi di uguale altezza, invece, richiede l'osservabilità del Sole sia prima che dopo il meridiano, ed è per questa ragione che Bianchini usò una meridiana ausiliaria.

La deviazione dal Nord, verso l'Est, è di 4′ 28.8″ ± 0.6″della retta che parte dalla verticale originale del foro e giunge al punto estremo della linea nel Capricorno, e riflette probabilmente l'errore già presente nella meridiana ausiliaria. Altre meridiane storiche mostrano deviazioni anche maggiori, che tendono a diminuire man mano che il calcolo delle effemeridi Kepleriane si stabilizza.

Le misure della seguente tabella sono state prese per la prima volta o ricontrollate dall'autore, salvo per la meridiana di Tycho di cui esiste una testimonianza della spedizione all'isola di Hven in Danimarca organizzata appositamente dall'Accademia di Francia guidata dall'astronomo abate Jean Picard.

Questa stessa misura viene citata da Ruggero Boscovich, astronomo gesuita che già nel 1750 aveva valutato correttamente la deviazione dal Nord della linea clementina.

| | |
|---|---|
| Toscanelli 1475 (Duomo Firenze) | -19′ 54″ |
| Tycho Brahe 1571 (Uraniborg, Hven) | -17′ |
| Cassini 1655 (San Petronio, Bologna) | +1′ 36.6″ |
| Bianchini 1702 (Santa Maria degli Angeli) | +4′ 28.8″±0.6″ |
| Celsius 1734 (Santa Maria degli Angeli) | +2′ |
| Boscovich e Maire 1750 (S. M. degli Angeli) | +4′ 30″ |
| Ximenes 1761 (Duomo Firenze) | -27″ |
| De Cesaris 1786 (Duomo Milano) | +0′ 07″ |
| Gigli 1817 (Piazza S. Pietro, Roma) | -5′36″ |

**Tabella 1.** Deviazione dal Nord delle principali linee meridiane storiche

Nella tabella ho inserito anche le ricognizioni storiche del Celsius e dei gesuiti Boscovich e Maire, che studiarono a fondo la meridiana. Celsius confrontava i dati della Basilica con una sua meridiana allestita in una sala del Quirinale. Essi dovevano confrontare gli istanti dei passaggi del Sole con delle effemeridi (tabelle di passaggi del Sole al meridiano), che evidentemente diventavano sempre più accurate col passare degli anni. Bianchini usava quelle pubblicate a Bologna da Flaminio Mezzavacca, dove i metodi di calcolo dell'equazione di Keplero erano ancora un po' imprecisi. Boscovich e Maire riportarono che i transiti solari ai solstizi estivi ed invernali avvenivano rispettivamente con 5 e 17 secondi di ritardo rispetto al tempo segnato da una meridiana realizzata in modo accurato partendo da principi primi, come il metodo dei cerchi di uguale altezza, corretto per le variazioni orarie di declinazione del Sole.

La figura seguente riassume questa sinossi: la precisione dell'allineamento delle meridiane con il vero Nord migliora esponenzialmente con il tempo, e già Ruggero Giuseppe Boscovich aveva raggiunto il valore migliore.

Rispetto alla tabella precedente sono aggiunti i dati della meridiana della Torre dei Venti in Vaticano; di Le Monnier nella Chiesa di Saint-Sulpice a Parigi; di Cassella al Museo Archeologico Nazionale di Napoli e di Calandrelli nella Torre omonima all'Osservatorio del Collegio Romano ora UCEA a Roma.

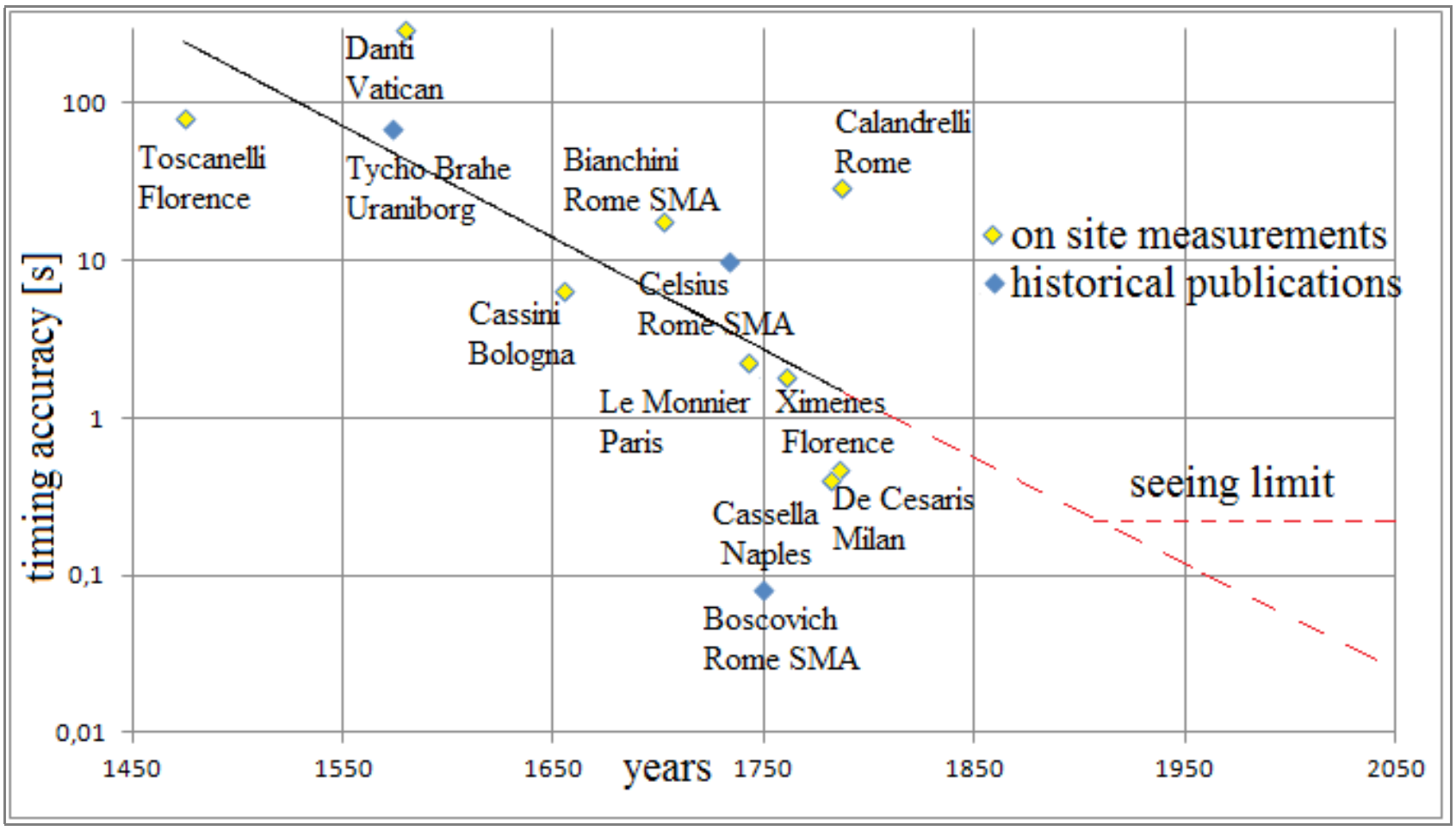

I ritardi della meridiana di Santa Maria degli Angeli possono essere verificati anche oggi confrontando gli istanti dei passaggi calcolati con le effemeridi con quelli misurati con un orologio radio controllato: si ritrovano valori vicini a quelli presi dai due padri gesuiti.

I dati di S. Petronio sono in Paltrinieri (2001), quelli del Duomo di Milano in Ferrari da Passano, Monti e Mussio (1977). Quelli di S. Maria del Fiore sono stati rilevati dall'autore il 20 giugno 2008. La ricognizione del Celsius è riportata in Monaco (1990) mentre quella di Boscovich in una sua lettera a Vallisneri del 1773.

# Latitudine

Per conoscere la latitudine del luogo Bianchini dovette realizzare la meridiana boreale, da cui osservare direttamente da dentro la Basilica la stella Polare.

Con il telescopio, durante la prima settimana di Gennaio del 1701, grazie alla lunghezza delle notti e al tempo siderale, egli poté osservare e misurare entrambi i transiti meridiani della stella Polare: quello superiore alle 6 di sera e quello inferiore alle 6 del mattino.

Utilizzando poi le tavole di Cassini per la correzione della rifrazione atmosferica, Bianchini poté valutare la posizione del Polo Nord Celeste apparente, che sta a metà tra la culminazione –o transito– superiore e quella inferiore della Polare, con una precisione che oggi sappiamo essere stata di 1 secondo d'arco, il meglio possibile per gli strumenti dell'epoca.

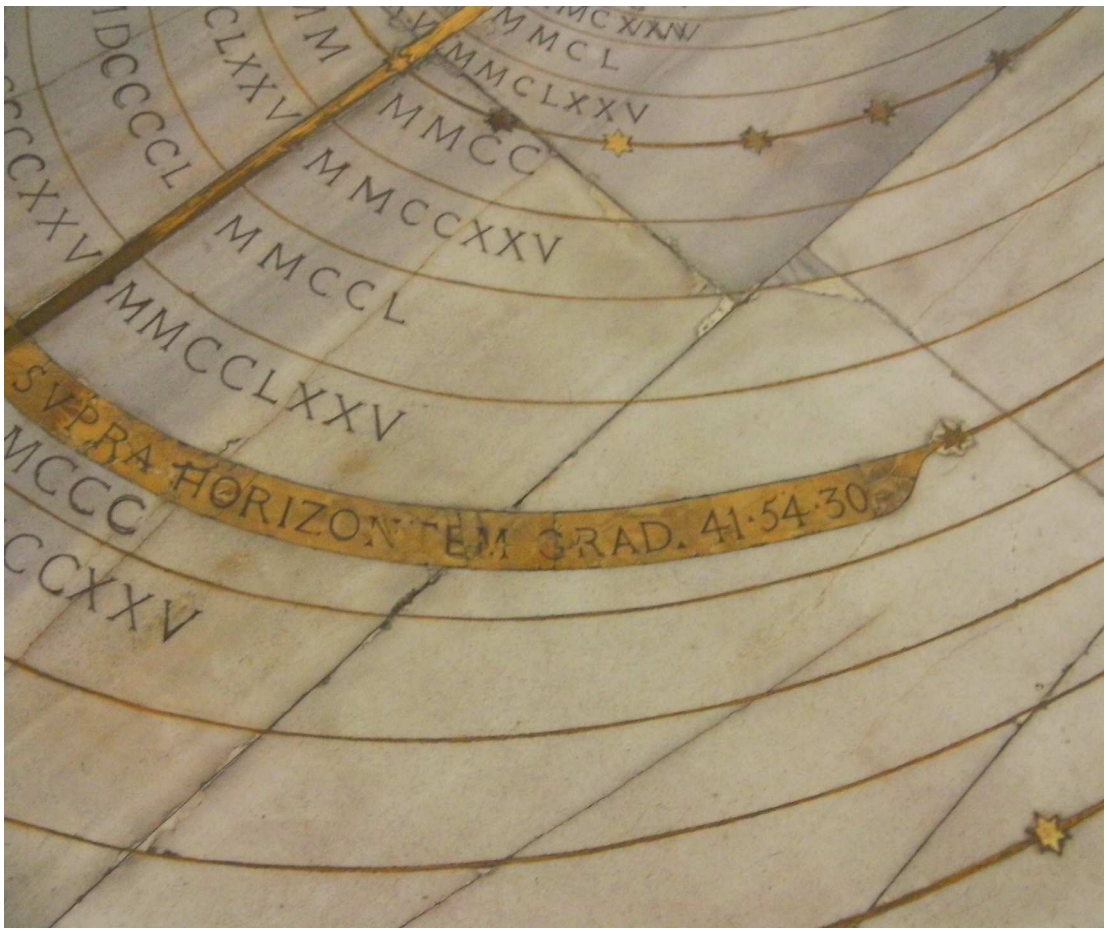

Il polo del mondo si eleva sopra l'orizzonte di 41° 54′ 30″, così fece scrivere in Latino sul marmo che rappresenta le orbite quotidiane della stella Polare durante tutti gli anni giubilari dal 1700 al 2500. Il dato che pubblicò nel suo testo De Nummo et Gnomone Clementino (1703) era leggermente inferiore 41° 54′ 27″, e corrisponde entro 1″ alla posizione del polo apparente per la prima settimana di gennaio.

Ho parlato di polo apparente poiché, a causa dell'aberrazione della luce, scoperta da James Bradley nel 1727, ogni stella compie in un anno un'orbita di raggio 21″ attorno alla sua posizione media, che sarebbe quella che vedremmo se la Terra fosse ferma al centro dell'Universo come nel modello di Tolomeo. Bianchini non conosceva ancora questo effetto, che fu la prima vera prova dinamica della rivoluzione terrestre.

L'aberrazione della luce è il maggiore degli effetti di Relatività speciale. Le misure allo Gnomone Clementino lo evidenziano.

Oggi con il GPS (Garmin II plus, valori mediati su numerosi dati) le coordinate, riferite all'ellissoide WGS84 sono:

Latitudine 41° 54′ 11.2″ N

Longitudine 12° 29′ 50.9″ E

La quota sul livello del mare del foro stenopeico è 70 m, dato di origine topografica, più preciso del dato GPS.

Se Bianchini avesse fatto le misure dei transiti della Polare 6 mesi dopo avrebbe trovato un valore della latitudine diverso, inferiore a quello del GPS della stessa quantità di cui il dato di gennaio lo superava. Ma il primo luglio il transito superiore cadeva alle 17:38 in pieno giorno, e quello inferiore alle 5:36, al mattino, condizioni difficili per osservare la Polare.

## Parti Centesime e Tangenti

Per stabilire un sistema di riferimento sulla linea meridiana, seguendo l'esempio della meridiana di Bologna, Bianchini riportò le parti centesime dell'altezza del foro stenopeico, di modo che in corrispondenza del numero 100 la linea e la verticale del foro formano un triangolo rettangolo isoscele, ed il foro visto da quel punto si eleva con un angolo di 45°.

In questo modo la linea arriva fino al numero 220, che corrisponde all'altezza del bordo inferiore del Sole nella sua culminazione più bassa di tutto l'anno: al solstizio d'Inverno.

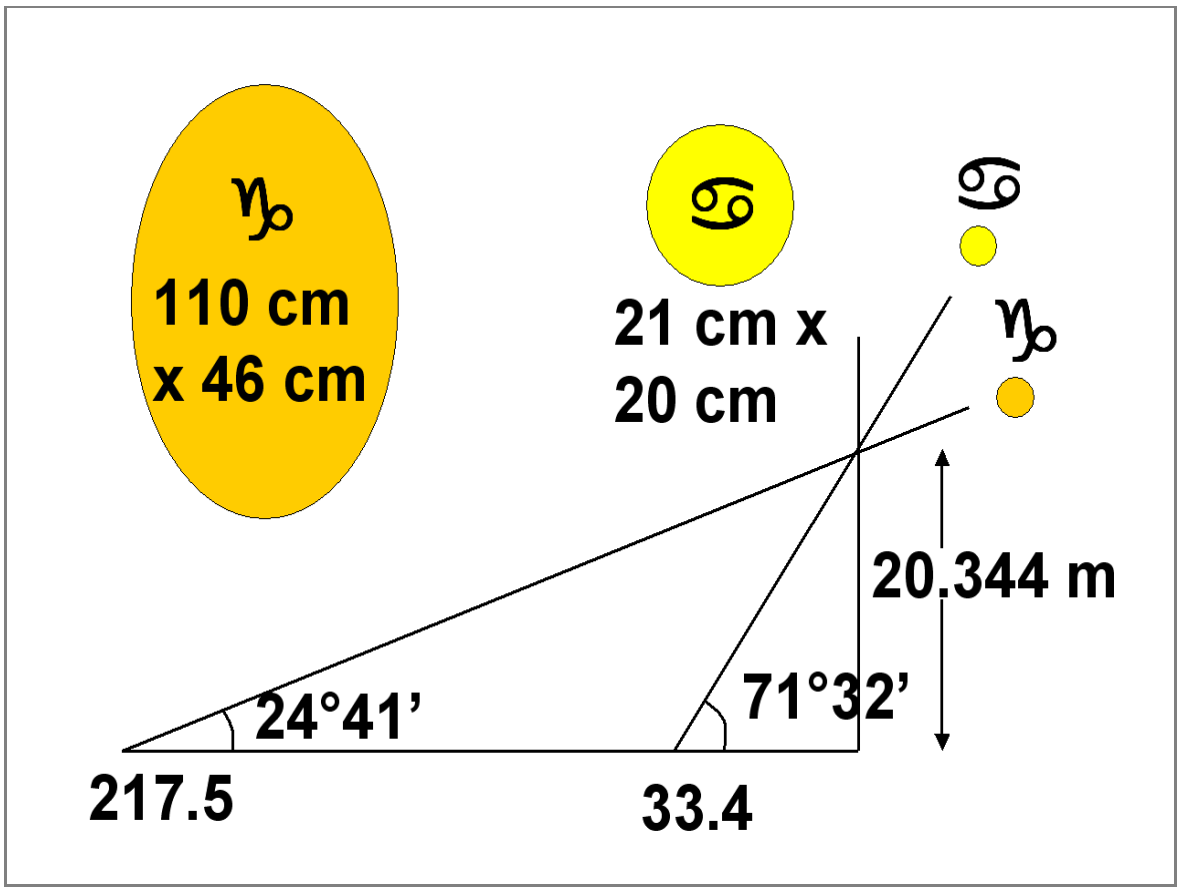

L'angolo formato con il pavimento, dall'ipotenusa del triangolo che ha come cateti la verticale del foro alta 100 e la linea lunga 217.5, vale arcotangente(100/217.5)=24° 41'.

La corrispondente distanza dallo zenit vale invece arcotangente(217.5/100)=65° 19'.

Le misure allo Gnomone Clementino si basano tutte sulla funzione trigonometrica della tangente, per questo lo gnomone è un vero e proprio monumento a questa funzione trigonometrica, insieme al corrispondente triangolo rettangolo di riferimento che è pienamente materializzato quando al momento del transito meridiano i raggi del Sole ne descrivono l'ipotenusa. Sul lato Ovest della linea meridiana (verso l'ingresso della Basilica) è tabulata proprio la distanza zenitale in corrispondenza dei gradi interi.

# Posizione dell'Equatore Celeste

Una volta determinata la latitudine del luogo, la posizione dell'equatore è una linea perpendicolare alla meridiana che la interseca ad una distanza ben precisa.

La distanza zenitale dell'equatore è pari alla latitudine, perciò la linea dell'equatore si trova a 100*tangente(41°54'27")=89.75 parti centesime dal piede della verticale del foro.

Bianchini tenne conto anche della correzione cassiniana dovuta alla rifrazione atmosferica, che innalza di circa 1' la posizione apparente dell'equatore, per cui la posizione esatta risulta 100*tangente(41°55'27")=89.80 parti centesime.

L'equatore è lo zero della declinazione. Declinazioni boreali (a nord dell'equatore) sono verso il piede della verticale del foro, declinazioni australi sono verso il Capricorno.

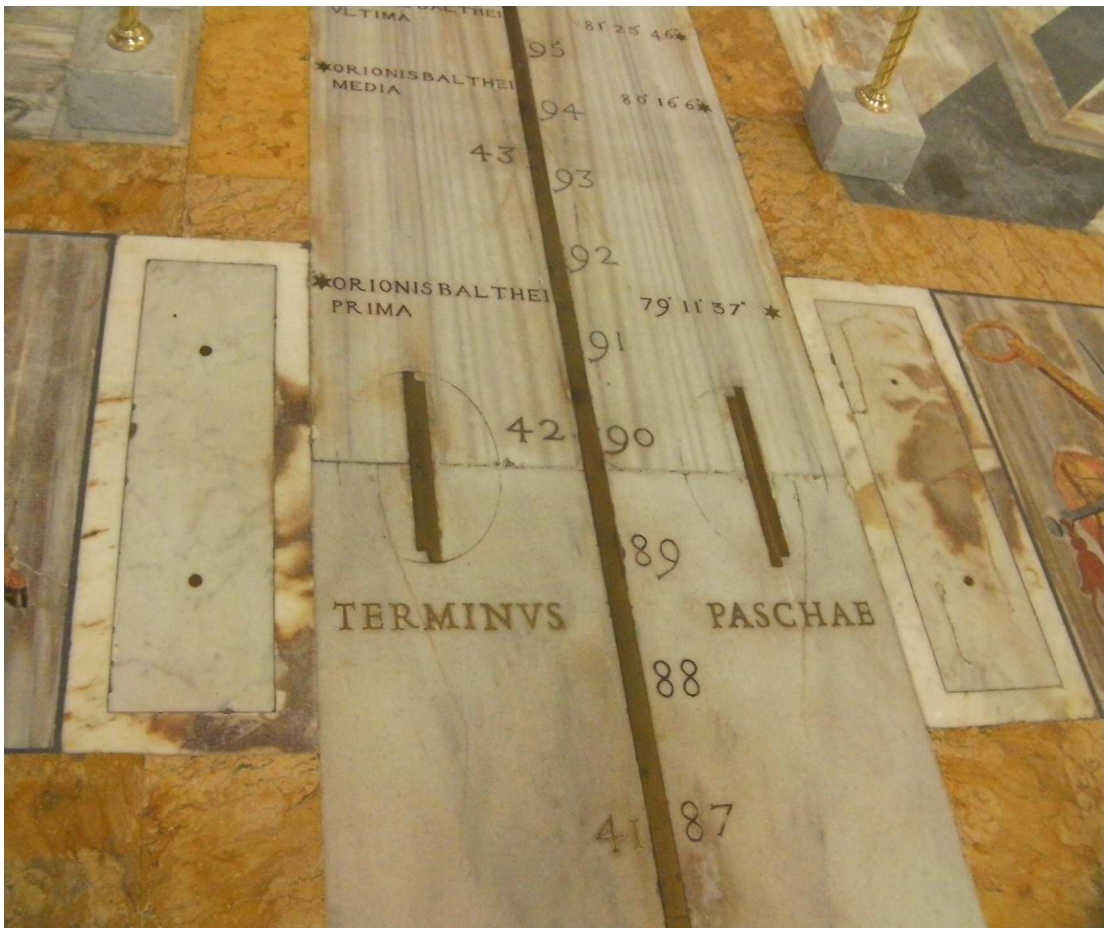

La prima stella della cintura di Orione nel 1701 era a Sud dell'equatore, che è la linea trasversale che passa ad 89.80 parti centesime. L'ordine delle famose tre stelle della cintura di Orione, chiamate anche "i tre Re", o "las tres Marias" in America latina, è quello del passaggio al meridiano, la prima è quella che passa per prima, quindi la più a Ovest delle tre, e così via. Oggi la prima stella della cintura di Orione si è avvicinata all'equatore celeste per effetto della precessione degli equinozi dimezzando quella che era la sua distanza del 1701, tra altri tre secoli sarà passata nell'emisfero boreale.

# Posizioni dei Segni Zodiacali

I magnifici segni zodiacali, realizzati con la tecnica delle tarsie marmoree sotto la guida di Carlo Maratta (1656-1713) a riprodurre le costellazioni della celebre Uranometria di Johannes Bayer (1603).

Le posizioni dei segni sono calcolate sempre con formule trigonometriche e sono quelle corrispondenti a 12 settori uguali, di 30° ciascuno, dell'orbita che il Sole compie nel Cielo.

L'eclittica è l'orbita del Sole nel Cielo, e si chiama così perché su questo percorso avvengono le eclissi.

Per convenzione, mantenuta dal Bianchini e valida ancora oggi, si chiama Ariete il settore di 30° che segue l'equinozio di primavera, Toro il successivo e così via.

Questo fa sì che i segni siano accoppiati a due a due, salvo Capricorno e Cancro, e sul lato Ovest della meridiana sono rappresentati i segni percorsi dal solstizio invernale fino a quello estivo e sull'altro lato quelli dal solstizio estivo verso quello invernale.

Con il programma di effemeridi Ephemvga, disponibile sul sito web della Basilica www.santamariadegliangeliroma.it menù meridiana, calcolo delle effemeridi, ci si può divertire a vedere i valori della declinazione che corrispondono all'inizio dei vari segni zodiacali, avendo l'accortezza di fissare la longitudine eclitticale (Hlong) del Sole a multipli interi di 30°.

Dalla declinazione si passa poi alle distanze zenitali, aggiungendo quella alla latitudine del luogo.

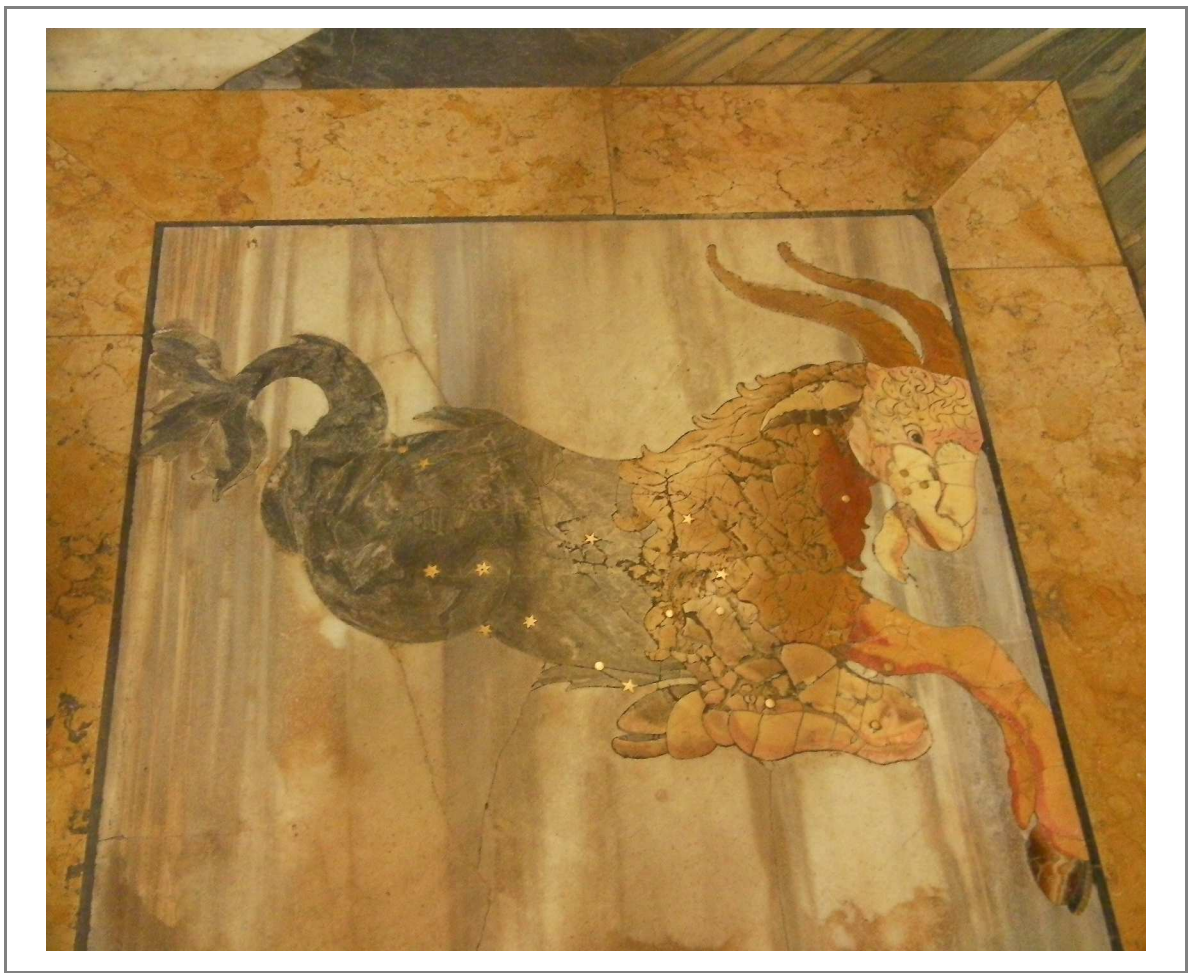

In queste immagini abbiamo il confronto tra il Capricorno in Basilica e quello di Bayer.

Mancano soltanto le stelle α e β vicino all'occhio e al corno sinistro dell'animale, ma le altre sono tutte riprodotte fedelmente al loro posto.

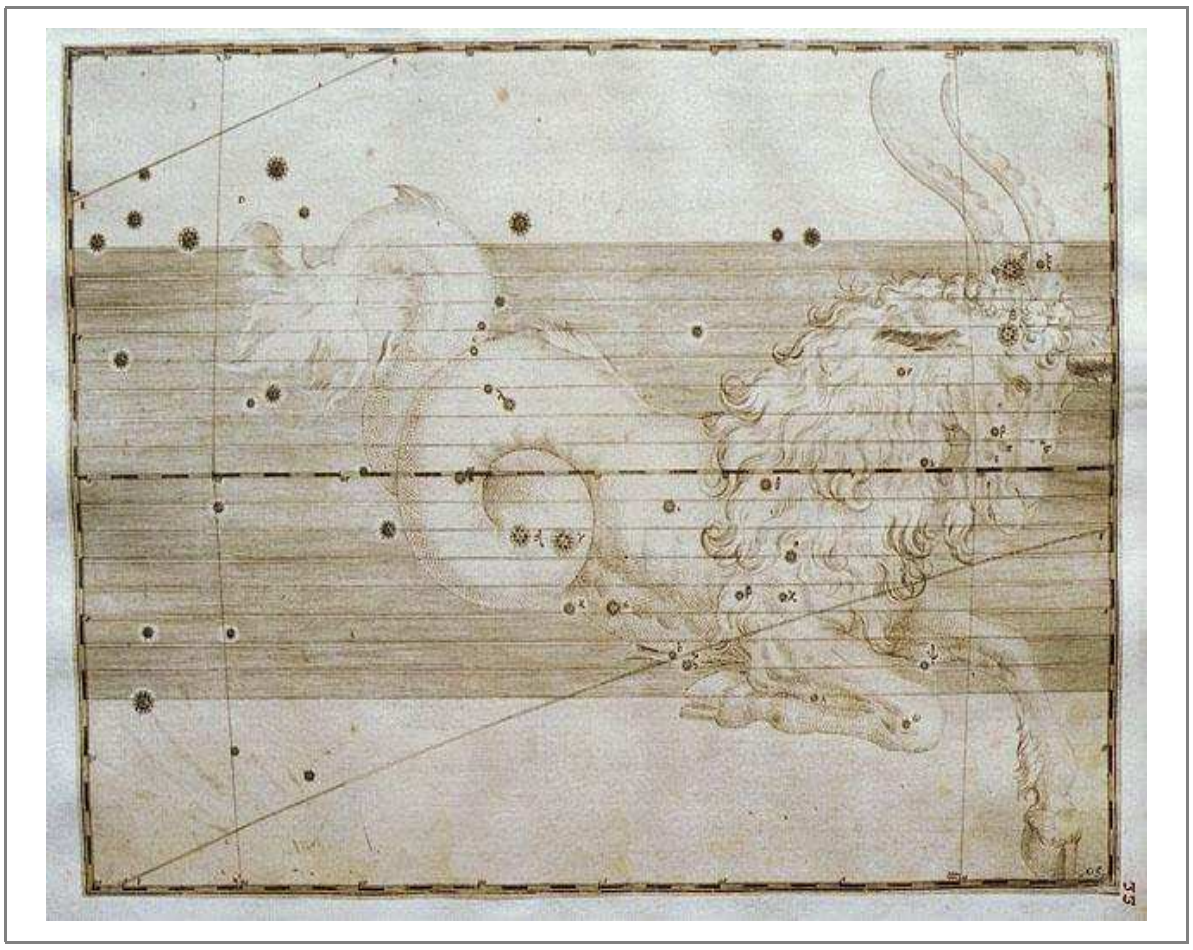

# Posizioni delle Stelle fisse

Bianchini utilizzò il catalogo stellare di Philippe de La Hire, allievo di Jean Picard, aggiornato al 1701 per selezionare le stelle di cui rappresentare le coordinate sulla meridiana.

In realtà non ne fa menzione nel testo De Nummo et Gnomone Clementino quanto alla costruzione della meridiana, dove peraltro omette di descrivere molti altri dettagli, tuttavia le nomina in occasione di alcune loro osservazioni.

Nel recente restauro in occasione del grande Giubileo del 2000 le stelle presso la linea meridiana sono state ripristinate, anche se è stato commesso qualche errore di trascrizione. In particolare per l'omissione di un segno negativo le stelle: Leonis Cauda, Pegasi Os, Geminorum Pes Lucidus e Arietis Lucida sono finite nell'emisfero opposto.

Le 22 stelle rappresentate sulla meridiana sono indicate tramite i loro nomi, e le loro coordinate celesti al 1701.

La loro posizione sulla linea dipende dalla declinazione, che distanza angolare dall'equatore celeste. Il loro nome latino è scritto a destra guardando verso Nord, mentre la loro ascensione retta è riportata in gradi sulla sinistra della linea, accanto alla stella.

Sono stati individuati 5 errori, commessi probabilmente nell'ultimo restauro, quando gli studi disponibili erano ancora scarsi e le tracce delle scritte antiche erano andate perdute.

Quattro stelle si trovano nell'emisfero sbagliato: Arietis Lucida, Leonis Cauda, Pegasi Os, Geminorum Pes Lucida, per un errore di segno. La quinta stella con errore è Procione, Canis Minor, ha come ascensione retta quella di un'altra stella del Cane Maggiore, Canis Maior in dorso, che segue immediatamente Procione nel catalogo di Philippe de la Hire, utilizzato da Francesco Bianchini, spiegandosi così l'origine della svista. Quest'ultimo dato non era stato ancora trovato e pubblicato prima del lavoro svolto da Magdalena Bedynski e Monica Nastasi come tesina d'esame del corso "La Terra nel Sistema Solare" A. A. 2006/7 al dipartimento di Geografia della Sapienza.

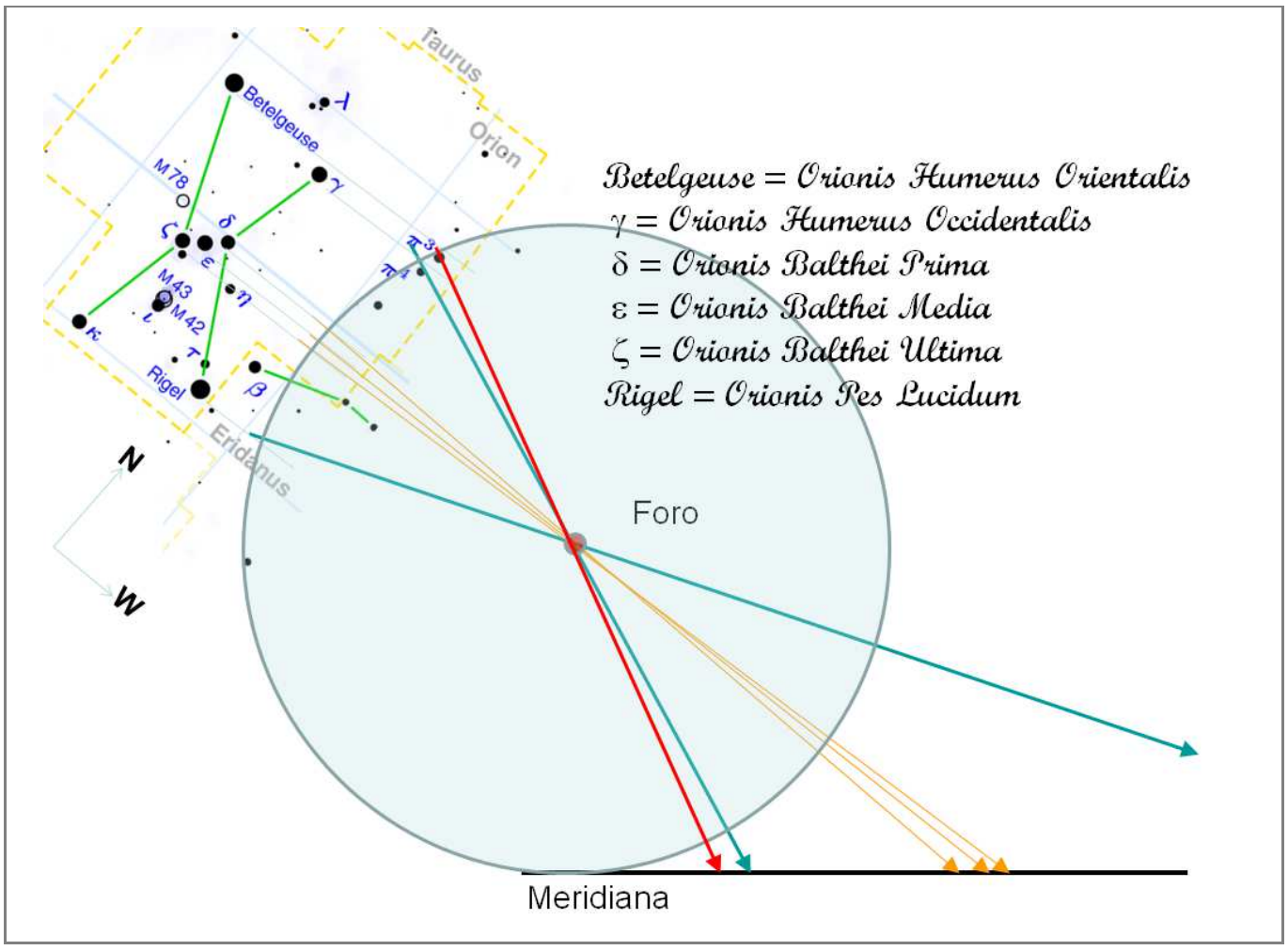

Schema semplificato del posizionamento delle stelle sulla meridiana. Vengono prima trasportate sul meridiano locale con una rotazione del sfera celeste, mediante le linee tra loro parallele che partono dalle stelle verso il cerchio. Poi vengono proiettate attraverso il foro stenopeico nella loro posizione sulla Linea Meridiana. Le stelle della cintura di Orione ad esempio sono nominate come prima, media ed ultima secondo l'ordine con cui passano al meridiano: prima quella più ad Ovest e così via. Con questo esempio si vede bene che le stelle a Nord dell'equatore (quasi esattamente su δ Orionis, ovvero Orionis Balthei Prima nella nomenclatura latina) finiscano più vicine alla verticale del foro ed il contrario accade per quelle meridionali.

Le due coordinate per individuare una stella nel Cielo sono declinazione ed ascensione retta nel sistema equatoriale e l'origine delle coordinate è nell'equinozio di Primavera.

La precessione degli equinozi modifica gradualmente queste coordinate, poiché sposta l'equinozio di Primavera rispetto alle stelle fisse.

Inoltre, essendo rigorosi, anche le stelle hanno un loro moto proprio, che, proprio negli anni in cui lo Gnomone Clementino cominciava a funzionare, veniva scoperto da Edmund Halley sulle coordinate di alcune delle stelle più brillanti.

Oggi Sirio, per effetto del suo solo moto proprio in declinazione di 1.211" per anno si è spostata di 373" verso Sud, cioè 0.65 parti centesime, più di mezza tacca.

La precessione negli ultimi 3 secoli ne ha modificato molto di più la declinazione. Se "de-precessioniamo" le coordinate attuali di Sirio, ed in particolare la sua declinazione che oggi vale -16°45' e corrisponde al numero 163.93 (con la correzione cassiniana), troviamo che nel 1701 sarebbe dovuto essere al numero 162.12, sempre con la correzione cassiniana. La stella invece è riportata nella posizione 161.2. La differenza, pari a quasi una tacca intera è dovuta al suo moto proprio.

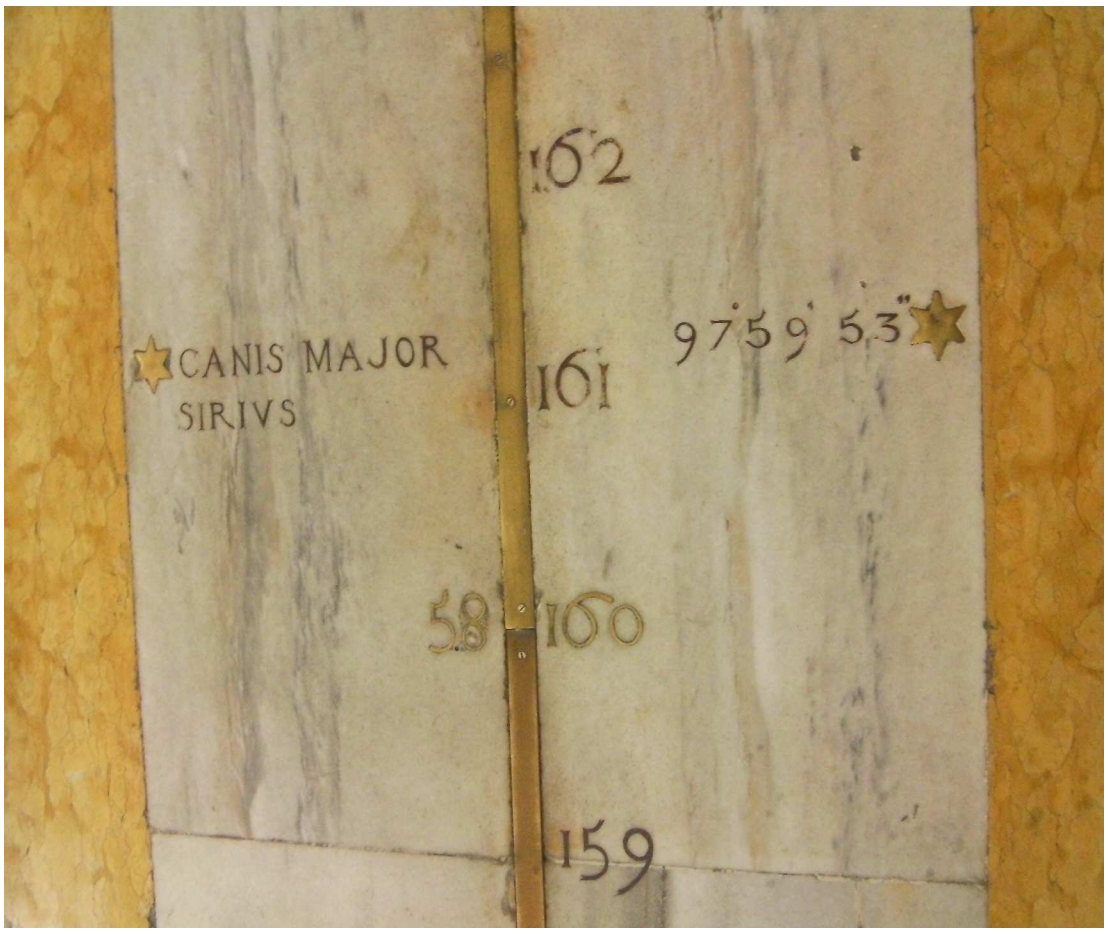

Bianchini, nel suo testo, riporta le osservazioni dei transiti di Sirio fatte anche durante il giorno, ed il loro valore in parti centesime andava da 161.28 (d'Inverno) a 161.38 (d'Estate) a causa della diversa densità dell'atmosfera.

| Data osservazione | Posizione di Sirio |
|---|---|
| 01/08/1702 | 161340 |
| 08/08/1702 | 161358 |
| 23/09/1702 | 161330 |
| 8/1/1703 | 161280 |
| 3/2/1703 | 161280 |
| 14/3/1703 | 161280 |
| 12/5/1703 | 161320 |
| 11/7/1703 | 161380 |

Riassumendo: nel caso di Sirio abbiamo che la precessione in tre secoli ha spostato la coordinata lungo la linea meridiana di 2 parti centesime verso i numeri più alti, mentre il moto proprio (differente da stella a stella) di poco più di metà nella stessa direzione.

Il moto proprio di Sirio è particolarmente grande perché Sirio si trova a 8.6 anni luce da noi, tutte le altre stelle, essendo molto più lontane manifestano moti propri ancora più piccoli.

Bianchini poteva osservare le stelle anche di giorno grazie all'oscuramento di tutta la Basilica mediante delle tende, di cui Catamo e Lucarini hanno ritrovato i ganci all'esterno. Egli usava un telescopio per traguardarne il passaggio al meridiano attraverso la finestrella aperta sopra al foro stenopeico, dove veniva materializzato con un filo il meridiano.

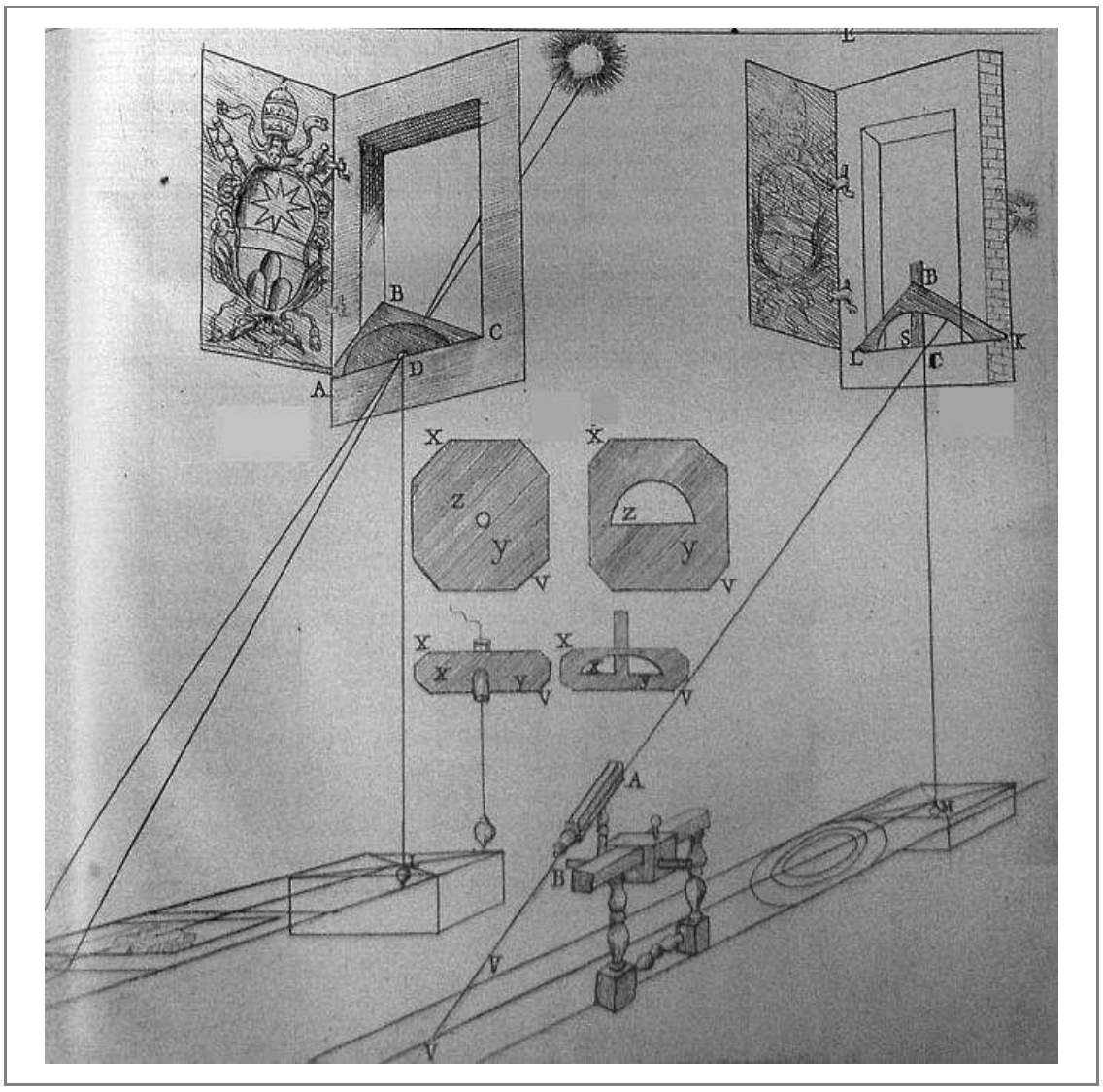

# La formazione dell'immagine solare

A differenza di tutti i telescopi che osservano in luce visibile, questo non ha elementi ottici rifrangenti (lenti) né riflettenti (specchi). Gli stessi principi oggi li troviamo in alcuni strumenti che osservano il cielo nei raggi X, chiamati collimatori.

Il foro stenopeico della meridiana è collocato in un vano apposito scavato nel muro sud-est della Basilica. La strombatura, visibile anche dalla stazione Termini, consente l'ingresso dei raggi solari fino al solstizio d'Estate, ma non quelli della Luna ai "lunistizi" più estremi, come nel 2009 dove arriva a 76° di altezza. All'interno il foro di 16 mm di diametro è incluso nello stemma di papa Clemente XI, sotto una finestrella che si poteva aprire per osservare anche le stelle.

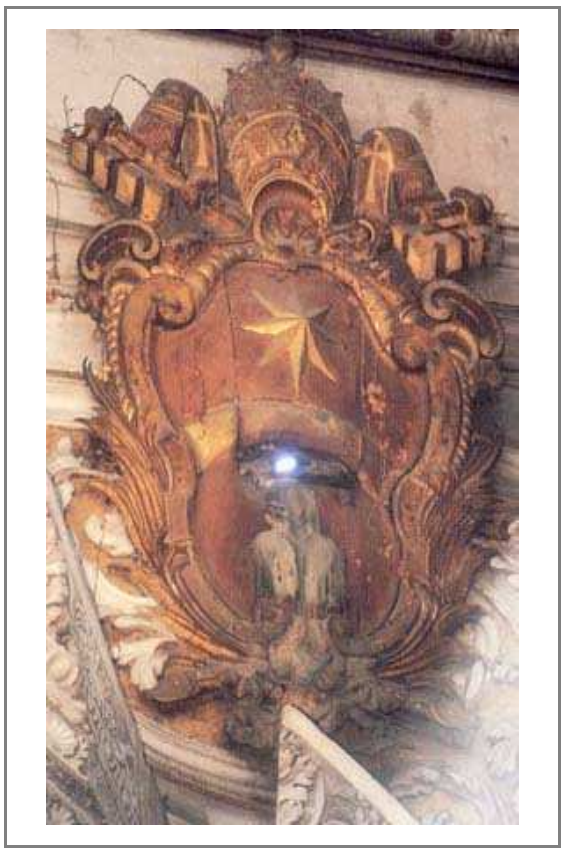

Per studiare la formazione dell'immagine da un foro stenopeico dobbiamo considerare le leggi dell'ottica geometrica, secondo cui i raggi luminosi si propagano in linea retta, e cominciamo dallo schermo su cui l'immagine è proiettata: il pavimento della Basilica.

Il Sole, con il suo diametro di 1.5 milioni di km e la sua distanza media da noi di 150 milioni di km ci appare con un rapporto D/F=1/100. Per questo ogni immagine stenopeica del Sole ha un diametro pari a circa 1/100 della distanza focale:

esempio 10 metri di focale, 10 centimetri il diametro dell'immagine.

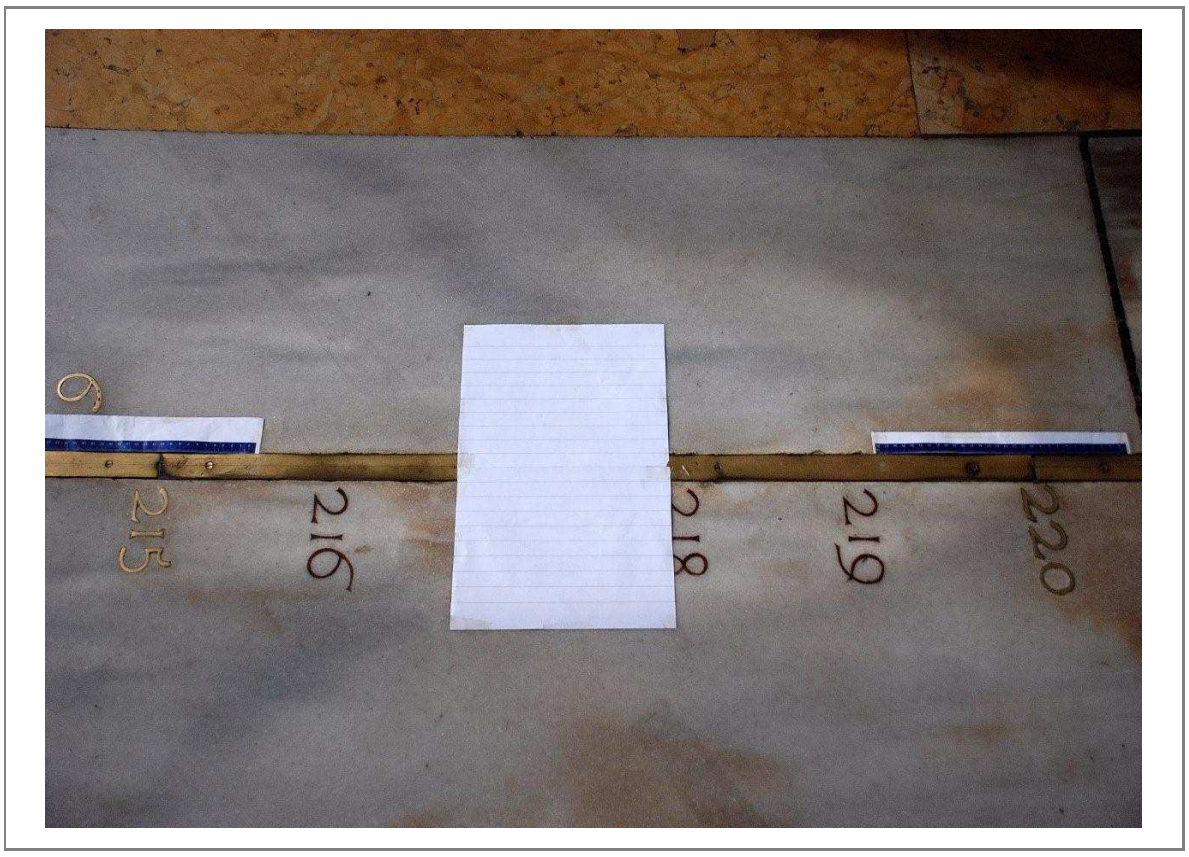

La foto qui sopra è del 24 dicembre 2007, e mostra l'apparato per la misurazione dei transiti paralleli delle posizioni dei lembi del Sole sulla linea meridiana. L'ellissi solare è lunga un metro e 10 cm.

Nelle settimane attorno al solstizio d'Inverno la distanza foro-immagine, che chiamiamo distanza focale, supera i 48 metri, il foro è di 16 mm, pari ad un rapporto D/F = 1/3000.

Possiamo dire che attraverso il foro stenopeico della Basilica si può vedere un'area di Sole di diametro pari ad 1/30 del diametro del Sole, e, al contrario, in ogni punto dell'immagine arriva luce da un disco ampio 1/30 dell'intero disco solare. In pratica la risoluzione angolare di questo strumento è 1/30 del diametro del Sole, all'incirca un (minuto) primo d'arco. L'informazione proveniente da dettagli sulla superficie del Sole più piccoli di un minuto primo d'arco vengono

"convoluti" con tutto ciò che hanno intorno entro quel minuto d'arco risultando sempre meno contrastati.

Per questo con lo Gnomone si possono osservare anche le macchie solari, come quella del 1 luglio 2006, con il foro ridotto a 6 mm di diametro per migliorare il potere risolutivo.

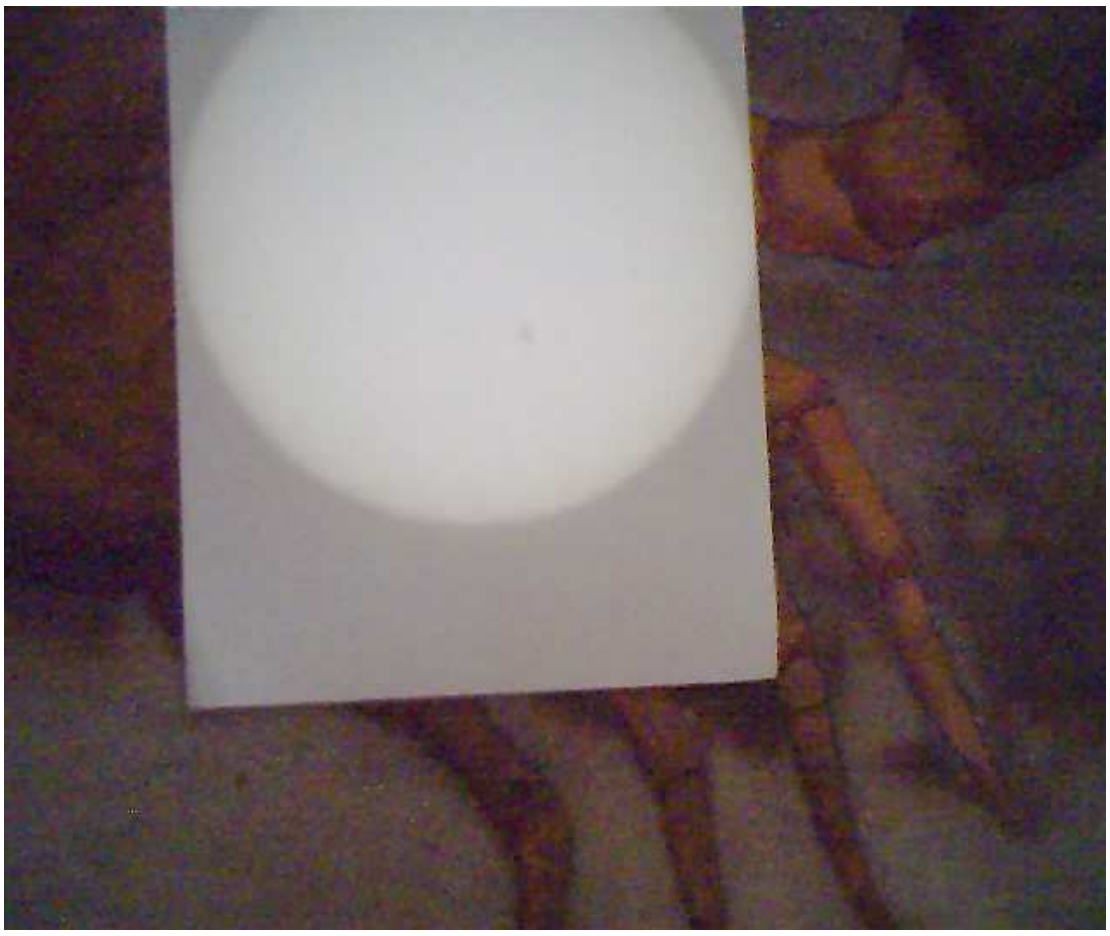

Questa immagine mostra la macchia n. 55 del 2006, osservata in Basilica , il giorno 1 luglio 2006.

La numerazione è quella della Specola Solare Ticinese.

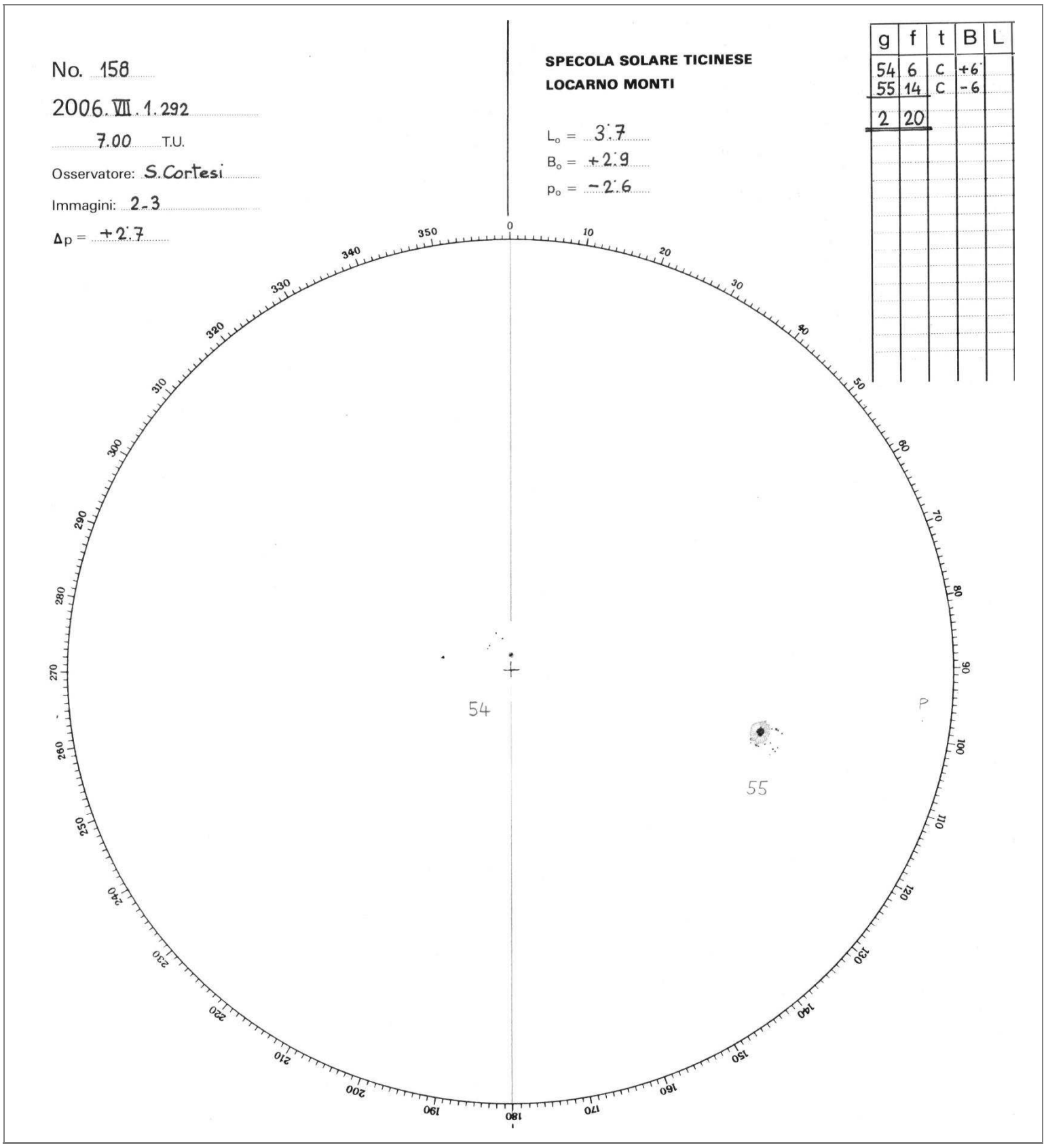

Questo effetto di convoluzione entro un disco di un minuto d'arco è efficiente anche quando si arriva al bordo dell'immagine: la luce proveniente dal bordo del Sole, che già di per sé è più oscuro del centro del disco, viene convoluta con quella del cielo che si trova immediatamente all'esterno, provocando così una regione di penombra attorno al lembo solare. La dimensione di questa penombra è pari ad un primo d'arco tutto attorno al disco del Sole, ma se ne percepisce solo la metà poiché il contrasto dell'immagine non è molto forte, a causa della luce diffusa dentro la Basilica. La penombra allarga quindi l'immagine del Sole, e già Bianchini correggeva i diametri solari da lui misurati per questo effetto di ottica geometrica: "dempta penumbra", sottratta la penombra otteneva il valore corretto del diametro angolare del Sole.

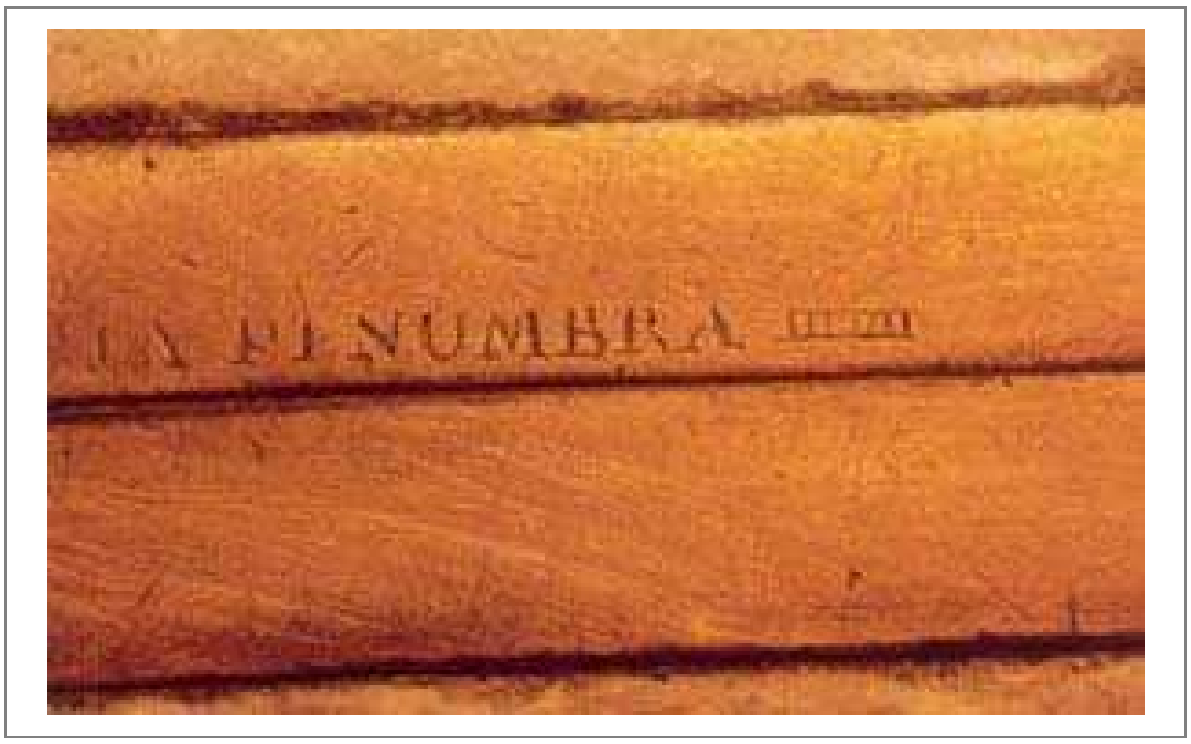

Presso il "cronometro degli equinozi" l'ampiezza della penombra da sottrarre è rappresentata da questa piccola scala, dopo la scritta nella foto sopra, di circa 5 mm di lunghezza.

Uno dei grandi meriti di Catamo e Lucarini è stato quello di riportare in auge il "cronometro degli equinozi". Egnazio Danti (1576) aveva già formulato la regola che attorno agli equinozi la velocità in declinazione del Sole è di 1′ all'ora, poco più di 8 mm sulla Linea Clementina. Bianchini, a scopo dimostrativo, collocò due dispositivi prima e dopo la Linea, dove calcolare a vista il tempo trascorso dall'equinozio o ancora mancante.

Il dettaglio di sopra è ripreso dal dispositivo a sinistra nella foto seguente.

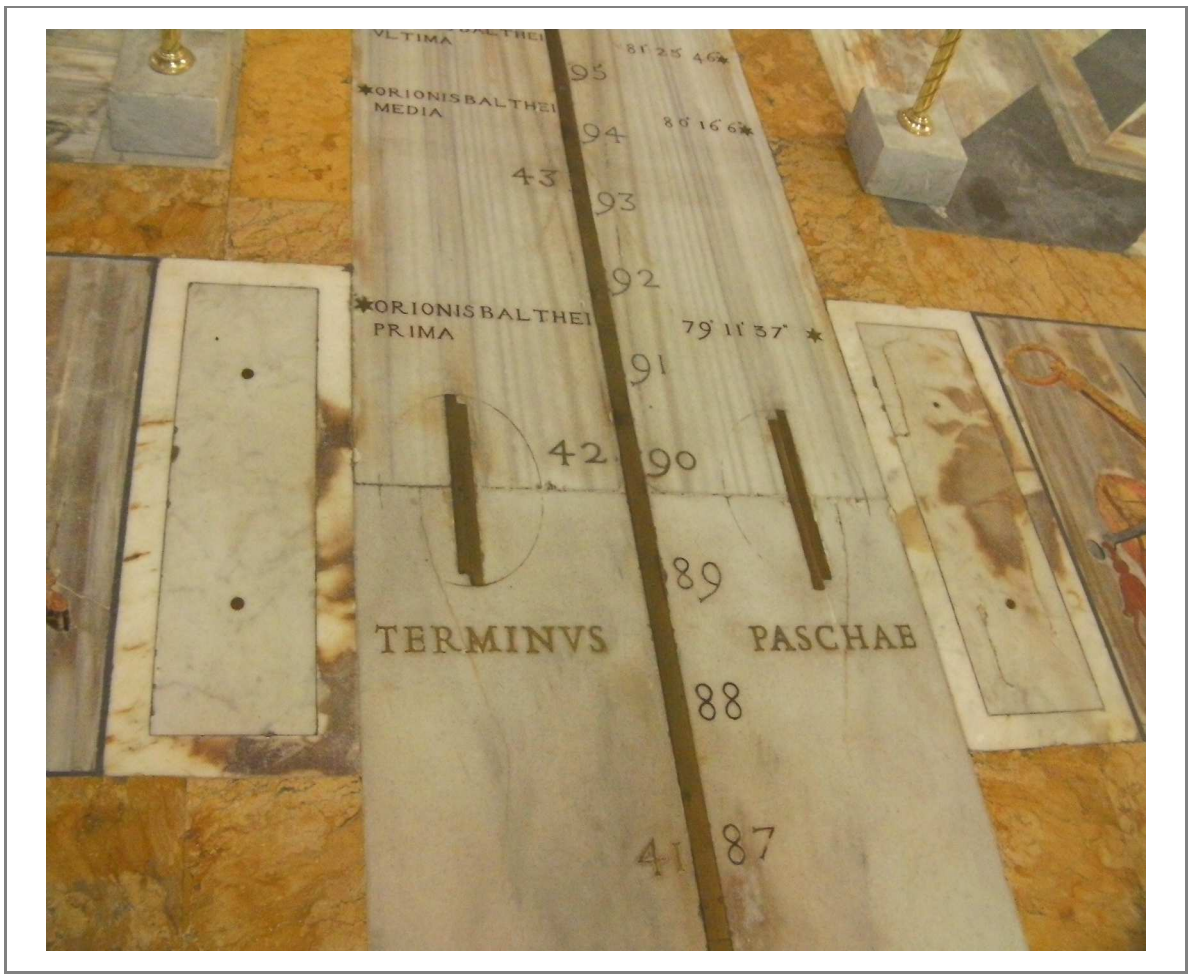

Riassumendo, il potere risolutivo di questo telescopio è tanto migliore quanto maggiore è il rapporto F/D, che varia nel corso dell'anno da 1250 (2.4 minuti d'arco) al solstizio estivo a 3000 (1 minuto d'arco) a quello invernale. La luminosità dell'immagine invece si riduce di quasi 6 volte dall'Estate all'Inverno perché l'area su cui è proiettata la luce aumenta dello stesso valore, per questo l'immagine invernale, più grande, è anche più pallida e meno contrastata. L'immagine seguente è stata ripresa il giorno che il Sole aveva la stessa declinazione di Arturo nel 1701.

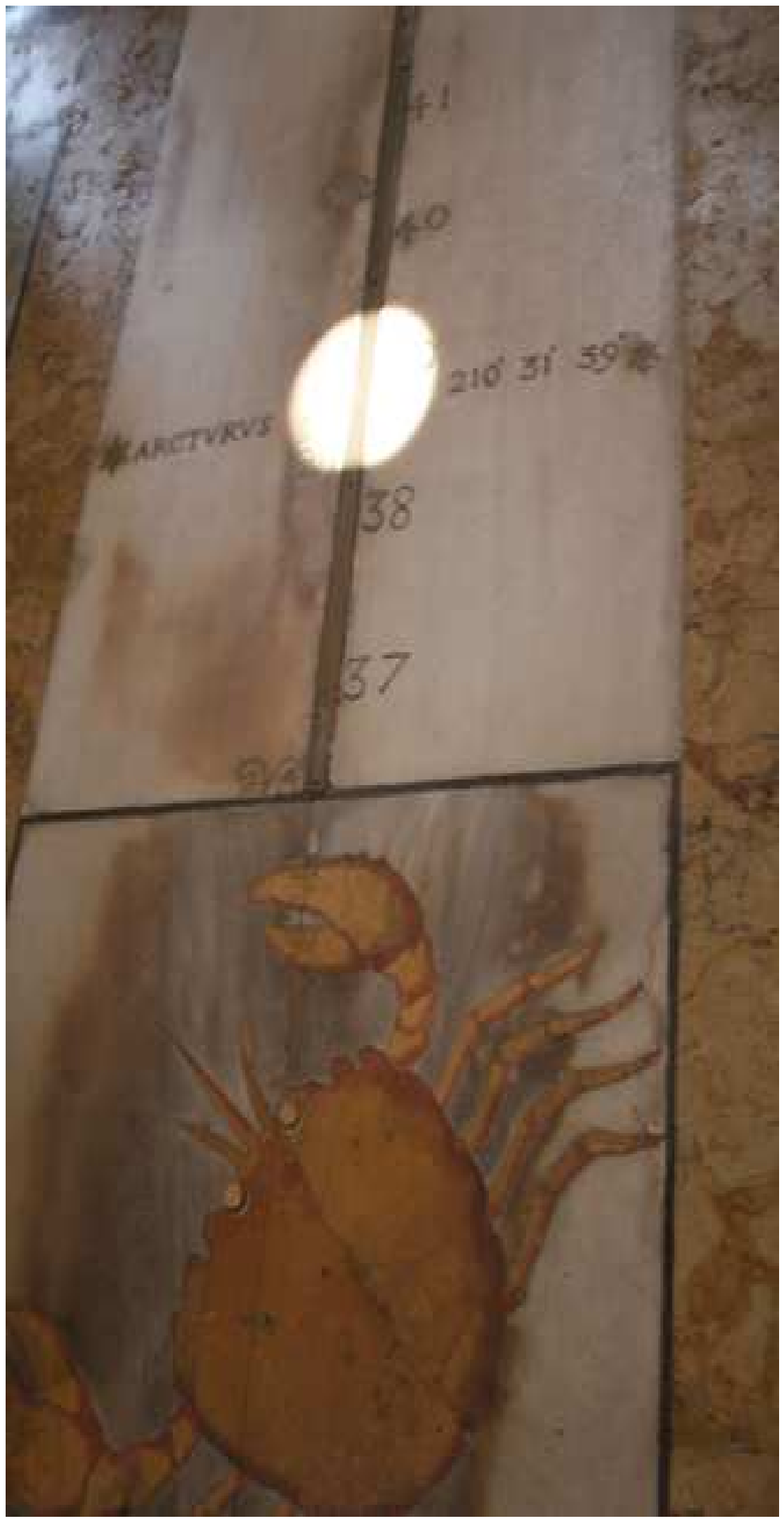

Quando Bianchini osservava il Sole e le stelle in Basilica, tutte le grandi finestre venivano oscurate con tende, ed il problema del contrasto dell'immagine non si presentava.

# Misura della posizione dei lembi solari

Queste proprietà ottiche del telescopio a foro stenopeico sono applicabili solo all'osservazione del Sole e, con grande difficoltà, anche della Luna, tuttavia i vantaggi che si possono trarre dalle misure di posizione del Sole fatte con questo strumento ne giustificano la costruzione ed il finanziamento da parte del papa Clemente XI.

Infatti è vero che l'immagine del Sole non ha una risoluzione paragonabile a quella ottenuta con i telescopi rifrattori e riflettori, ma poiché la penombra è simmetrica rispetto al centro del Sole, la misura della posizione del centro del disco solare –che si ottiene dalla media delle posizioni dei lembi– dipende solo dalla precisione con cui possono essere individuati e misurati questi ultimi.

Rispetto alla linea meridiana distinguiamo quattro lembi: quelli perpendicoli alla linea che sono il meridionale (più lontano dal foro) ed il settentrionale (più vicino al foro), e quelli che sono tangenti alla linea, il lembo precedente (occidentale) e quello seguente (orientale).

I lembi meridionale e settentrionale venivano misurati con il compasso ed il regolo "ticonico" sul pavimento, con una precisione fino a mezzo millimetro, mentre quelli precedente e seguente mediante i loro tempi di contatto, con una precisione temporale fino al mezzo secondo. Vediamo di seguito i due casi distinti.

# Misure di posizione col regolo ticonico

Le misure di posizione dei lembi perpendicolari servono per ricavare la declinazione del Sole. Disponendo dei fogli di carta in corrispondenza dei lembi perpendicolarmente alla linea, abbiamo fatto diverse serie di misure per verificare la precisione raggiungibile sulla determinazione della posizione dei lembi. L'immagine seguente è del 31 dicembre 2007 con le fasce millimetrate per effettuare la misurarne i bordi.

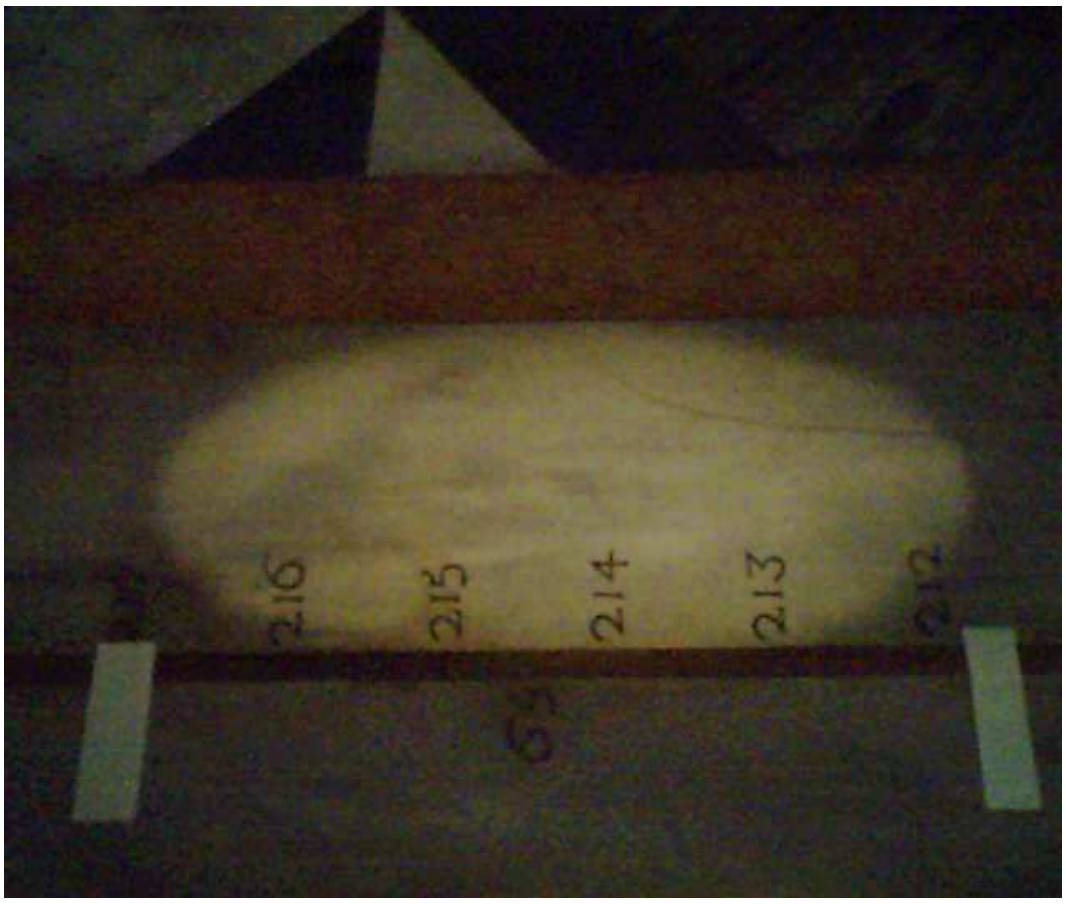

L'immagine solare è agitata da una continua turbolenza, generata dall'aria calda in movimento sul piano, fatto di mattoni, adiacente al foro stenopeico che si trova all'aria aperta. I mattoni sono scaldati dal Sole di mezzogiorno che anche d'Inverno li porta ad una temperatura maggiore di quella ambiente, da cui i moti turbolenti. Il passaggio di una bolla d'aria calda di dimensioni simili all'apertura del foro produce un sussulto dell'intera immagine del Sole sul piano del pavimento. Questo fenomeno si ripete in modo casuale con disturbi a frequenze superiori alle 10 volte al secondo.

La nostra vista tende a compensare questi disturbi, facendone una media nel tempo, ed il punto corrispondente al lembo solare può essere individuato con una accuratezza del mezzo millimetro, così come per i lembi tangenti l'istante di tempo può essere determinato al meglio del mezzo secondo.

Bianchini usava un compasso di cui una punta veniva collocata su una tacca sulla linea meridiana, corrispondente ad una parte centesima esatta, e l'altra individuava il lembo del Sole, poi portava il compasso presso il regolo "ticonico" per leggere il risultato.

Facciamo un esempio in cui la distanza del lembo solare dalla tacca centesima intera 206 sia 0.235 parti centesime.

La punta del compasso che era stata presso la tacca centesima 206 si piazza sul regolo in corrispondenza della linea verticale corrispondente a 20-200: 20 sono le parti centesime dell'unità fondamentale e 200 sono le parti millesime della stessa unità, unità che corrisponde a 203.44 mm. Poi si fa scorrere questa punta sulla linea verticale finché l'altra punta non incontra la linea diagonale che va su e giù. Se la diagonale è quella che sale dal 30 (parti millesime) al 40, e la linea orizzontale intersecata in quel punto è la quinta, perciò la misura risulta 35, e la somma complessiva 235 parti millesime.

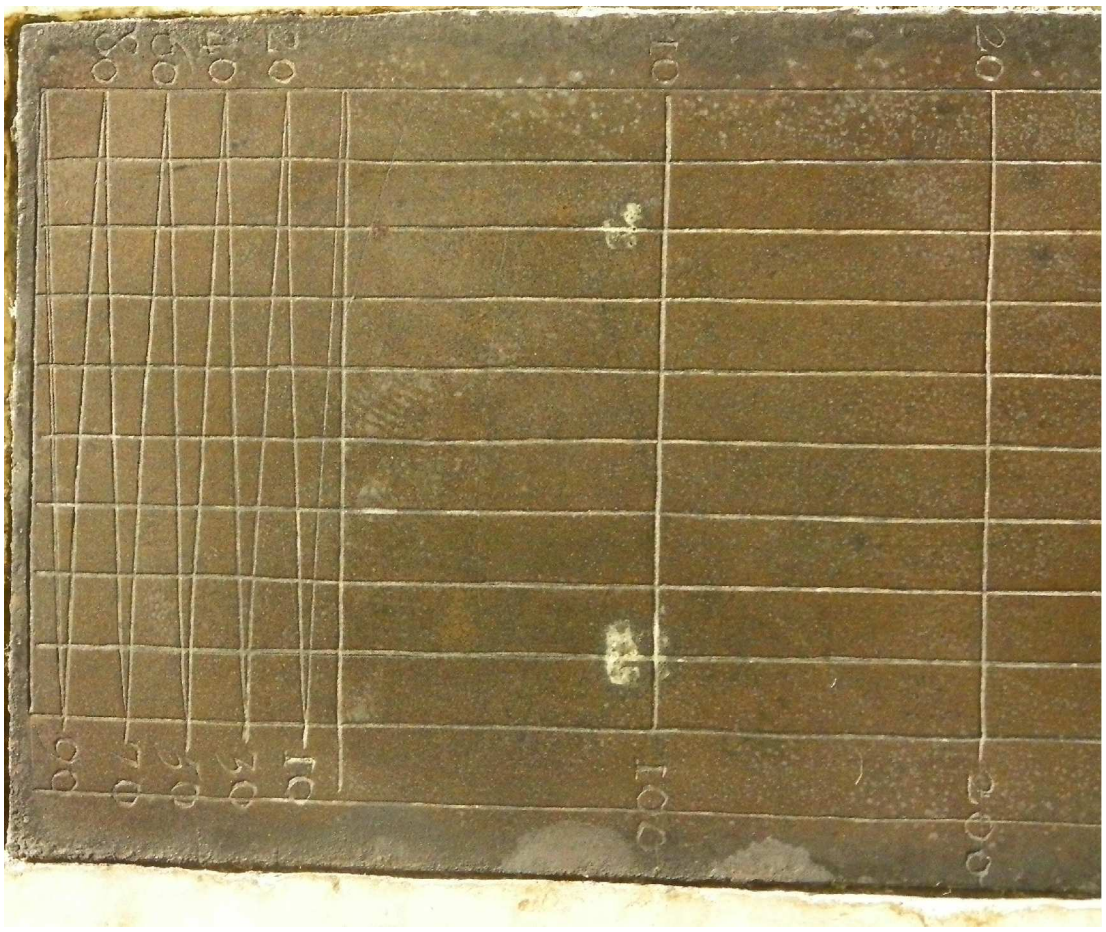

Nel muro settentrionale del presbiterio è murato un regolo "ticonico" che dimostra il tipo di strumento usato. Ha 9 righe orizzontali invece di 10, comunque aiuta a capire il suo funzionamento. Fu il grande astronomo danese Tycho Brahe (1546-1601) ad inventare questa scala a zig zag che perciò si chiama "ticonica".

## Misure di tempo

Queste sono le misure cruciali per la determinazione della durata dell'anno tropico, lo scopo dichiarato "ad usum Kalendarii" sulla moneta coniata in occasione dell'inaugurazione dello Gnomone Clementino – secondo anniversario della prima messa solenne di papa Clemente XI.

Bianchini si era fatto inviare da Parigi un orologio dotato di pendolo e cicloide di Thuret, il migliore sul mercato. Un assistente si peritava di caricarlo periodicamente, e, durante le osservazioni dei transiti solari leggeva gli istanti di tempo corrispondenti alle tangenze tra immagine del Sole e linea meridiana.

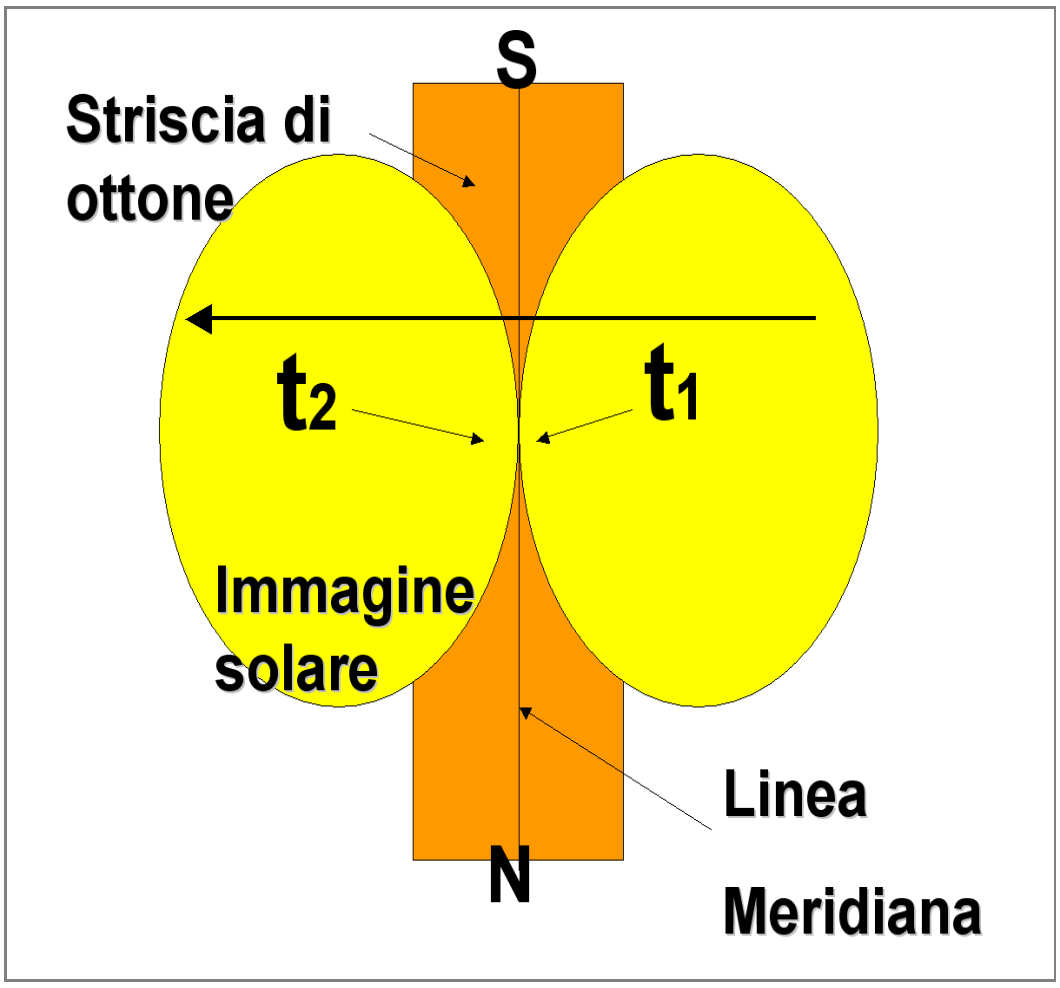

Nei resoconti delle osservazioni di transiti solari e stellari, abbiamo la possibilità di valutare le prestazioni di questo orologio che veniva continuamente calibrato ogni volta che si osservava una stella al meridiano.

Una serie di misure dei transiti di Sirio ci fa capire come veniva adoperato questo orologio.

Sappiamo che il giorno siderale è il tempo tra un transito di una stella al meridiano ed il successivo, e vale 23 ore 56 minuti e 4.09 secondi, cioè 86164.09 secondi. Per essere più precisi possiamo dire 86164 parti minute seconde di ora equinoziale.

L'ora equinoziale è la 24ª parte del giorno solare medio.

Anche i minuti primi, sono parti minute prime dell'ora (o del grado angolare che pure si misura nel sistema sessagesimale).

Il pendolo veniva tarato in modo da fare 86164 oscillazioni complete tra un passaggio ed il successivo di una stessa stella, tuttavia gli astronomi sapevano bene che quando la temperatura scendeva le componenti metalliche subivano una contrazione e il pendolo aumentava il ritmo delle oscillazioni, mentre accadeva l'opposto con l'aumentare della temperatura.

Dal 26 giugno all'11 luglio 1703 Bianchini osservò sia il Sole che Sirio al meridiano. Il moto dell'orologio fu mantenuto continuo ed il valore medio del giorno siderale su 16 giorni consecutivi (15 dati) fu di 86163.13±0.64 oscillazioni. La deviazione standard di 0.64 s ci dà un'idea sia della bontà della meccanica, che della effettiva risoluzione temporale del metodo di misura.

Se 86163.13 oscillazioni corrispondevano a 86164.09 secondi reali, questo significa che durante quei 16 giorni l'orologio tendeva ad andare in ritardo di 0.96 s al giorno, mostrando che la meccanica e la regolazione fine dell'orologio era davvero eccellente. Ma la cosa più importante è che questo minimo ritardo era misurabile e quindi correggibile.

L'osservazione dei transiti stellari era dunque un modo per sincronizzare continuamente l'orologio con il ritmo, ritenuto immutabile, della rotazione terrestre. L'equivalente di avere a disposizione un orologio radio controllato oggi.

La possibilità di apprezzare il mezzo secondo dipendeva poi dalla tecnica adoperata. Se l'astronomo ascoltava il ticchettìo dell'orologio e contemporaneamente osservava il fenomeno poteva stabilire se il contatto dei lembi era avvenuto nella prima parte di un certo minuto secondo o nella seconda.

Bianchini pubblicò dati di soli secondi interi, dunque non utilizzava questa tecnica. La sua precisione era già molto superiore a quella di tutti gli osservatori solari precedenti.

Le misure di tempo erano dunque garantite con una precisione di un secondo. L'effetto della penombra non influenza la misura della posizione del centro del Sole, e quindi Bianchini poté disporre delle misure di posizione del Sole lungo l'eclittica più precise mai fatte.

A suo favore giocava anche il fatto che effettivamente durante il XVIII secolo il ritmo della rotazione terrestre restò pressoché costante. Tuttavia l'azimut della linea meridiana non esattamente orientato a Nord determinò alcuni errori sistematici nella misura di solstizi ed equinozi, quando il Sole e la stella di riferimento passavano al meridiano a diverse distanze zenitali. Ad esempio d'Estate il Sole passa sulla Linea Clementina con 3 secondi di ritardo rispetto al suo vero passaggio meridiano, mentre il transito di Sirio essendo osservato da una posizione più a Est attraverso la finestrella meridiana, viene registrato 14 secondi dopo del suo passaggio in meridiano reale. Il Sole estivo, quindi, sembra sistematicamente in anticipo rispetto alla sua posizione reale tra le stelle fisse. L'ascensione retta del Sole estivo è sistematicamente inferiore al vero, e conseguentemente il solstizio calcolato arriva dopo di quello vero. Queste differenze di tempo cambiano con le stagioni, poiché cambia la posizione del Sole sulla linea Clementina.

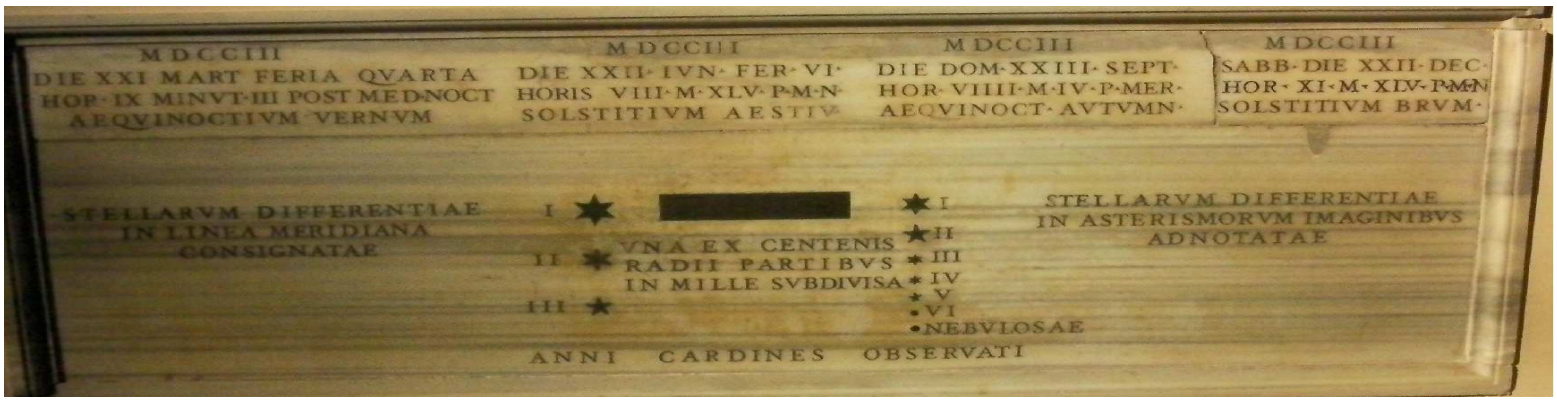

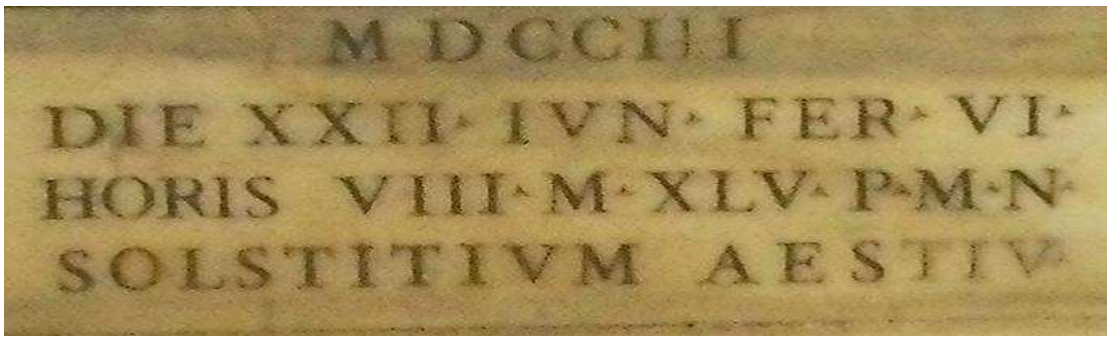

Bianchini calcolò il solstizio estivo del 1703 il giorno 22 di giugno, 8 ore e 45 minuti dopo la mezzanotte (ora locale), mentre con le effemeridi al computer troviamo che il solstizio capitava alle 8 e 23. La situazione si rovescia al solstizio invernale.

Bianchini confrontò i suoi risultati con quelli di Bernardo Walther (1501) in modo da spalmare su 200 anni gli errori di una singola misura, ed ottenne una precisione di 2x22/200=0.22 minuti, cioè ±13 secondi sulla durata dell'anno tropico.

# Come domare i "Cavalli del Sole"

Avendo a disposizione un catalogo stellare accurato, con le posizioni delle stelle tali da avere 1 secondo di ascensione retta (15″ angolari) di precisione è presumibile di raggiungere la medesima accuratezza nella misura dell'ascensione retta del Sole. La declinazione con la posizione del centro del Sole determinata entro 0.2 mm poteva essere accurata fino a 2″ angolari.

Tutte le possibili descrizioni del moto del Sole, chiamate anche teorie solari, devono soddisfare, a questi livelli di precisione, le osservazioni, per poter essere accettate.

Il primo esempio, dovuto a Cassini, che usò questa colorita espressione equestre, è la valutazione della rifrazione atmosferica, il secondo quello dell'eccentricità dell'orbita terrestre, il terzo è la discriminazione tra orbita tolemaica ed orbita kepleriana, noto come il problema della bisezione dell'eccentricità. Li vediamo nel seguito.

## La rifrazione "cassiniana "

Giandomenico Cassini (1625-1712) divenne celebre proprio con questo primo importante risultato conseguito alla meridiana di S. Petronio a Bologna.

Già Tycho Brahe in occasione delle misure di posizione della supernova del 1572, che stabilirono che la stella nuova era nel cielo delle stelle fisse, si era accorto che queste misure differivano in modo sistematico da un osservatorio all'altro.

In particolare per osservatori più a Sud la stella (boreale e circumpolare alle nostre latitudini) tendeva ad avere declinazioni più a nord.

Tolomeo aveva notato il fenomeno in cui la Luna piena in eclissi sorgeva contemporaneamente al tramonto del Sole, ed entrambi erano contemporaneamente interamente visibili. Sapendo che le eclissi avvengono solo se Luna-Terra e Sole sono su una retta, è chiaro che l'atmosfera "tira su" un oggetto che si trova al di là, in misura tanto maggiore quanto più questo è vicino all'orizzonte.

Cassini, e poi anche Bianchini, potevano studiare quantitativamente il fenomeno, con il Sole che si allontana fino a 23 gradi e mezzo dall'equatore celeste durante i solstizi. Per capire come ciò sia concettualmente possibile occorre fare alcune considerazioni.

Prima: l'orbita del Sole (o della Terra, che dir si voglia) è un cerchio massimo nel cielo, l'Eclittica, di cui noi siamo il centro. Questa legge dell'astronomia sferica classica è conservata anche nella prima legge di Keplero, che recita "le orbite dei pianeti sono piane".

Seconda: l'Eclittica intercetta l'equatore celeste in due punti, chiamati nodi. I nodi sono diametralmente opposti e i solstizi sono a 90° dai nodi e si discostano in modo simmetrico dall'equatore, tanto a Nord 23°30′ tanto a Sud -23°30′.

Al solstizio d'Inverno invece abbiamo che il centro del Sole passa circa 2′ più indietro, cioè quasi 3 cm, rispetto alla posizione che dovrebbe avere simmetrica al solstizio estivo. Cassini elaborò una tavola empirica di correzioni valide per tutti gli angoli, che lo stesso Bianchini usava, e che oggi possiamo riassumere nella formula

$$c=60''\cdot\tan(z)$$

valida fino a z=75°, dove z è la distanza zenitale. Presso l'orizzonte la correzione c=35′ così da "tirare su" il Sole e la Luna del loro intero diametro.

Nella tabella seguente si riportano le correzioni di Cassini ed il "fit" della formula, nel grafico sono rappresentate sia i dati in tabella che la formula.

| Distanza zenitale | Correzione Cassini ["] | 60"·tan(z) |
|---|---|---|
| 0 | 0 | 0 |
| 5 | 5 | 5.2 |
| 10 | 10 | 10.6 |
| 15 | 16 | 16.1 |
| 20 | 21 | 21.8 |
| 25 | 27 | 28.0 |
| 30 | 34 | 34.6 |
| 35 | 41 | 42.0 |
| 40 | 50 | 50.4 |
| 45 | 59 | 60.0 |
| 50 | 70 | 71.5 |
| 55 | 83 | 85.7 |
| 60 | 102 | 103.9 |
| 65 | 126 | 128.7 |
| 70 | 159 | 164.9 |

**Tabella 2.** Correzione Cassiniana dovuta alla rifrazione atmosferica, riportata in grafico nella figura seguente.

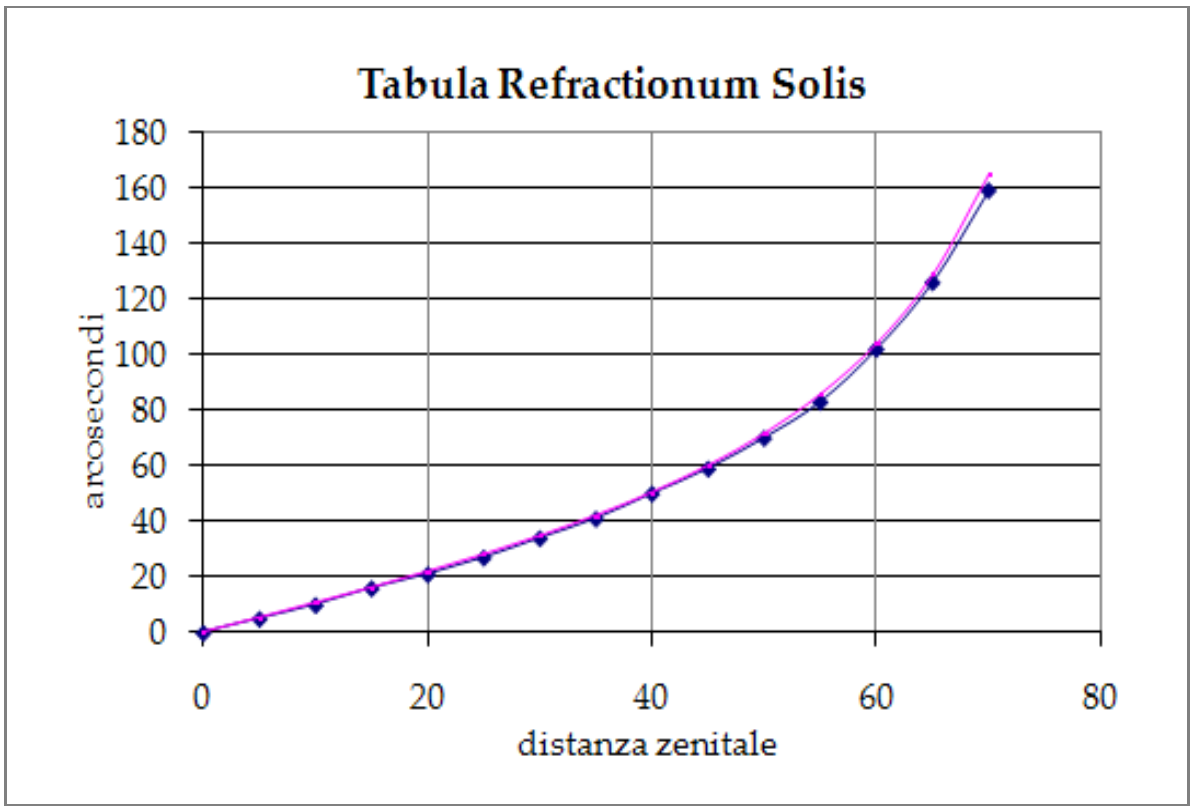

# Eccentricità tolemaica dell'orbita

La misura dell'eccentricità dell'orbita dipende da due fattori: uno pratico-sperimentale ed un altro teorico.

Quello pratico è la misura degli istanti di equinozi e solstizi, in cui Bianchini poteva assicurare la massima precisione possibile. Quello teorico è la distinzione tra modello tolemaico e modello kepleriano, entrambi indifferentemente geocentrici o eliocentrici.

Una volta fissati gli istanti di equinozi e solstizi si calcolano le durate delle stagioni. Se tutto l'anno equivale ad un giro ogni stagione equivale ad un settore di circa 90°. Disegnando su un cerchio di raggio unitario questi settori proporzionali alla durata delle stagioni si individuano 4 punti. Unendoli a due a due si traccia una croce il cui centro corrisponde al centro della Terra. La distanza di questo centro dal centro del cerchio si chiama eccentricità e vale e=0.0334. Se tutte le stagioni fossero di uguale durata avremmo un'orbita circolare senza alcuna eccentricità e=0.

Nell'immagine successiva ho rappresentato un'eccentricità tale da avere l'Estate di 94 giorni, l'Autunno di 90, l'Inverno di 89 e la Primavera di 92.

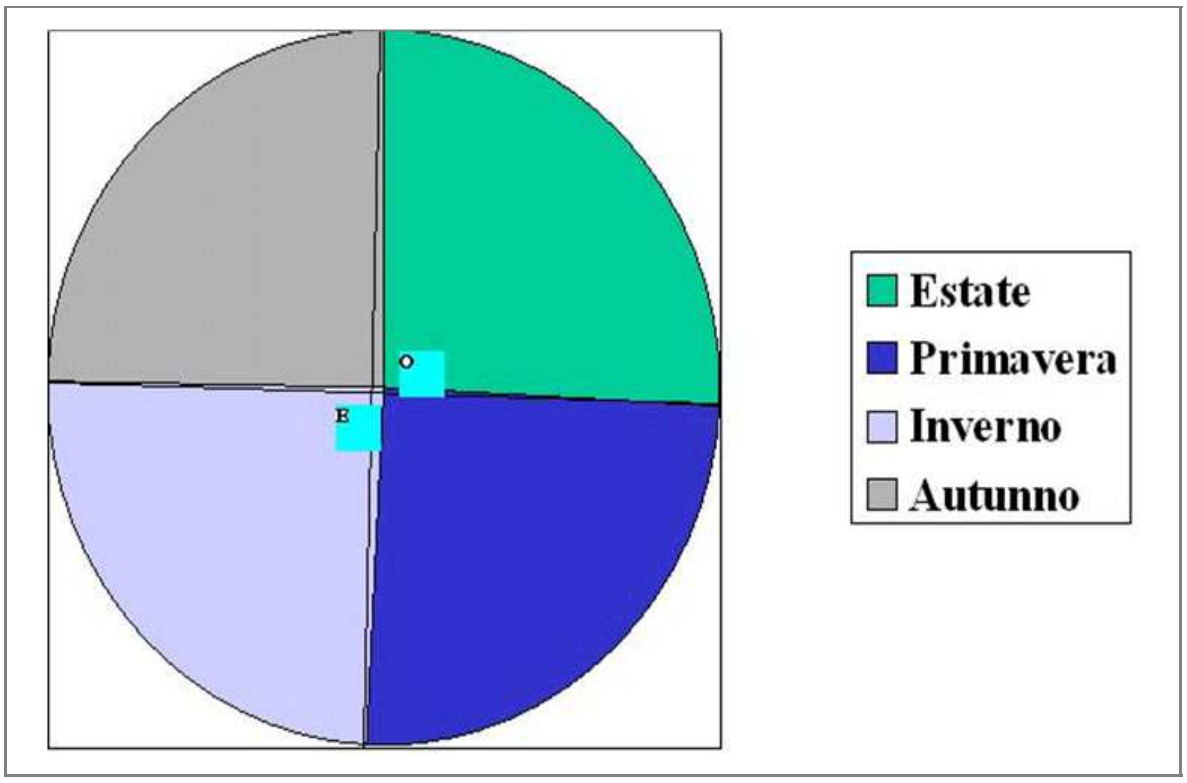

Il modello di Tolomeo prevede che il Sole (o la Terra) si muovano a velocità uniforme su un'orbita circolare e la Terra (o il Sole) non sia al centro di questa orbita, ma in posizione eccentrica a distanza e=0.0334.

È anche al bimillenario successo di questo modello che dobbiamo l'insegnamento del moto circolare uniforme nei corsi di fisica delle scuole superiori.

# Bisezione dell'eccentricità

Il modello tolemaico prevede anche che la distanza Sole-Terra cambi del ±3.34% rispetto al valore medio, e quindi anche il diametro angolare del Sole deve mostrare la stessa variazione percentuale.

Con le misure del passaggio al meridiano sia del diametro lungo la linea meridiana, sia dei tempi di contatto dei lembi paralleli alla linea, si poteva vedere se dal perigeo all'apogeo il diametro del Sole fosse variato complessivamente della quantità prevista da Tolomeo oppure no.

Il risultato delle osservazioni fu consistente con una variazione esattamente parti alla metà: ±1.67%, da questo il nome di "bisezione dell'eccentricità" che era sulla bocca di tutti gli astronomi del XVII secolo.

Lo Gnomone Clementino naturalmente confermava i dati del secolo precedente, sulla bontà dell'orbita kepleriana, ellittica, rispetto a quella tolemaica, circolare. La risoluzione angolare inferiore al minuto d'arco sulle misure del diametro consentiva questo proprio questo tipo di misura.

La prova definitiva della rivoluzione terrestre, che anche Galileo aveva cercato, sarebbe arrivata in due tempi: nel 1727, due anni prima della morte di Bianchini con la scoperta dell'aberrazione della luce (20.45") da parte di James Bradley, e nel 1838 con la prova osservativa della prima parallasse stellare (0.71") di 61 Cygni da parte di Friedrich Wilhelm Bessel.

# La variazione secolare dell'obliquità

L'obliquità è l'inclinazione dell'asse terrestre sul piano dell'orbita. È quello che determina il carattere delle stagioni: oggi vale poco più di 23°26', al tempo di Keplero e Tycho 23°30', al tempo di Tolomeo 23°50' ed in quello di Arato di Soli 24°. Maggiore è questo valore più estreme sono le stagioni, col Sole molto basso d'Inverno e molto alto d'Estate. L'effetto si somma alla lenta variazione secolare dell'eccentricità, ed alla

"rotazione degli apsidi" per cui l'apogeo si muove in avanti, ed era il 24 giugno al tempo di Tolomeo ed ora è il 4 luglio.

All'inizio del terzo millennio abbiamo eccentricità relativamente piccola ed apogeo nell'Estate, con obliquità moderata. Ci sono stati periodi preistorici in cui l'eccentricità era maggiore, l'apogeo cadeva in Inverno e l'obliquità era maggiore di quella attuale: gli inverni erano più lunghi e più freddi e gradualmente le terre emerse cominciarono a coprirsi di ghiacci. Durante le brevi estati il ghiaccio e le nevi, riflettendo bene la luce solare, rimandavano nello spazio una buona percentuale della radiazione solare incidente, contribuendo ad un progressivo raffreddamento del pianeta.

Questa teoria astronomica delle ere glaciali è quella di Milankovich, e lo Gnomone Clementino consente di verificare l'andamento dei parametri eccentricità, obliquità e data del perigeo e dell'apogeo (6 mesi dopo) con buona precisione da tre secoli a questa parte.

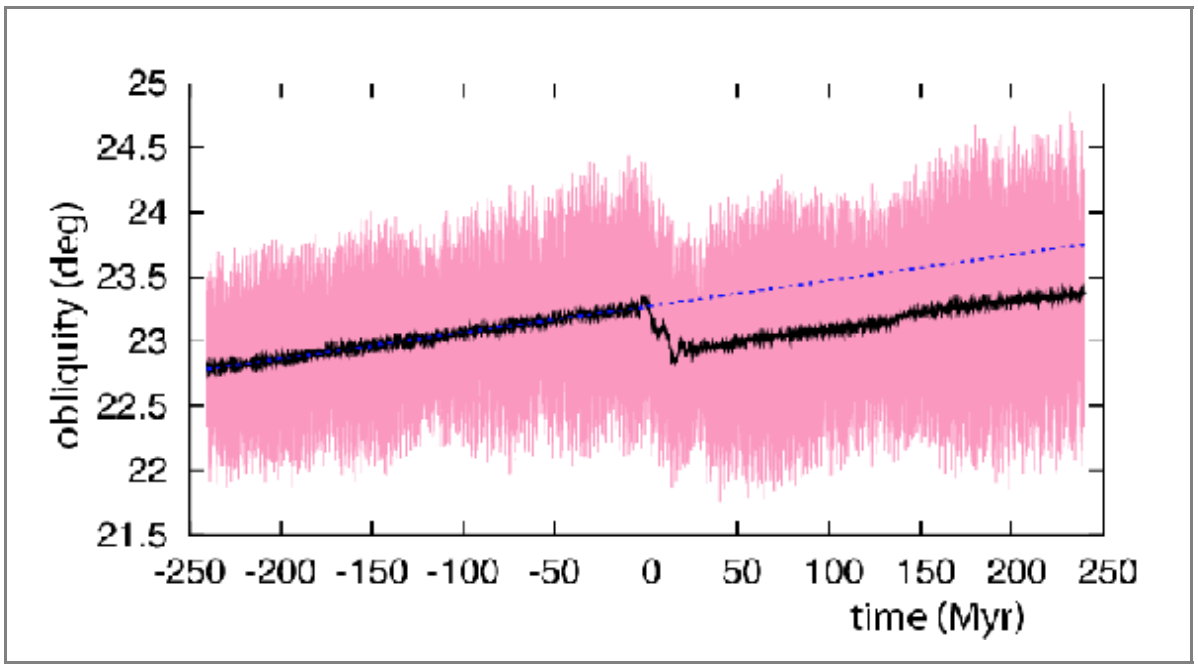

Qui vediamo l'andamento di obliquità ed eccentricità secondo il modello numerico di J. Laskar, et al. (2004).

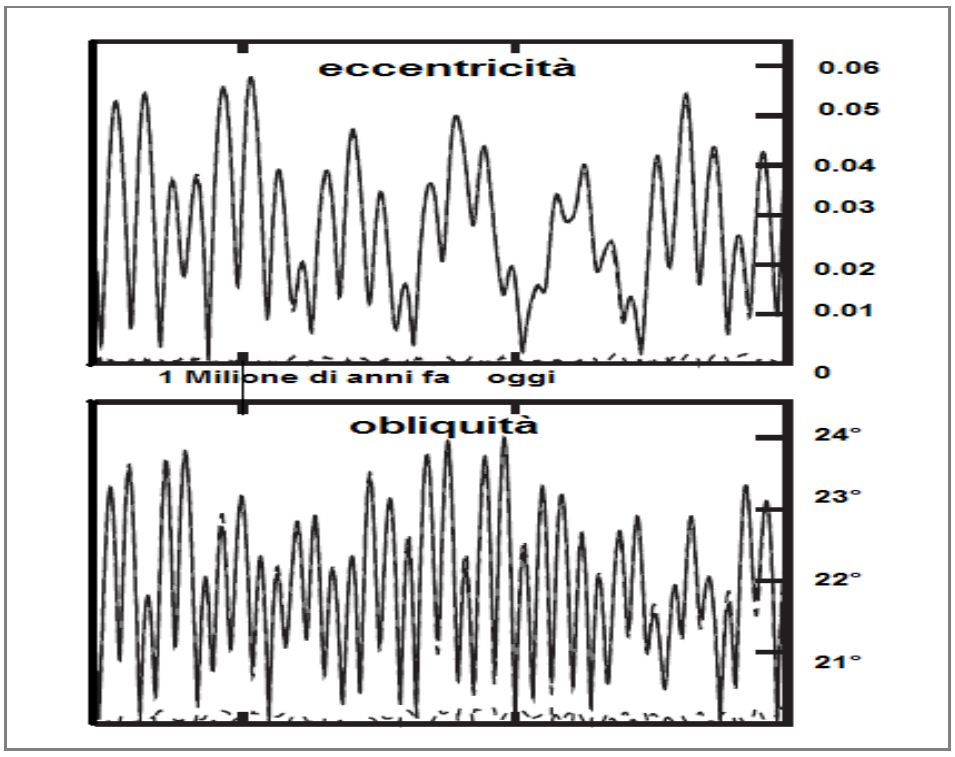

Bianchini aveva curato la stabilità del foro gnomonico, collocato nelle mura romane, e l'orizzontalità del piano della linea meridiana mediante una gigantesca livella idraulica estesa per tutta la lunghezza della linea. Ancora oggi possiamo vedere queste condutture presso il piede del foro stenopeico.

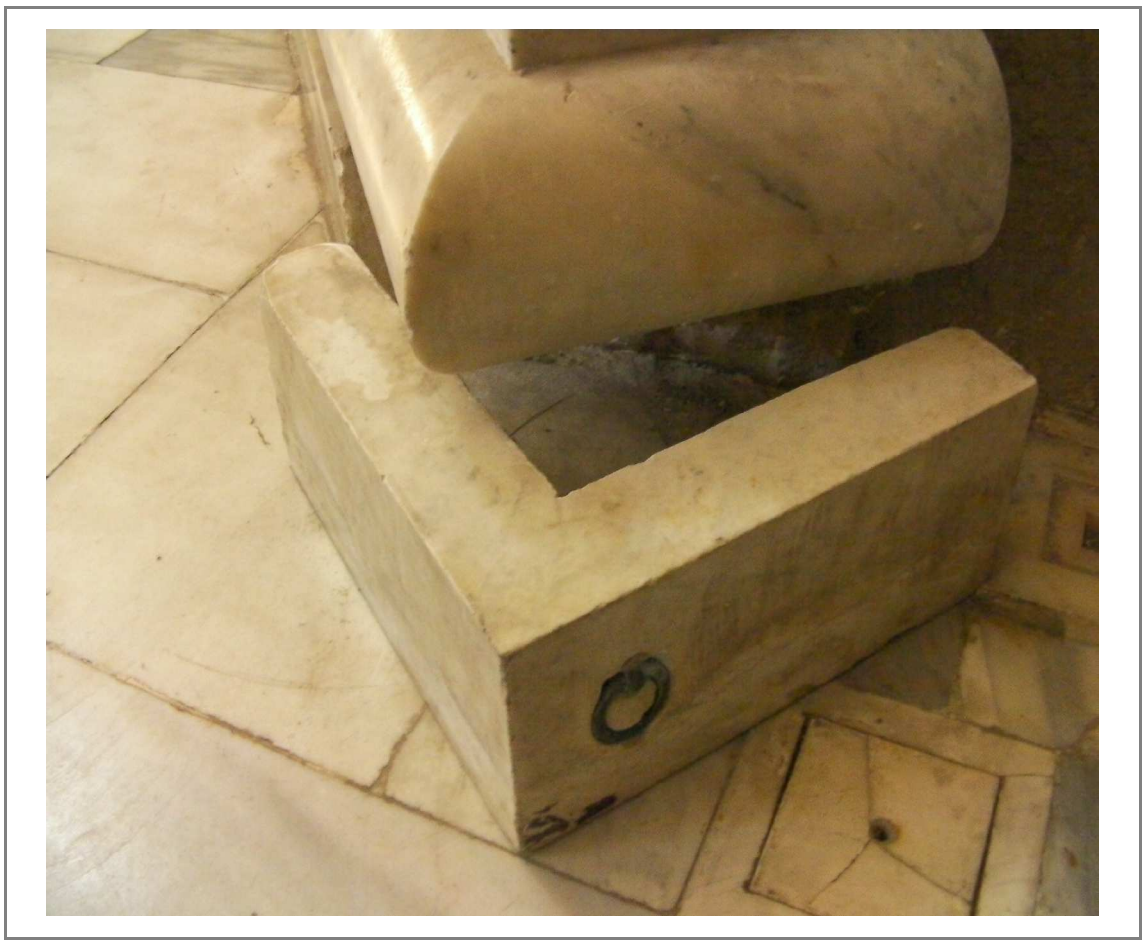

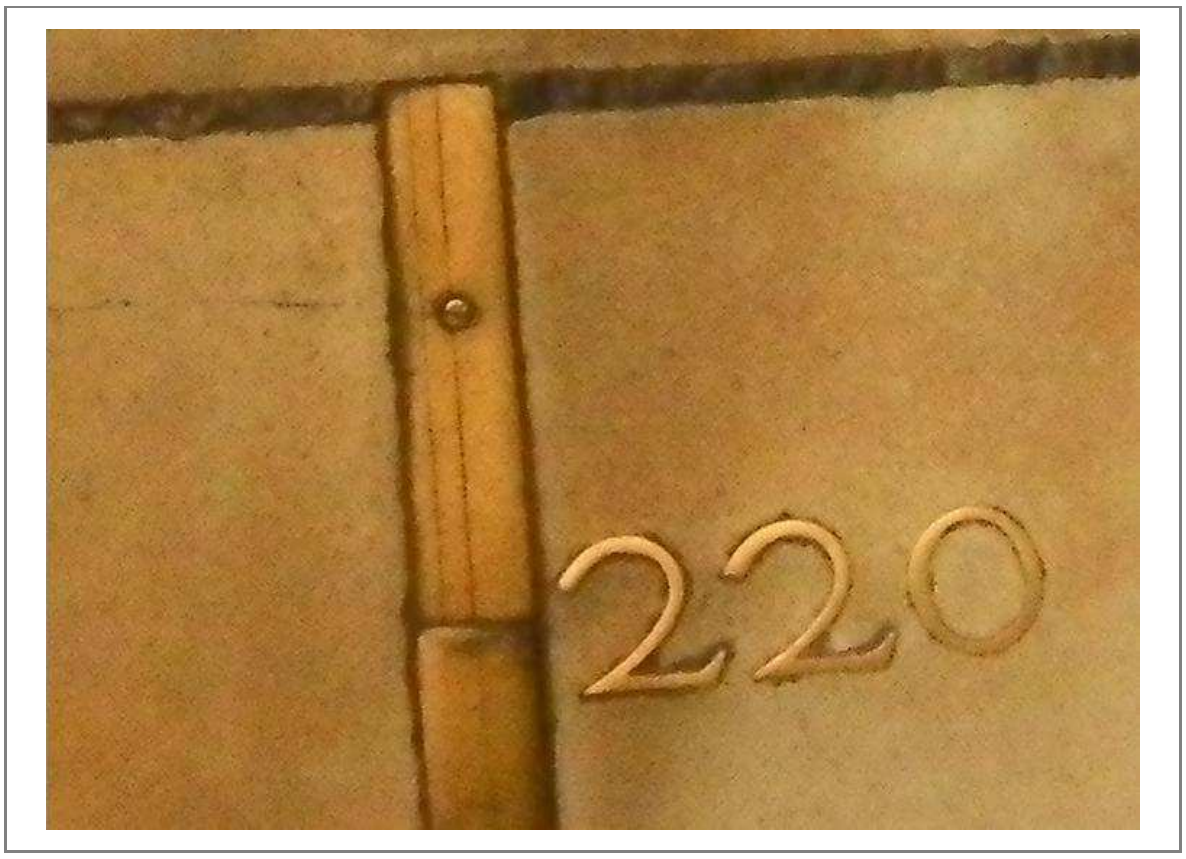

La distanza tra la tacca corrispondente alla parte centesima numero 220 e la fine della linea meridiana è di 13 cm. Oggi il Sole al solstizio d'Inverno lambisce la vite che vediamo nella foto. Tre secoli fa, lambiva il bordo nero superiore, 7 cm più in là poiché l'eccentricità valeva circa 23°28′.

A questo livello di precisione, dove un minuto d'arco corrisponde a 14 mm, l'oscillazione periodica dell'asse terrestre dovuta alla nutazione di circa ±9″ in 18.6 anni corrisponde a ± 2 mm, alla portata di questo strumento.

La nutazione fu scoperta solo nel 1737, ancora dall'astronomo reale James Bradley, che era succeduto ad Halley all'osservatorio di Greenwich. Il suo effetto ha influenzato i dati del Bianchini, il quale aveva misurato l'obliquità apparente ancora una volta con una precisione di 1″, degna di un osservatorio in piena regola.

# Misure di tempo ad alta risoluzione

Se oggi usassimo ancora il metodo di Bianchini per misurare il tempo, ovvero usando le stelle come riferimento, non potremmo accorgerci che la Terra sta rallentando il suo ritmo di rotazione. Infatti sia il Sole che le stelle ci appaiono transitare in ritardo della stessa quantità. Il ritardo accumulato tra il primo gennaio del 2006 ed il 31 dicembre 2008 è di circa 0.6 secondi rispetto ad un ritmo uniforme.

Per vedere questo fenomeno in Basilica dobbiamo avere un orologio atomico di riferimento, che per l'Italia è dato dal segnale dell'Istituto Elettrotecnico Nazionale Galileo Ferraris di Torino, e delle effemeridi molto precise sull'istante del transito del Sole al meridiano, ne esistono diverse disponibili sulla rete: Horizon, della NASA, quelle francesi dell'IMCCE, SPA del National Renewable Energy Laboratory, oltre al già citato Ephemvga a minore risoluzione.

L'istante del passaggio al meridiano si può determinare con una precisione molto alta usando una videocamera. Si dispone un set di linee parallele stampate su un foglio di carta, in modo che quella centrale corrisponda con la linea meridiana. Estendendo il metodo di Bianchini che media tra primo e secondo contatto, si fanno N medie incrociate che convergono tutte allo stesso istante in cui il centro del Sole è sulla linea.

# Misure di turbolenza atmosferica

La turbolenza atmosferica, produce un rapido movimento dell'immagine. Con una sequenza di foto durante il transito del 24 giugno 2006 abbiamo valutato l'ampiezza di queste deformazioni.

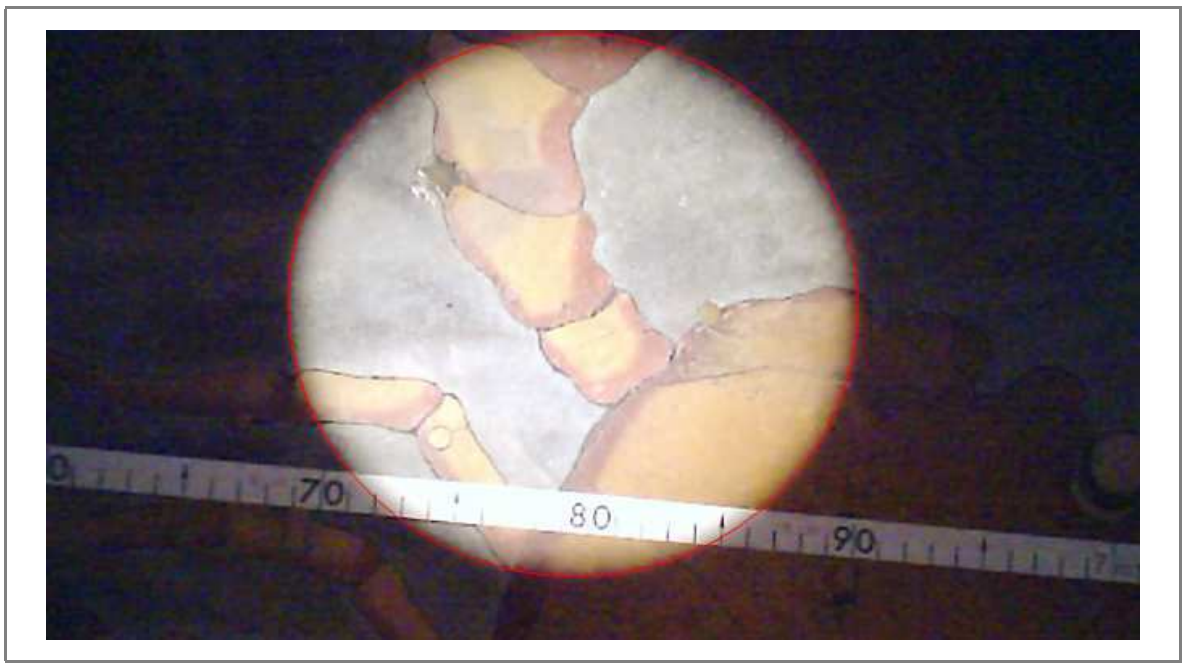

L'idea è molto semplice: durante il transito il disco solare intercetta la linea meridiana in due punti che si distanziano tra loro fino all'istante del mezzogiorno locale, per poi riavvicinarsi fino all'ultimo contatto del disco solare con la linea. Il punto medio tra questi due punti d'intersezione corrisponde sempre allo stesso punto che è il centro del Sole.

Le foto hanno messo in evidenza che l'immagine si muove in modo da dare un valore leggermente differente del centro ogni volta. La variazione di questo punto su 27 immagini ha mostrato una nuvola di punti di raggio 0.59 mm (deviazione standard) che corrispondono a 5.6 secondi d'arco.

In termini di tempo il contatto dell'immagine del lembo solare, turbolento, ha degli scatti casuali in avanti e indietro corrispondenti a ±0.37 s, visto che la velocità tipica con cui il

Sole si muove nel cielo, e quindi sul pavimento, è pari a 15″ al secondo. A causa della turbolenza c'è una incertezza ineliminabile di ±0.37 s sulla singola misura di transito, che diventa ±0.26 s = $\sqrt{(0.37^2+0.37^2)}/2$ sulla media di due contatti. E' da notare che l'uso di una videocamera su due soli tempi di contatto non migliora di molto la stima fatta ad occhio nudo di contatti al meglio di ±0.5 s, e di ±0.35 s sulla media dei due contatti.

Usando i transiti paralleli si può ridurre questa incertezza statisticamente. Con 10 linee parallele si scende a ±0.08 s, giungendo ad una precisione tale da poter vedere la derotazione terrestre.

Con la videocamera Sanyo CG9 che dà 60 fotogrammi al secondo abbiamo raggiunto questa precisione.

Per verificare la derotazione terrestre, una volta trovate le effemeridi affidabili, occorre un'ultima calibrazione, quella della rettilineità della linea meridiana.

## Misure della rettilineità con il LASER

Usando un LASER si crea una linea retta tramite la luce. Facendo in modo che il primo e l'ultimo punto coincidano con la linea meridiana, si possono vedere quanto i punti intermedi della linea si discostano dalla retta.

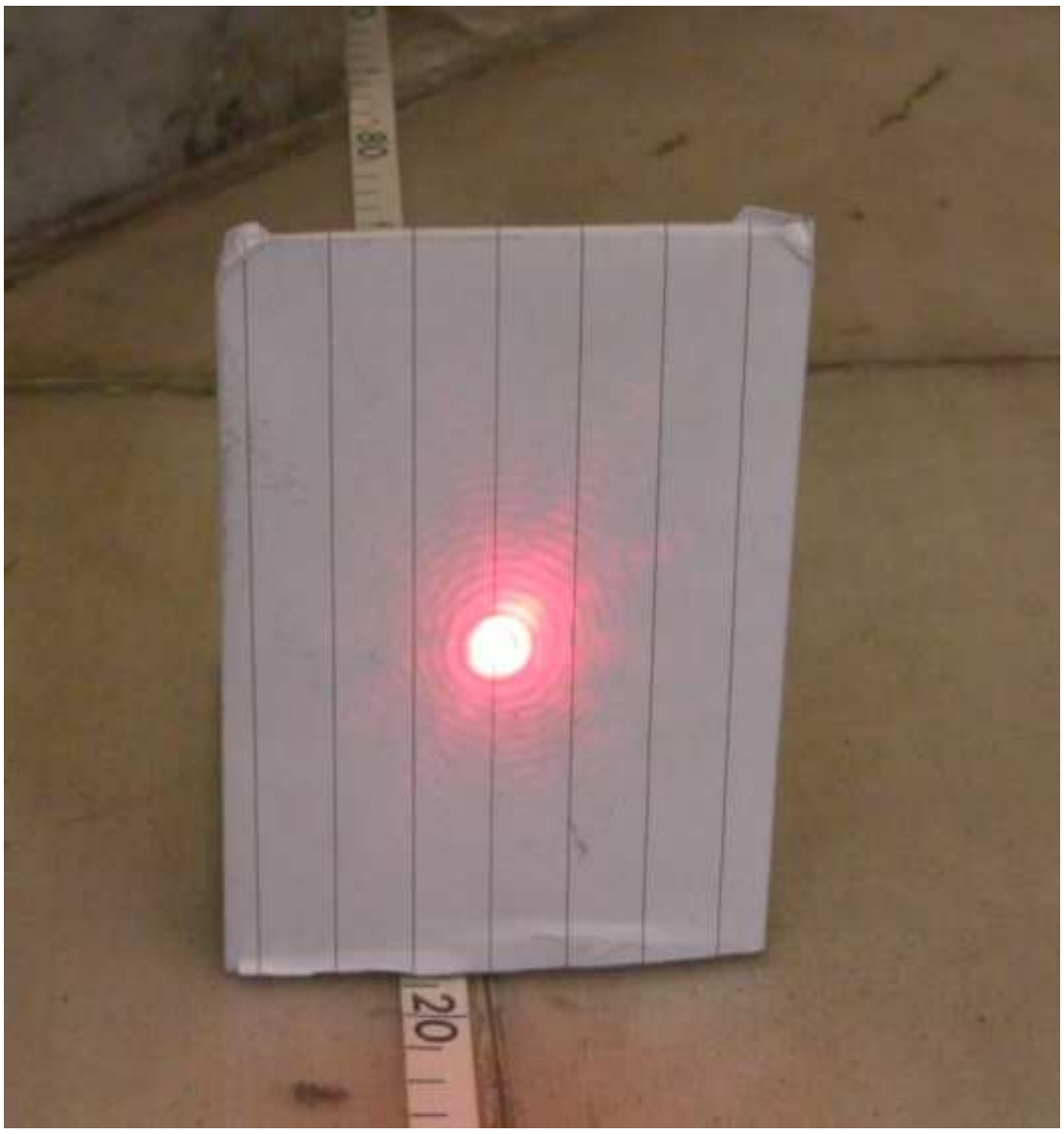

Infatti la Linea Clementina è composta da segmenti lunghi circa un metro fissati con delle viti al pavimento. Nei vari restauri questi segmenti non sono sempre stati ricollocati allo stesso punto, e alle giunzioni si trovano spesso delle piccole discontinuità.

Ad esempio presso le parti centesime 91-92 si trova una tacca esattamente in posizione intermedia, che testimonia il fatto che sia stata ricollocata in modo erroneo.

Già i Padri Gesuiti Ruggero Giuseppe Boscovich e Christopher Maire avevano fatto una ricognizione della Linea attorno al 1750, trovando un ampio "seno". La Linea Clementina è stata verosimilmente fin dall'inizio affetta da questa imprecisione.

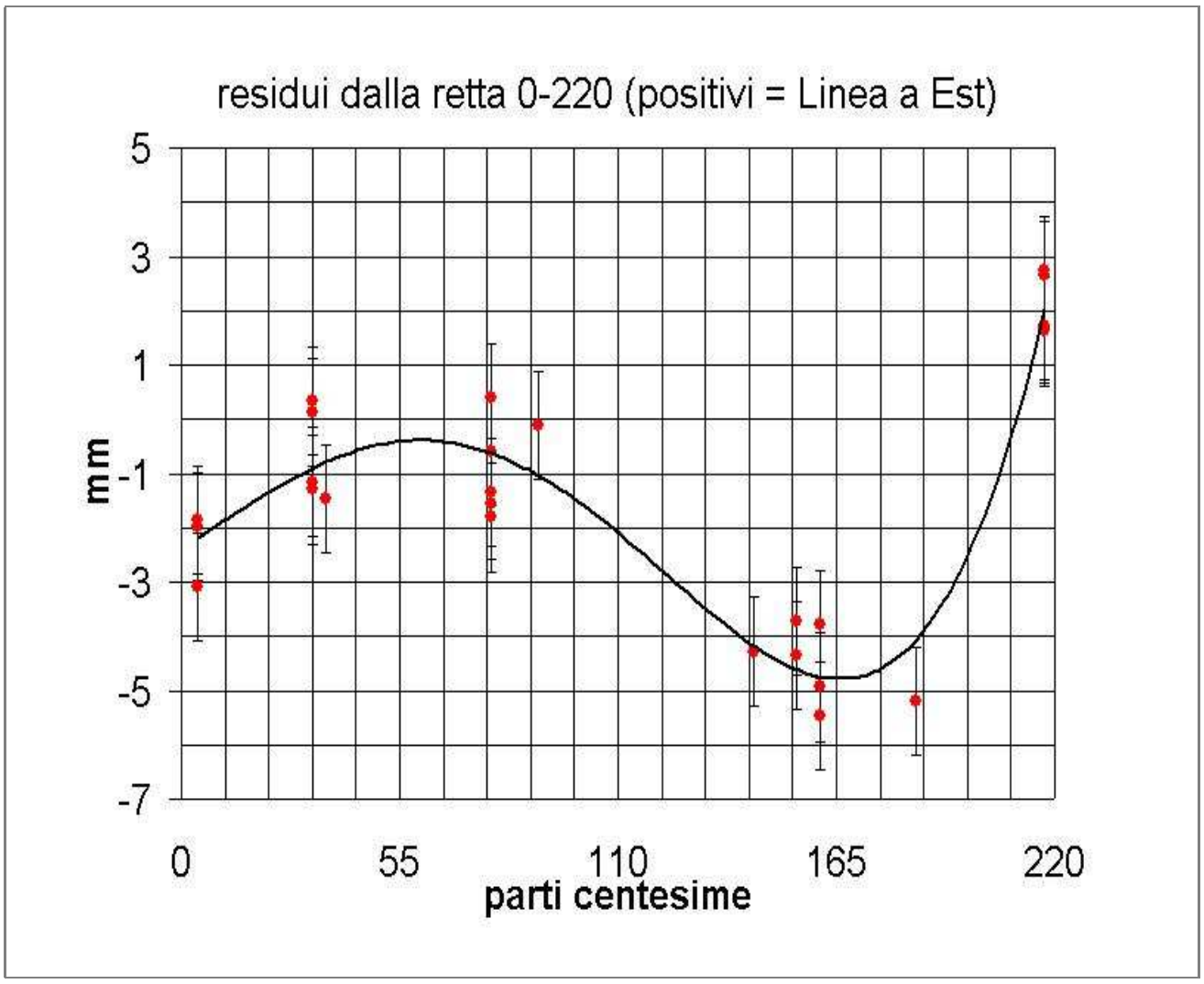

Deviazioni dalla retta, positive verso Est, negative ad Ovest.

Dopo il Boscovich queste sono le prime misure che mostrano quanto egli affermava. Il segno negativo indica la Linea che giace ad Ovest della retta LASER, mentre il positivo ad Est.

La retta LASER descrive la direzione 0-220 orientata 4′28.8″±0.6″ verso Est, individuata con il teodolite astro-referenziato con la stella Polare.

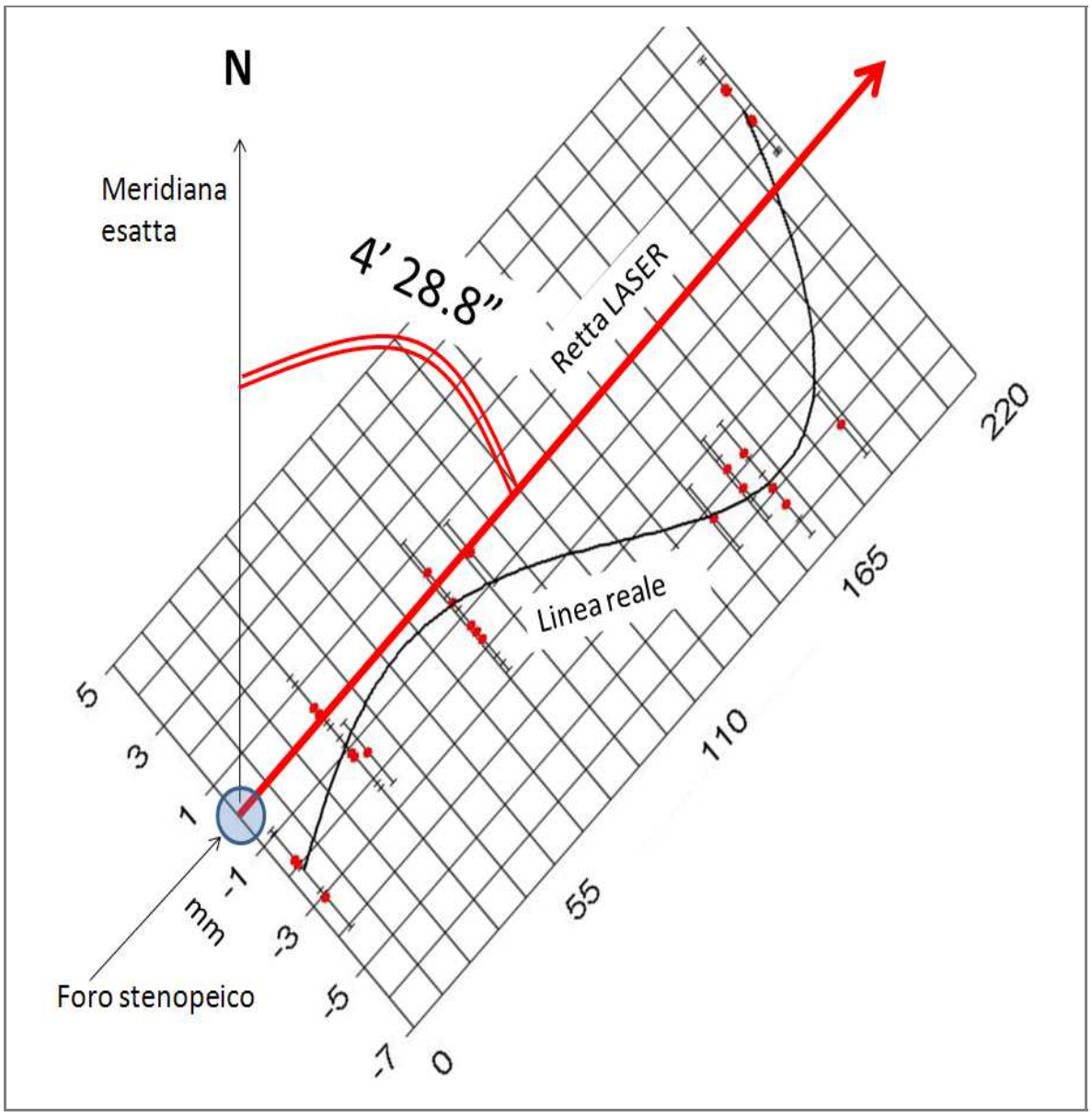

Questo schema è per distinguere la meridiana esatta dalla retta LASER e dalla linea reale.

Per poter fare delle misure astrometriche con lo Gnomone Clementino dobbiamo riportare le misure prese sulla Linea meridiana "reale" a quella esatta, diretta precisamente lungo il meridiano locale Nord-Sud.

Per poter svolgere questa operazione dobbiamo conoscere l'azimut della retta LASER e le deviazioni locali dalla retta LASER della linea reale.

# La misura del ΔUT1, dovuto alla derotazione terrestre

Prendiamo come esempio il 3 dicembre 2008, quando il centro del Sole è transitato sulla linea reale al punto 205.5 dello Gnomone Clementino alle 12:00:15.65 UTC. Le effemeridi IMCCE prevedevano il passaggio al meridiano esatto alle 11:59:57.87, 17.78 s prima. Localmente la linea reale si trova 1.35 mm a Ovest della retta LASER.

La velocità dell'immagine solare, diretta perpendicolarmente alla linea da Ovest verso Est, è per quel punto 3.13 mm/s, a tale velocità la deviazione locale si copre in 0.43 s. Dunque il ritardo complessivo misurato in due tempi (osservazione + deviazione locale) è 18.21 s, corrispondente a 57 mm percorsi a 3.13 mm/s.

La retta LASER, di cui conosciamo l'azimut esatto, a 205.5 parti centesime dell'altezza del foro, cioè 41807 mm, si discosta di 41807·tan(4'28.8")=54.5 mm. Restano 2.5 mm di differenza che alla velocità di 3.13 mm/s corrispondono a 0.80 s.

Questi 8 decimi di secondo sono il ritardo che il Sole ha accumulato sulle effemeridi, basate su un ritmo di rotazione terrestre costante, corrisponde al valore cercato di ΔUT1 ed è negativo a causa del ritardo.

Usando ephemvga risultano 0.05 s in più, mentre con le effemeridi della NASA 0.21 s in meno, le differenze dipendono dalle approssimazioni del moto solare adottate nelle varie effemeridi.

Il valore del ritardo di UT1 rispetto ad UTC per il 3 dicembre 2008 pubblicato dall'IERS è di ΔUT1=-0.564 s, compatibile con i risultati trovati in Basilica con il metodo qui esposto.

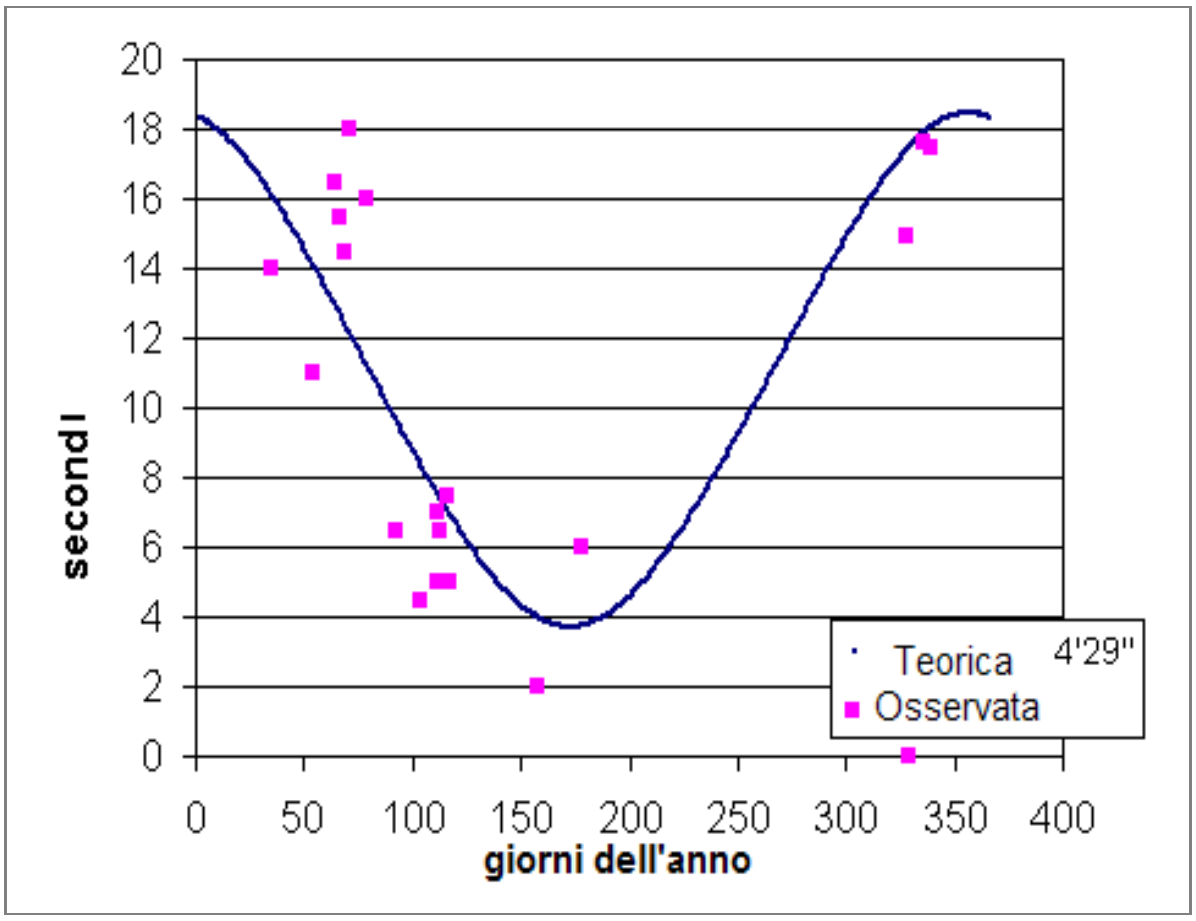

Questa figura rappresenta il ritardo del passaggio del Sole sulla Meridiana rispetto alle effemeridi. Nella fase discendente della curva si passa dalla posizione 217 al solstizio d'Inverno alla 33 al solstizio d'Estate. Rispetto alla curva corrispondente a 4' 29" di deviazione verso Est (retta LASER) abbiamo tempi maggiori nella prima parte (Linea ad Est della retta) ed inferiori nella seconda, con la Linea ad Ovest della retta, a causa delle deviazioni locali dalla rettilineità. Le misure in figura spaziano sull'anno 2005. Le misure riportate in figura sono state prese ad occhio nudo.

# La Meridiana Boreale

Uno dei tanti meriti degli studi di Catamo e Lucarini sulla meridiana è stato quello di puntare l'attenzione sulla meridiana boreale, e di aver individuato il piede della verticale del foro gnomonico boreale.

Ai rilievi effettuati nel 2005/2006 è risultato che la meridiana boreale, di cui esiste solo un breve tratto all'interno delle ellissi della stella Polare negli anni giubilari, non è parallela a quella australe, come invece sarebbe da aspettarsi.

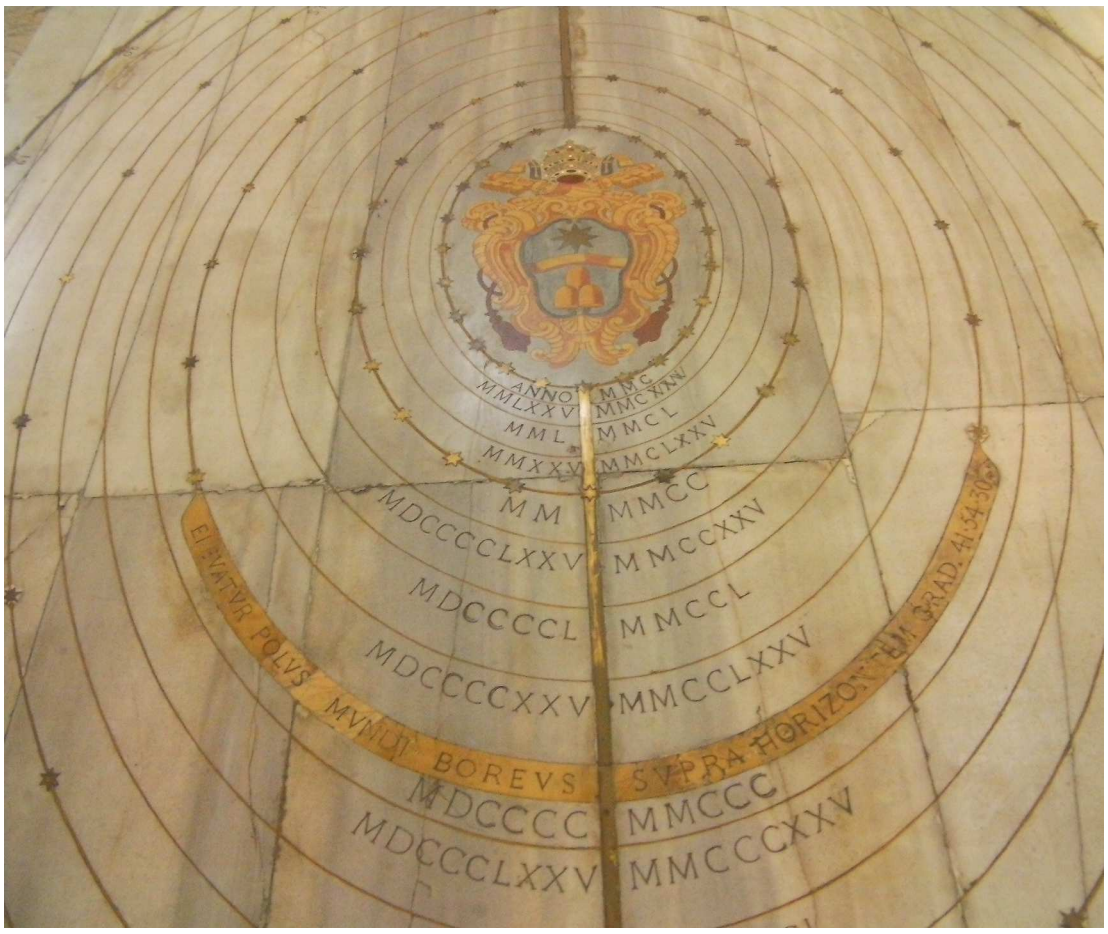

La linea boreale ha un azimut che devia di circa 3′ dal Nord, verso Est.

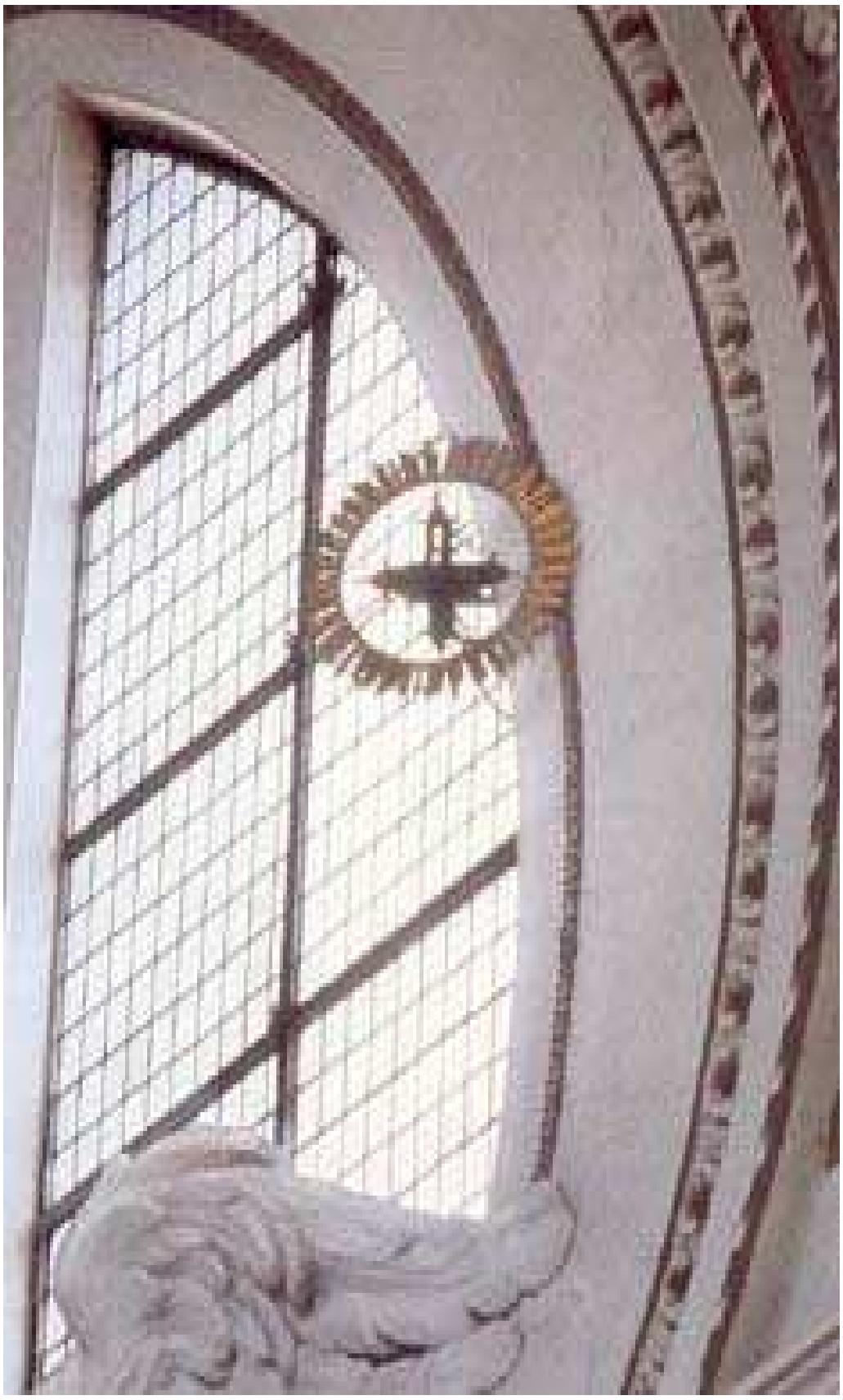

La ragione di questo errore non è nota. Boscovich attribuì alla debole vista di Bianchini gli errori che aveva riscontrato sulla linea, tuttavia questa ragione è da escludersi alla luce delle analisi svolte sulle misure di Bianchini relative alla latitudine e all'obliquità, precise al secondo d'arco.

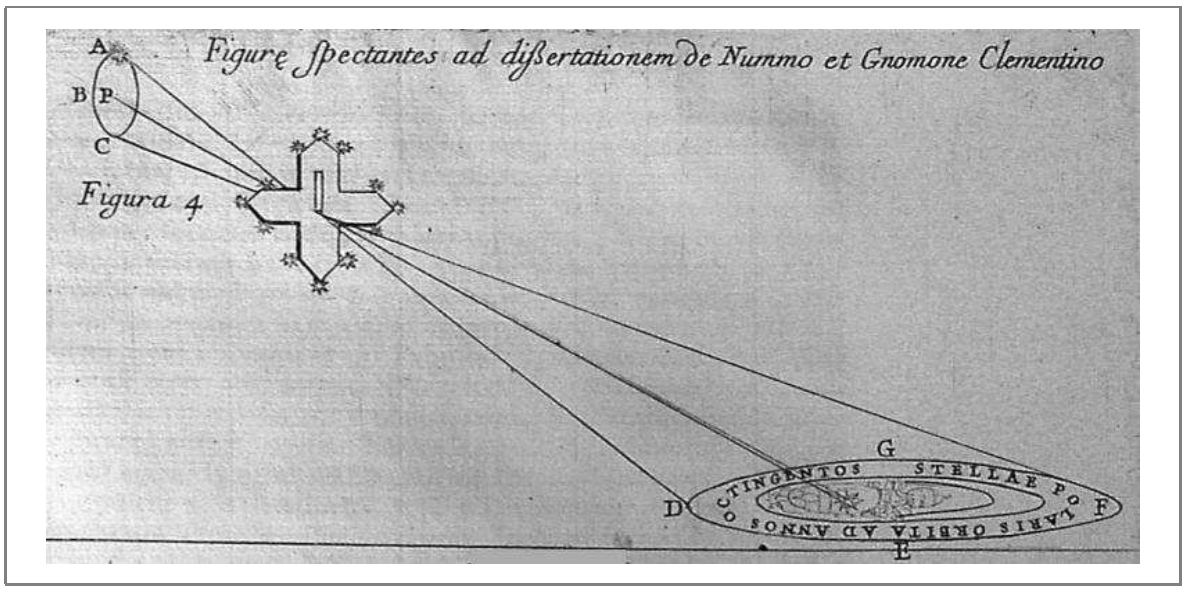

Forse fu un errore da attribuire semplicemente alle maestranze, oppure –ipotesi meno realistica– riflette errori di calcolo sulle effemeridi di Flaminio Mezzavacca o su quelle usate nella meridiana ausiliaria di Maraldi a S. Marco.

Attraverso la meridiana boreale si poté misurare la latitudine con il metodo sopra descritto dei transiti superiori ed inferiori della Polare, le cui altezze poi erano corrette per il coefficiente tabulato dal Cassini. Non sembra che questo gnomone sia stato usato per verificare con le massime elongazioni Est ed Ovest della Polare, all'epoca di oltre 2°, l'azimut della Linea.

# La data della Pasqua

L'alibi per la sopravvivenza dell'astronomia durante il medioevo è stato proprio il Computo Pasquale. Si sono cimentati su questi calcoli personaggi come Dionigi il Piccolo (VI secolo), al quale dobbiamo la cronologia ante-post Christum natum, al posto dell'era di Diocleziano, oppure Beda il Venerabile (725) che implementò l'uso.

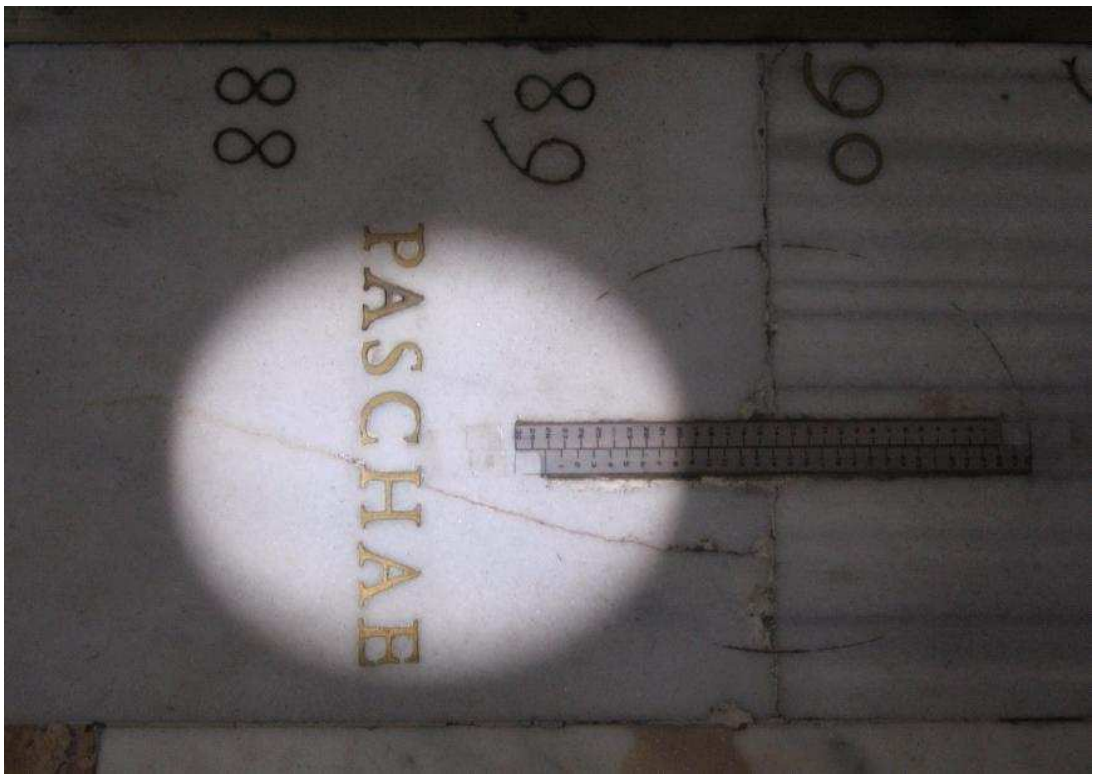

Immagine del 22 settembre 2007.

Con il Concilio di Nicea del 325 si era promulgato un solo criterio per il calcolo della Pasqua valido per tutta la Chiesa Universale: "Dominica proxima sequente lunam decimaquartam post vernum aequinoctium illuscentem, ab omnibus Ecclesiis Pascha celebratur."

La data dell'equinozio era stata fissato al 21 marzo dagli astronomi alessandrini chiamati in causa dai padri conciliari.

Dopo il fatto che l'anno giuliano di 365.25 giorni medi si discostava da quello tropico 365.2422 di 0.0078 giorni ogni anno, e quindi cumulativamente di 1 giorno ogni 128 anni, fece sì che nel 1582 si erano accumulati 10 giorni di anticipo dell'equinozio sulla data del 21 marzo.

La precessione degli equinozi, contrariamente ad un'opinione piuttosto diffusa, non ha nulla a che vedere con questo slittamento progressivo dell'equinozio, è solo una questione di approssimazione dell'anno tropico con l'anno civile.

Gregorio XIII con la bolla "Inter gravissima" del 24 febbraio 1582 decretò che si passasse dal 4 direttamente al 15 ottobre di quell'anno e, con l'eliminazione dei bisestili degli anni secolari non multipli di 400, si portasse l'anno gregoriano medio a 365.2425 giorni, più vicino al valore dell'anno tropico.

Bianchini volle verificare proprio il valore dell'anno tropico, e quello del mese sinodico, cruciali per conoscere l'equinozio vero. Ed il 1700 fu proprio il primo anno secolare a non essere bisestile, perciò fu di stimolo anche per Innocenzo XII Pignatelli ed il suo successore Clemente XI Albani.

L'indicazione del "Terminus Paschae" corrisponde al 22 marzo ed al 25 aprile, le date estreme possibili per la Pasqua. Quanto alla misura del mese sinodico, Bianchini le aveva eseguite con grande precisione, al telescopio.

## La Luna sulla meridiana

È possibile osservare sul pavimento della Basilica il transito della Luna, quando avviene di notte. Le prime foto con successo sono state effettuate il 19 dicembre 2007 da Pietro Oliva, tuttavia è chiaro che Bianchini non avrebbe potuto raggiungere la precisione del secondo sulla misura del mese sinodico osservando la Luna in proiezione come il Sole. Segue una sequenza di tre scatti durante il passaggio in meridiano della Luna. La Luna passava sul n. 52 a 26.5° dallo zenit. Sono le prime immagini di questo evento mai ottenute. Le pose durano alcuni secondi durante i quali la Luna si muove, la maggiore intensità è al centro.

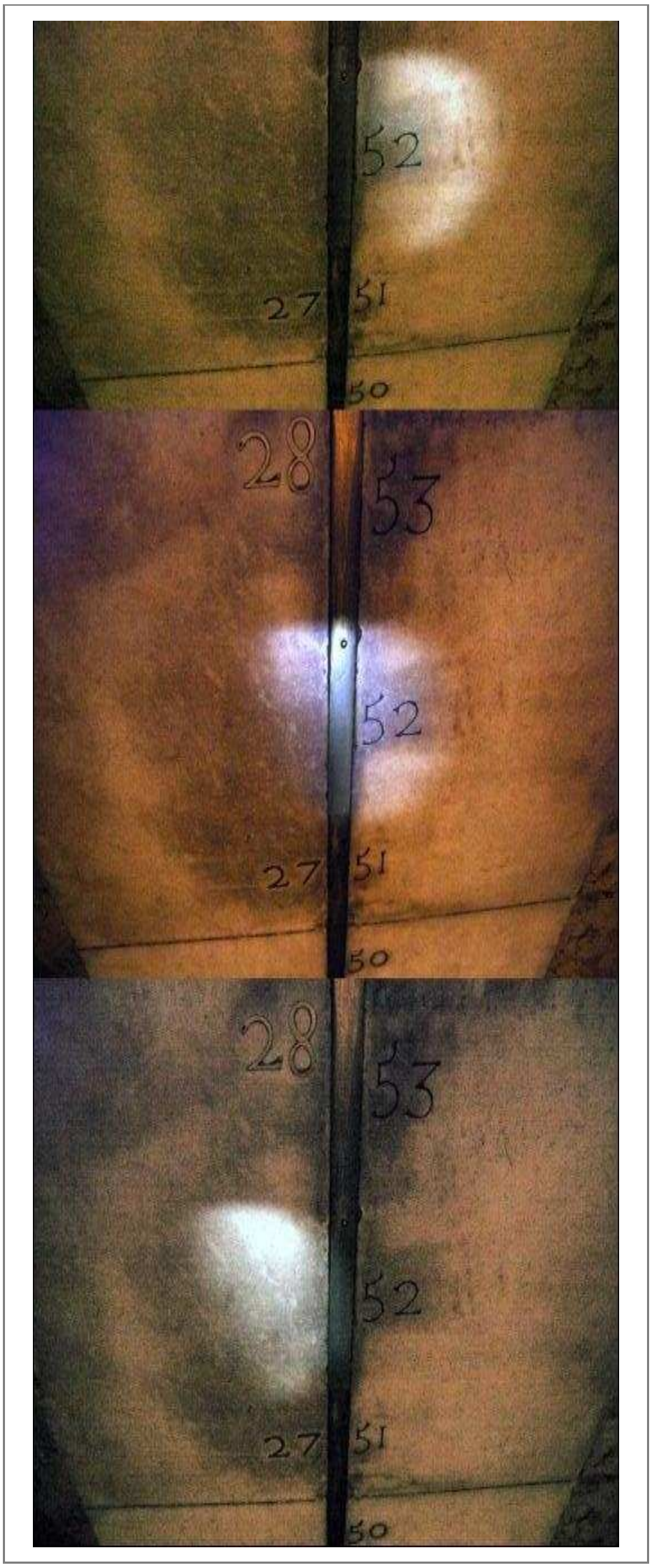

# L'equazione del tempo

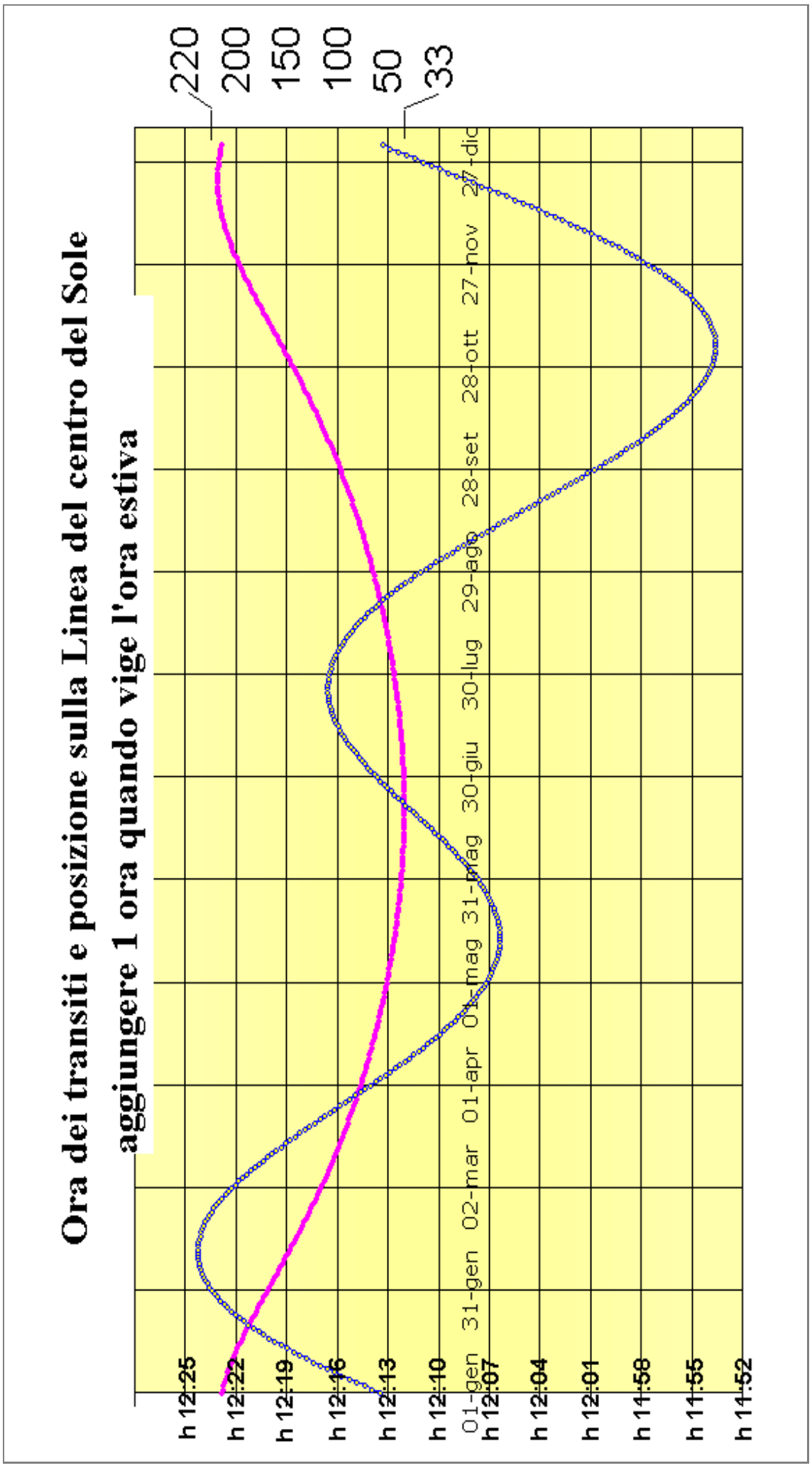

# Restauro del foro stenopeico

Per poter lavorare con lo Gnomone Clementino al pieno delle sue potenzialità sarebbe auspicabile completare il restauro dello strumento con il ripristino del foro stenopeico secondo le dimensioni originali, pari a 20 mm. Attualmente infatti il foro non è circolare, ed con uno schermo circolare di diametro 15.9 mm posto all'esterno si cerca di rimediare al problema.

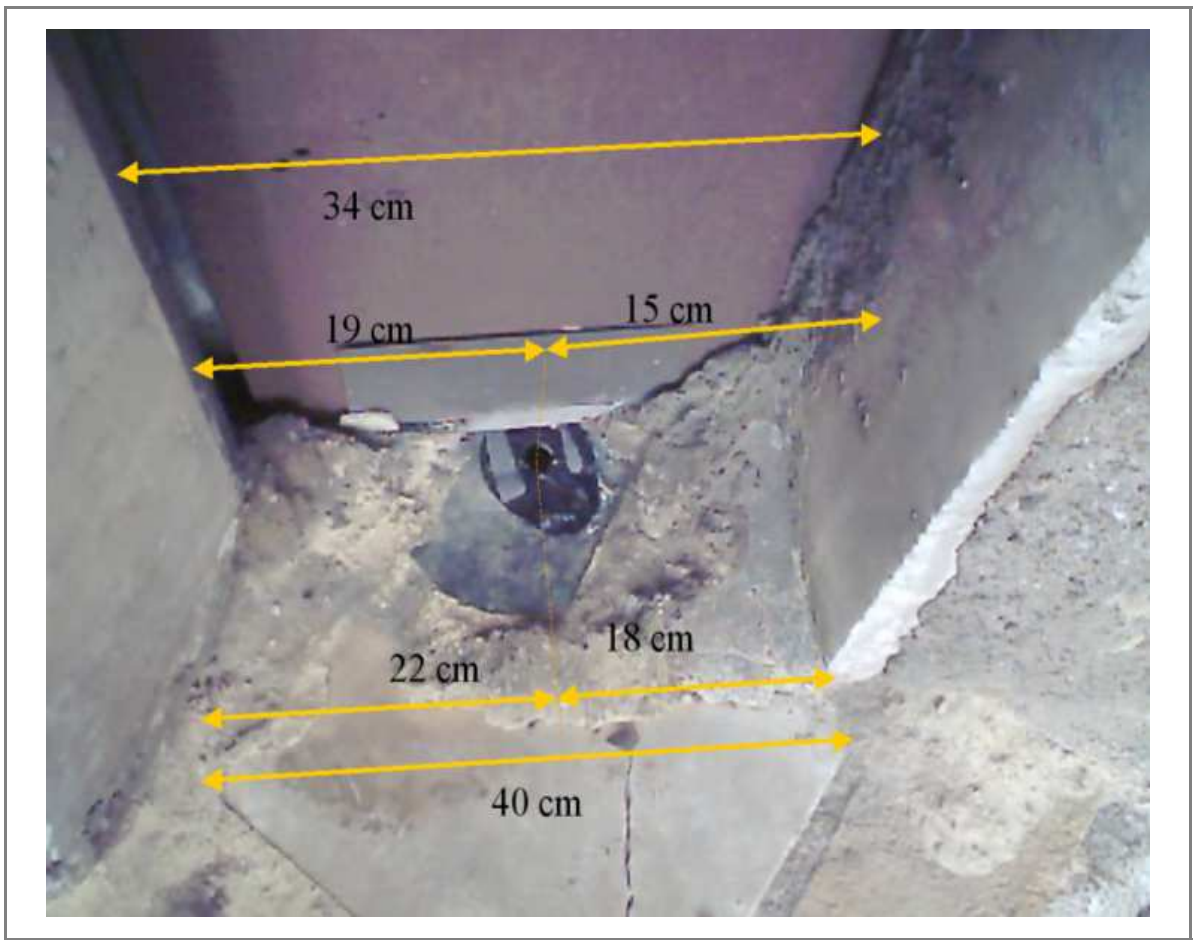

Lo schermo è completamente amovibile, ma proprio per quest ragione non si conosce con la precisione necessaria la proiezione a terra del suo centro, che è il parametro fondamentale per poter condurre gli studi di astrometria qui presentati.

Dato che lo Gnomone Clementino è uno strumento perfettamente funzionante, questo intervento insieme alla sua successiva calibrazione, ne garantirebbe un uso anche scientificamente corretto.

Con l'occasione il piano esterno del foro stenopeico potrebbe essere verniciato di bianco al biossido di titanio, come negli osservatori solari, per ridurre la turbolenza atmosferica locale.

Una soluzione avveniristica potrebbe essere quella di realizzare un revolver con fori coassiali, per poter tornare ad osservare anche le stelle con lo Gnomone Clementino attraverso il foro maggiore, e le macchie solari col minore.

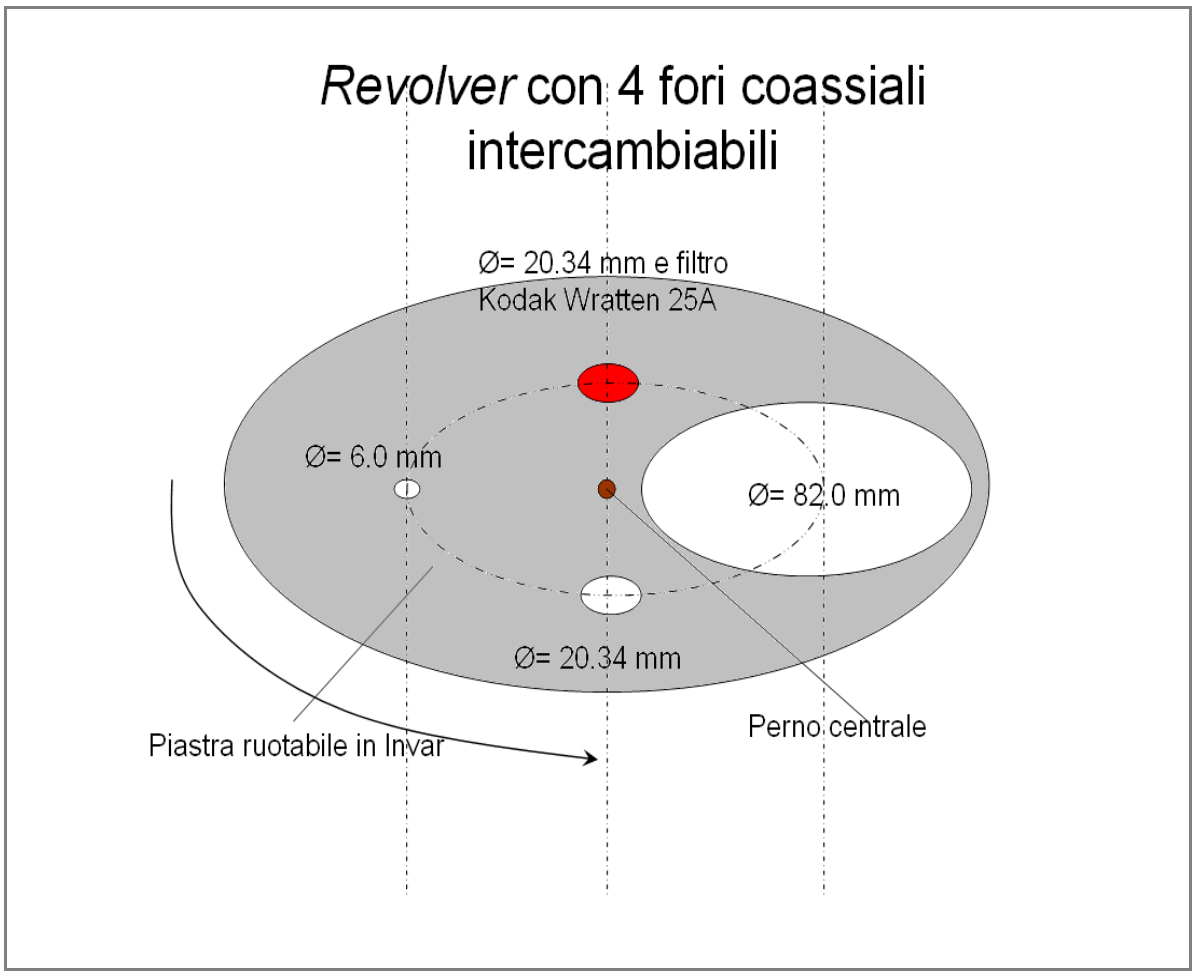

# Conclusioni e prospettive

All'inizio dei miei studi sulla meridiana ero stimolato dalla possibilità di osservare le macchie solari senza ottica, era il 2000, con i lavori di restauro ancora in corso. Il libro di John Heilbron, The Sun in the Church fu un'ottima guida. Dopo, progressivamente, è stato l'aspetto astrometrico e sperimentale a conquistarmi, e l'approccio storico-documentale è stata una conseguenza necessaria.

Così ho affrontato il problema dell'azimut della Linea Clementina, mettendo in evidenza le analogie con strumenti analoghi in Italia. Questo problema non è ancora risolto definitivamente, così come resta insoluta la ragione dell'esistenza di un settore di cerchio a 34 m 80 cm dal piede del foro stenopeico, individuato da 7 stelle sul pavimento. E' la proiezione attraverso il foro di una parte del cerchio d'altezza, o almucantarat, corrispondente a 30°18' sull'orizzonte.

Il metodo dei transiti paralleli ed il "pinhole solar monitor" di S. Maria degli Angeli sono i prototipi del progetto CLAVIUS –presentato nel 2008– di misura del diametro del Sole da transiti su cerchi orari, in collaborazione con l'IRSOL, Istituto Ricerche Solari di Locarno, la SUPSI, Scuola Universitaria Professionale della Svizzera Italiana di Lugano, l'Università di Como e la Sapienza, che si affiancherà alla missione spaziale PICARD. Le tecniche di ripresa video sviluppate in Basilica, sincronizzate con il tempo universale coordinato, sono state esportate in occasione delle missioni osservative delle eclissi totali e anulari di Sole in Spagna, Egitto e Guyana Francese, dedicate alla misura del diametro solare a precisioni del centesimo di secondo d'arco.

La meridiana di Santa Maria degli Angeli è dunque stata una palestra per tutti questi studi di astrometria solare di precisione, ed ancor più lo sarà per gli studenti delle Università e delle scuole secondarie e primarie di Roma e di tutta Italia che vengono a visitare sempre più numerosi ed interessati questo strumento.

Tutti i movimenti della Terra possono essere visualizzati facilmente con lo Gnomone e la comprensione dei fondamenti dell'astronomia diventa così solidamente basata anche sull'esperienza personale.

## Ringraziamenti

A Mario Catamo e Cesare Lucarini che mi hanno preceduto ed illuminato nello studio di questo magnifico strumento.

A Mons. Giuseppe Blanda e Mons. Renzo Giuliano che mi hanno consentito ed incoraggiato tante misure in Basilica, fino a collocare la S. Messa quotidiana alle 12:30, che in ora solare segue sempre il transito meridiano ed in ora legale lo precede.

Allo Studio di Archittettura MCM di Monica Cola, ed ad Alessandro Lupi, che hanno realizzato le misure di calibrazione dell'azimut della Linea Clementina con la stella Polare nel febbraio 2006, con un teodolite elettronico Leica TCR703.

A gli studenti dei corsi tenuti alla Sapienza a partire dal 2002, ai professori Cosimo Palagiano e Paolo de Bernardis della Sapienza, che hanno accolto con entusiasmo il mio contributo nei corsi di Laurea in Geografia e Fisica -indirizzo Astrofisica.

Al Preside dell'Istituto di Studi Superiori Alessandro Volta, Ing. Giorgio Sordello, per aver consentito alle mie classi di visitare ripetutamente lo Gnomone ed esercitarsi nei rilievi metrici e cronometrici.

A Christopher Schaefer e Lisa Hoffer che mi hanno ospitato a New Haven durante l'editing di questo libro, nel maggio 2014.

# Referenze

Aristotele, Problemata, III, 15

Plinio il Vecchio, Naturalis Historia, XXXVI , 72

Claudio Tolomeo, Almagesto, Traduzione di G. J. Toomer (1998), Ptolemy's Almagest, Princeton University Press, NJ USA.

Egnazio Danti (1576), Usus et Tractatio Gnomonis Magni, Bononiae.

Flaminio Mezzavacca (1701) Otia sive ephemerides Felsinae, Bonomiae.

Francesco Bianchini (1703), De Nummo et Gnomone Clementino, Romae.

Ruggero Giuseppe Boscovich (1773), Lettera a ad A. Vallisneri del 25 agosto 1773, in Germano Paoli, Ruggero G. Boscovich nella scienza e nella storia del '700, Accademia Nazionale delle Scienze, Roma 1988, p. 283.

Carlo Ferrari da Passano, Carlo Monti, Luigi Mussio (1977), La Meridiana Solare del Duomo di Milano, Verifica e Ripristino nell'anno 1976, Veneranda Fabbrica del Duomo di Milano.

Girolamo Fantoni (1979), La Grande Meridiana di S. Maria degli Angeli, in Atti dell'Istituto Italiano della Navigazione, Roma.

Dorrit Hoffleit (1982), The Bright Star Catalogue, Yale University Observatory, New Haven CT USA.

Peter Duffett-Smith (1983), Astronomia Pratica, Sansoni (Firenze) p. 38.

Ian Ridpath ed. (1989), Norton's 2000.0 Star Atlas and Reference Handbook, Longman, Essex GB.

Giuseppe Monaco, L'Astronomia a Roma (1990), dalle origini al novecento, Osservatorio Astronomico di Roma.

Elwood Charles Downey (1992) Ephemvga 4.27, software free su www.santamariadegliangeliroma.it menù meridiana, calcolo delle effemeridi.

John L. Heilbron (1999), The Sun in the Church, Harvard University Press, Boston MS USA ora anche in Italiano "Il Sole nella Chiesa" (2005) Compositori, Bologna.

Cesare Barbieri (1999), Lezioni di Astronomia, Zanichelli (Bologna)

Alberto Peratoner (2000), L'Orologio della Torre di San Marco in Venezia, Cafoscarina, Venezia.

Giovanni Paltrinieri (2001), La Meridiana della Basilica di San Petronio in Bologna, Centro Editoriale S. Stefano, Bologna.

Mario Catamo e Cesare Lucarini (2002), Il Cielo in Basilica, ARPA AGAMI, Roma.

Costantino Sigismondi (2002), Measuring the angular solar diameter using two pinholes, American Journal of Physics **70**, p. 1157-1159.

Costantino Sigismondi (2003), Solstizi equinozi e bisestili in Basilica, Astronomia nella Basilica di Santa Maria degli Angeli, Geografia, 101-102, p. 2-8.

Jacques Laskar et al., (2004), A long-term numerical solution for the insolation quantities of the Earth, Astronomy and Astrophysics **428**, 261-285.
http://www.aanda.org/content/view/87/42/lang,en/

Costantino Sigismondi ed. (2006), Meridiani e Longitudini a Roma, Università La Sapienza, Roma.

Michel Rougé (2006), Le gnomon de l'église Saint-Sulpice, Paroisse Saint-Sulpice, Paris.

Costantino Sigismondi (2006), Le Meridiane nella Chiesa, in Amici dei Musei 105-106 p. 196-209, Polistampa, Firenze.

Fabrizio Bònoli, Gianluigi Parmeggiani e Francesco Poppi (2006), Atti del Convegno Il Sole nella Chiesa: Cassini e le grandi meridiane come strumenti di indagine scientifica, Giornale di Astronomia 32, Istituti Editoriali e Poligrafici Internazionali, Pisa-Roma.

Simone Bartolini (2006), I Fori Gnomonici di Egnazio Danti in Santa Maria Novella, Polistampa, Firenze.

Costantino Sigismondi (2006), Pinhole Solar Monitor Tests in the Basilica of Santa Maria degli Angeli in Rome, in Solar Activity and its Magnetic Origin – Proc. IAUC233 Cairo, March 31-April 4, 2006. V. Bothmer and A. Hady eds. Cambridge University Press (2006) p. 521-522.

Costantino Sigismondi (2006), Misura della declinazione magnetica a Roma usando l'azimut del Sole*, e La Meridiana e la Relatività, in Meridiane e Longitudini a Roma, Semestrale di Studi e Ricerche di Geografia, Università "La Sapienza", p. 164-169 e 194-203.
*Metodo per misurare l'anomalia magnetica confrontando misure di bussole e Sole.

Magdalena Bedynski, Monica Nastasi (2007), Le Stelle e la Meridiana, www.santamariadegliangeliroma.it menù meridiana, stelle e meridiana.

Juergen Giesen (2007)
http://www.jgiesen.de/SiderealTimeClock/index.html è un Applet Java per il calcolo del Tempo Siderale per ogni luogo con precisione al centesimo di grado di longitudine.

Licia Mangione (2007), Il Sistema Tolemaico e quello Copernicano al vaglio delle grandi meridiane, Tesi di Laurea di primo livello, Facoltà di Lettere e Filosofia Dipartimento di Geografia Umana.

Costantino Sigismondi (2007), La Terra come Osservatorio Astronomico: la Correzione al Raggio Solare Medio nell'Eclissi Anulare del 22 Settembre 2006, Semestrale di Studi e Ricerche di Geografia 2007 (1), Università "La Sapienza", 151-164.

US Naval Observatory Time Service (2008)
http://tycho.usno.navy.mil/sidereal.html permette il calcolo del TS apparente istantaneo (include l'equazione degli equinozi).

Roberto Buonanno (2008), Il cielo sopra roma, i luoghi dell'astronomia, Springer-Verlag Italia, Milano.

Simone Bartolini (2008), Gli Strumenti Astronomici di Egnazio Danti e la Misura del Tempo in Santa Maria Novella, Polistampa, Firenze.

Costantino Sigismondi (2008), Effemeridi, Introduzione al Calcolo Astronomico, Ateneo Pontificio Regina Apostolorum.

Costantino Sigismondi (2008), Astronomia di Posizione per muoni, Pontificio Regina Apostolorum Roma, insieme al file EEEtest.xls Simulatore di dati EEE,
http://www.upra.org/articulo.phtml?se=11&id=2644

I parametri utilizzati sono quelli delle coordinate geografiche dell'Aquila. Lo stesso file è presente anche su
www.santamariadegliangeliroma.it menù Meridiana, Astronomia di Posizione: l'obiettivo dello Gnomone Clementino è fisso ed il cielo che vi si può vedere attraverso cambia ogni istante. L'identificazione delle coordinate assolute e del tempo siderale è utile anche per ricavare le coordinate del Sole al momento del transito meridiano.

IERS International Earth Rotation Service, bollettino C del 4/7/2008 http://hpiers.obspm.fr/iers/bul/bulc/bulletinc.dat in futuro questo link terminerà con /bulletin.37

http://www.iers.org/products/13/1388/orig/finals2000A.daily
su questo sito sono i dati giornalieri del ΔUT1.

Costantino Sigismondi (2008), Stellar and Solar Positions in 1701-3 observed by F. Bianchini at the Clementine Meridian

Line e Solar Radius at the End of Cycle 23, in Proc. XI Marcel Grossmann Meeting on General Relativity, Berlin July 22-29, 2006 R. Ruffini R. T. Jantzen and H. Kleinert eds. World Scientific Publisher, p. 571-573 e 605-607 (2008).

Costantino Sigismondi, Michele Bianda e Jean Arnaud (2008), European Projects of Solar Diameter Monitoring, American Institute of Physics Conference Proceedings **1059**, p. 189-198.

Ali Kilcik, Costantino Sigismondi, Jean Pierre Rozelot, Konrad Guhl (2008), Determination from Total Solar Eclipse of March, 29 2006, Solar Physics ms. SOLA639R2 accepted for publication.

Costantino Sigismondi (2009), Misura del Ritardo Accumulato dalla Rotazione Terrestre, ΔUT1, alla Meridiana Clementina della Basilica di Santa Maria degli Angeli in Roma, in Atti del Convegno Mensurā Caeli, ed. Manuela Incerti, University of Ferrara Press.

Costantino Sigismondi (2014), La Meridiana nella Torre dei Venti in Vaticano, GERBERTVS 7, 81.

# Indice